\documentclass[twocolumn]{aastex701}

\usepackage[utf8]{inputenc}
\usepackage{newunicodechar}
\newunicodechar{✱}{\ensuremath{*}}
\usepackage{pifont}
\usepackage{needspace}
\usepackage{amsmath} 
\usepackage{changepage}
\usepackage{float}

\newcommand{\mjup}{\ensuremath{\mathrm{M_{jup}}}}

\newcommand{\paperI}{Paper~I}

\usepackage{fontawesome5}

\newcommand{\nrc}{NRC Herzberg Astronomy and Astrophysics,
5071 West Saanich Road,
Victoria, BC, V9E 2E7, Canada}
\newcommand{\ucsd}{Department of Astronomy \& Astrophysics,  University of California, San Diego, La Jolla, CA 92093, USA}
\newcommand{\nmsu}{Department of Astronomy, New Mexico State University, 1320 Frenger Mall, Las Cruces, NM 88003, USA}
\newcommand{\uvic}{Department of Physics and Astronomy, University of Victoria,
3800 Finnerty Road, Elliott Building,
Victoria, BC, V8P 5C2, Canada}
\newcommand{\mac}{Department of Physics \& Astronomy, McMaster University, 1280 Main Street West, Hamilton, ON, L8S 4L8, Canada}
\newcommand{\caltech}{Department of Astronomy, California Institute of Technology, Pasadena, CA 91125, USA}

\newcommand{\ucsc}{Department of Astronomy \& Astrophysics, University of California, Santa Cruz, CA 95064, USA}
\newcommand{\ucsb}{Department of Physics, University of California, Santa Barbara, CA 93106, USA}

\begin{document}

\title{Joint Gaia-Hipparcos astrometry, RV, and Imaging posteriors from the G23H Hipparcos sample}

\author[0000-0001-5684-4593]{William Thompson}
\affiliation{\nrc}
\email[show]{william.thompson@nrc-cnrc.gc.ca}

\author[0000-0001-7402-8506]{Alexandre Bouchard-Côté}
\affiliation{Department of Statistics, University of British Columbia, Vancouver, BC V6T 1Z4, Canada}
\email{bouchard@stat.ubc.ca}

\author[0000-0001-9582-4261]{Dori Blakely}
\affiliation{\uvic}
\affiliation{\nrc}
\email{dblakely@uvic.ca}  

\author[0000-0003-1499-0191]{Trevor Campbell}
\affiliation{Department of Statistics, University of British Columbia, Vancouver, BC V6T 1Z4, Canada}
\email{trevor@stat.ubc.ca}

\author[0000-0002-6618-1137]{Jerry W. Xuan}
\altaffiliation{Heising-Simons Foundation 51 Pegasi b Fellow}
\affiliation{\ucsb}
\email{jxuan@ucsb.edu}  

\author[0000-0002-2696-2406]{Jingwen Zhang}
\affiliation{\ucsb}
\email{jwzhang@ucsb.edu}

\author[0000-0003-4557-414X]{Kyle Franson}
\altaffiliation{NHFP Sagan Fellow}
\affiliation{\ucsc}
\email{kfranson@ucsc.edu}

\author[0000-0003-3430-3889]{Jessica Speedie}
\altaffiliation{Heising-Simons Foundation 51 Pegasi b Fellow}
\affiliation{Department of Earth, Atmospheric, and Planetary Sciences, Massachusetts Institute of Technology, Cambridge, MA 02139, USA}
\email{jspeedie@mit.edu}

\author[0000-0001-5173-2947]{Clarissa R. Do \'O}
\affiliation{\caltech}
\email{cdoo@caltech.edu}

\author[0000-0002-6773-459X]{Doug Johnstone}
\affiliation{\nrc}
\affiliation{\uvic} 
\email{Douglas.Johnstone@nrc-cnrc.gc.ca}

\author[0000-0001-6975-9056]{Eric Nielsen}
\affiliation{\nmsu}
\email{nielsen@nmsu.edu}

\author[0000-0003-2649-2288]{Brendan P. Bowler}
\affiliation{\ucsb}
\email{bpbowler@ucsb.edu}

\author[0000-0002-9820-1884]{Simon Blouin}
\affiliation{\uvic}
\email{sblouin@uvic.ca}

\author[0000-0003-2233-4821]{Jean-Baptiste Ruffio}
\affiliation{\ucsd} 
\email{jruffio@ucsd.edu}

\author[0000-0001-5383-9393]{Ryan Cloutier}
\affiliation{\mac}
\email{ryan.cloutier@mcmaster.ca}

\author[0009-0007-6766-2040]{Andre Fogal}
\affiliation{\uvic}
\email{afogal@uvic.ca}

\author[0009-0009-2223-2404]{Kaitlyn Hessel}
\affiliation{\uvic}
\email{khessel@uvic.ca}

\author[0000-0002-4164-4182]{Christian Marois}
\affiliation{\nrc}
\affiliation{\uvic}
\email{Christian.Marois@nrc-cnrc.gc.ca}

\author[0009-0008-1229-3230]{Alexandra Rochon}
\affiliation{\mac}
\email{rochoa3@mcmaster.ca}

\author[0009-0008-9687-1877]{William Roberson}
\affiliation{\nmsu}
\email{wcr@nmsu.edu}

\begin{abstract}

We present uniform posteriors indicating the presence or absence of stellar and planetary companions over a sample of $10^5$ stars in Hipparcos and Gaia. We use the ``G23H'' catalog to fit a model that incorporates absolute astrometry constraints from Hipparcos, Gaia DR2, and Gaia DR3, as well as Gaia DR3 photometry, excess RV variability, and includes information from the presence or absence of common-proper motion companions. The resulting posteriors can be used directly to detect new companions, to target RV and imaging follow-up observations, and for downstream demographics studies. We highlight example detections of known and  new candidate companions and the companion detection rate across the HR diagram. The modelling approach we present is directly applicable to Gaia DR4 epoch astrometry and RVs. All posteriors are provided as a machine-readable catalog and user-accessible website ``exopost'' \href{http://exopo.st/}{\faDatabase}.

\end{abstract}


\section{Introduction}

The astrometry method is a way to detect companions through the Keplerian reflex motion they induce in their host star, in the plane of the sky. Astrometry was one of the first methods applied to the detection of exoplanets, dating back to, for example, efforts by van de Kamp in the 1960s \citep{vandeKamp1963}. Later, targeted programs with the Hubble Space Telescope Fine Guidance Sensors measured orbits and dynamical masses for a handful of known planets \citep[e.g.,][]{benedict_gl876,mcarthur_55cnc}. More recently, the comparison of Hipparcos and Gaia positions and proper motions has revealed the accelerations imprinted by unseen companions across the nearby stellar population \citep{hgca_dr2,hgca_dr3,kervella_stellar_2019,kervella_stellar_2022}; and the non-single-star solutions of Gaia Data Release 3 \citep[DR3; ][]{gaia_dr3} provided the first astrometric orbits derived from Gaia DR3's own time series, including substellar and planetary-mass companions \citep{gaia_dr3_multiplicity,holl_2023}, and enabling confirmations such as Gaia-4b \citep{gaia4b}.

Astrometry's strengths are complementary to those of the radial velocity (RV) method. Because the astrometric signal scales with the projected size of the star's orbit, sensitivity \textit{increases} with companion semi-major axis up to the time baseline of the observations, whereas the RV signal weakens.  Astrometry is also generally not impacted by stellar activity, and it does not rely on the presence of narrow spectral lines, which are sparse in early-type stars. Together, these properties make astrometry an excellent tool for probing companion populations at separations beyond the reach of RV surveys, and around a wide range of host spectral types.

Building on \citet{Thompson_2026}, hereafter \paperI, we present orbit posteriors sampled for over $10^{5}$ stars observed by both Hipparcos \citep{hipparcos_1,hipparcos_2} and Gaia \citep{gaia_dr3}. We employ the G23H catalog, which combines:
\begin{itemize}
    \item Hipparcos vs.\ Gaia DR3 proper motion anomaly \citep{hgca_dr3}
    \item A calibrated Gaia DR3 Renormalized Unit Weight Error (RUWE) constraint (UEVA; \citealt{pmex1}; \paperI)
    \item Calibrated Gaia Data Release 2 \cite[DR2; ][]{gaia_dr2} vs.\ DR3 proper motion anomaly (\paperI),
    \item Gaia radial velocity spectrograph (RVS) excess noise (\citealt{paired}; \paperI)
\end{itemize}
We expand on \paperI{} to model luminous companions, multiple companions, Gaia photometry, and information from neighbouring resolved Gaia sources. Because this model includes both Gaia RVS information and treats Gaia's detection (or non-detection) of resolved neighbours as an imaging-like completeness constraint, the results are effectively joint radial velocity (RV), absolute-astrometry, and imaging posteriors on a large scale.

Our goals with this work are threefold: to provide constraints that can be used to detect new companions; to provide the best available upper limits on the presence of companions around each star; and to deliver a large set of posteriors, sampled in a relatively uniform way, for downstream demographics work.

This last goal motivates us to sample posteriors even from systems that do not show strong evidence of a sub-stellar companion, which has not been the case for most large companion searches in the past.
Instead, companion searches across all detection techniques have generally proceeded through periodogram- or matched-filter-based search statistics \citep[e.g.,][]{BLS, TLS, GLS, Rosenthal2021}, which may then be followed by fitting a local posterior to prominent peaks. A detection threshold is applied and only companions exceeding it are reported \citep[e.g.,][]{Thompson2018, Mawet2014, Halbwachs2023}. Multi-companion systems are commonly handled by iteratively subtracting or masking a ``best-fit'' solution and re-running the search on the residuals \citep[e.g.,][]{Jenkins2010, Rosenthal2021}.

The thresholding and model selection is convenient and appropriate for a catalog of detections, but for demographics it imposes a selection function that depends on the details of the heuristic and must be reconstructed after the fact before the catalog can be used to infer occurrence rates. 
Signals whose period is not uniquely determined---single transits, or period aliases from gaps in coverage---are typically held as candidates rather than fit with full posteriors until follow-up pins the period down \citep[e.g.,][]{Dawson2010, Cooke2018, Osborn2022}, with notable exceptions such as \citet{ForemanMackey2016}.
For \textit{non-detections}, posteriors are likewise usually not produced or shared: for transits, such a posterior is uninformative for a single target (recovering the prior, minus a set of narrow period exclusions) and poorly suited to sampling.
The selection function is instead recovered by injection and recovery, which calibrates the sensitivity of the algorithm, data, and threshold through simulation.

For RV, the situation is different: full posteriors for non-detections trace out the $m\sin{i}$ that an undetected companion may have, and single-companion, potentially multi-modal, posteriors have been generated with the \texttt{Joker} \citep{joker} and reported threshold-free for over $2\times10^5$ stars  in \citet{Price-Whelan_2020} (with sensitivity to stellar companions and brown dwarfs).
The \texttt{kima} code \citep{kima} has proposed and demonstrated marginalization over a variable number of companions with simulated RV data.
Still, at the time of writing, the above approaches have not yet been applied at scale for precision radial velocity surveys capable of detecting planetary mass companions. For example, analyses of datasets like the California Legacy Survey \citep{Rosenthal2021} have proceeded through heuristic search algorithms like RVSearch, paired with injection-recovery to calibrate completeness.

An astrometric posterior for a star with no strongly detected companions is even more informative than it is from RV: it traces the regions in the full orbital parameter space---including eccentricity, orientation, and phase---where companions of a given \emph{true mass} could remain hidden. This makes these ideas, pioneered for instance by the \texttt{Joker} and \texttt{kima} worth applying at scale to this new type of data.
With Gaia DR4 poised to release epoch astrometry for the full Gaia catalog, we take this opportunity to demonstrate, on existing public data, the approach for which we advocate: fitting full orbit posteriors to every star in a sample, with no detection threshold or model comparison ladder.

The sample we consider in this work is a set of approximately $10^{5}$ stars cataloged by both Hipparcos and Gaia. 
We publish machine-readable posteriors for every system which can then be used for downstream analyses: companion detection, RV and imaging follow-up, and demographics. Depending on the evidence threshold selected (confident planetary massed companion, or circumstantial evidence of a companion whose minimum mass is in the planetary regime), the results contain dozens to thousands of new exoplanet candidates and a myriad of brown-dwarf and stellar companions. Some of these candidates will be vetted and interpreted in upcoming work.

\section{Methods}\label{sec:methods}

The G23H catalog presented in \paperI{} combines several complementary constraints from Gaia and Hipparcos: 
\begin{itemize}
    \item Hipparcos intermediate astrometric data (IAD; \citet{hipparcos_2}, measurements not used in this work)
    \item The Hipparcos vs.\ Gaia DR3 proper motion anomaly \citep{Brandt2021}
    \item The Gaia DR3 excess astrometric noise through the ``UEVA'' constraint from \citet{pmex1}, as calibrated and modelled in \paperI
    \item The Gaia DR2 vs.\ DR3 proper motion anomaly (\paperI)
    \item The Gaia DR3 radial velocity spectrograph (RVS) excess noise from the \texttt{paired} catalog \citep{paired}
\end{itemize}

\begin{figure*}
\centering
\includegraphics[width=\textwidth]{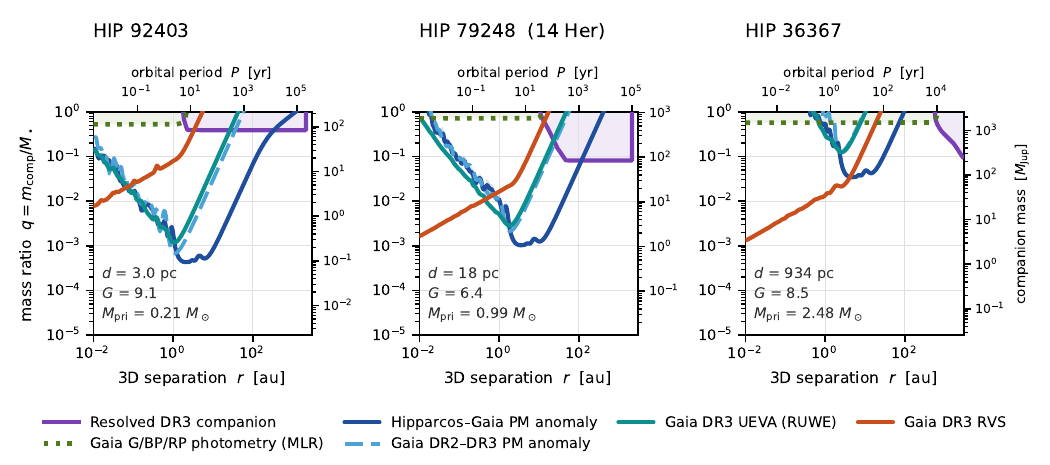}
\caption{Illustrative detection-limit curves for three example systems spanning the distance range of the sample. Each curve shows the mass ratio above which each data source would detect a companion on a circular orbit at an SNR of 3 as a function of the companion's 3D separation at epoch 2015.4 (top axis:   orbital period for $q\to0$) averaged over orbital phase and orientation. The curves are computed on each star's actual Gaia scan law; the thin bands show the 16--84\% spread over the latent choice of Gaia transits that the model marginalizes over. The astrometric constraints fall off linearly with distance to the system, while the RVS variability constraint does not. The ability to resolve a companion as an independent Gaia DR3 source depends on both the distance and brightness of the primary.}
\label{fig:detection-limits}
\end{figure*}

Here, we extend this method to support \textit{luminous} companions by modelling the photocentre motion Hipparcos and Gaia would measure for companions that approach the brightness of the primary star. To do so, we include a mass-to-light model based on MIST  isochrones \citep{mist_dotter,mist_choi}, which also unlocks two new data constraints from Gaia beyond those listed above: modelling and consistency of the primary and all unresolved companions' photometry with Gaia's measured $G$, $BP$, and $RP$ magnitudes,  given the system's parallax and extinction (as demonstrated in, \citet{2026A&A...708A.249B}, for instance); and the detection or non-detection of any purported wide, stellar companions with the full surrounding Gaia DR3 source catalog as in \citet{elbadry_binaries}, or for instance in \citet{2026arXiv260919249S}. Figure \ref{fig:detection-limits} visualizes how each of these constraints vary in their sensitivity to companions across three illustrative systems. We save the forward modelling of full Gaia RVS spectra until after the release of Gaia DR4.

\subsection{Astrometry and Gaia RV modelling}\label{sec:astrometry-rv}

We adopt the same Hipparcos versus Gaia DR2 \citep{gaia_dr2} and Gaia DR3 \citep{gaia_dr3} astrometric data, and modelling as \paperI. The only change to the astrometric data included in the model is to account for how the photocentre measured by Gaia and Hipparcos is displaced by the presence of luminous companions, and the exclusion of the Hipparcos IAD. We performed the same Gaia RV variability modeling as in  \paperI, but, as we will describe below, with uncertainties inflated to be more conservative across a large population.

\subsection{Photometry modelling}\label{sec:photometry}

Expanding from \paperI{} to luminous companions requires a mass-to-light model because Gaia and Hipparcos do not record the position of the primary star but the photocentre of any blended companions. We follow a similar approach as that of \citet{2026A&A...708A.249B}. We adopt the MIST isochrones \citep{mist_dotter,mist_choi} and marginalize over a single $\mathrm{[Fe/H]}$ and age for each system, assuming any luminous companions are coeval and have not evolved past the core-helium-burning stage. This assumption is violated in several regions of the Hipparcos colour-magnitude diagram (pre-main sequence or evolved hosts), and the posteriors for those stars should be interpreted with caution.

We linearly interpolate MIST v1.2 isochrones over our sampled  $(\log_{10} M, \mathrm{[Fe/H]}, \log_{10} t)$  parameters, with  ($v/v_\mathrm{crit}=0.4$, $[\alpha/\mathrm{Fe}]=0$). The grid is limited to $-1.75 \le \mathrm{[Fe/H]} \le 0.5$, $7.0 \le \log_{10}(t/\mathrm{yr}) \le 10.15$, and $0.079 \le M/\mathrm{M}_\odot \le 5.13$, with evolutionary phases from the pre-main sequence through core-helium burning. 
For stars with primary masses exceeding this upper limit, we fallback to the values at $5.13\,\mathrm{M}_\odot$, but the photometry modelling in these posteriors are likely unreliable regardless. Luminous companions above the top of the grid are rejected \textit{a-priori}. 
We require that every luminous component (primary or companion) fall within these bounds, including the evolutionary phase, at the shared sampled system age. As a result, this model does not admit any companions that are stellar remnants and may exhibit tension on these systems.

We include an interpolation-error estimate of $\sim5$ mmag on the main sequence,up to $\sim$0.1--1\,mag on the subgiant/giant ridge which is combined in quadrature with a shared 50\,mmag systematic floor (MIST versus the Gaia photometric system) and the adopted conservative photometric uncertainties of 3\,mmag in $G$ and 10\,mmag in $BP$ and $RP$.

The extinction in each BP, RP, and G depends on the effective wavelength of each source. We sample $A_0$, the monochromatic line-of-sight extinction at $\lambda \simeq 550$\,nm  (\citet{gaia_hrd_babusiaux}). We then calculate extinction in each band using each component's intrinsic colour from the MIST isochrone. A single $A_0$ therefore reddens a red companion and its bluer primary by different but mutually consistent amounts in $G$, $BP$ and $RP$.

We consider extinction using a single per-system monochromatic $A_0$, adopted from \citet{gaia_hrd_babusiaux}. This is the monochromatic line-of-sight extinction at $\lambda \simeq 550$\,nm  (\citet{gaia_hrd_babusiaux}). The extinction in each band follows from it as $A_X = k_X\!\left((BP-RP)_0,\,A_0\right)\,A_0$, where $$k_X = c_1 + c_2 X + c_3 X^2 + c_4 X^3 + c_5 A_0 + c_6 A_0^2 + c_7 X A_0$$ is the Gaia DR2 extinction polynomial of \citet{gaia_hrd_babusiaux}, evaluated at the intrinsic MIST colour $X = (BP-RP)_0$ of the component being attenuated. A single $A_0$ therefore reddens a red companion and its bluer primary by different but mutually consistent amounts in $G$, $BP$ and $RP$. 

We sample $A_0$ with a prior from the line-of-sight integrated dust map of \citet{edenhofer_dust} to the star's catalog parallax distance, truncated to $[0, 15]$\,mag, or fixed to zero for stars inside the Local Bubble ($d < 70$\,pc).

The apparent magnitudes of each companion enter our model in three ways: the primary and unresolved companions contribute to the total flux observed by Gaia in BP, RP, and G; the apparent magnitude of a companion that would be resolved by Gaia (see following section) is compared to that source's observed apparent magnitude; and the G band flux ratios of each unresolved companion versus the primary are incorporated into the (astrometric) model of Gaia's photocentre. 

To separate out a companion at each draw as resolved by Gaia, or unresolved,  we include a binary indicator variable for each companion, marginalizing over whether Gaia, \textit{on net}, resolved it as a separate source or a contribution to a blended photocentre. We assume that this resolved state was the same for both DR2 and DR3. In reality, the contribution of partially resolved companions differs depending on the orbital parameters, scan angle, and details of the Gaia astrometry extraction process that we do not yet model and may have changed between DR2 and DR3. As a result, we caution that orbit parameters inferred for bright companions with near unity flux ratios and projected separations between roughly 100--800 mas should be treated with caution. These companions are still reliably detected.

For Hipparcos, we instead employ the ``Hipparcos-centre'' prescription of the \texttt{BINARYS} tool \citep{binarys}, which forward-models the non-linear response of the Hipparcos grating to luminous companions at every epoch and scan angle recorded in the Hipparcos IAD. We note that we do not fit the Hipparcos IAD residuals themselves, and only use the epochs, scan angles, rejected scan flag, and uncertainties (up to a scale factor; \citealt{brandt_iad}) in our model. 
In the two Hipparcos reductions, most sources were fit with a 5-parameter solution, which were then cross matched and calibrated against Gaia in \citet{hgca_dr3}. Some Hipparcos sources were instead fit with alternative models, such as full orbit solutions. We model each source's position and proper motion from the HGCA assuming the standard 5-parameter model was fit---stars with different solution types are not well-modelled (see Section~\ref{sec:badfits}).
We adopt the primary and each companion's $H_p$ Hipparcos-band flux ratio from the MIST isochrones when forward-modelling the astrometry, but do not fit the actual Hipparcos photometry. The Hipparcos forward model does not consider the marginalized Gaia resolved state per companion, since this prescription naturally handles the modulated response of Hipparcos to multiple bright companions at any separation at every epoch. We anticipate that this type of transit-level blending model will, in future data releases, be possible with Gaia.

We do not model or consider photometric variability of the primary or of any luminous companion in this work.

We finally note that the above assumptions and limitations only impact the parameters of luminous companions, the determination of the primary mass, and determining if candidates are either dark or near equal-flux binaries. The inferred mass-ratios and periods of dark companions around single-stars is robust to these choices.

\subsection{Gaia companion matching}\label{sec:companion-matching}

To constrain or rule out the orbital signature of wide stellar companions, we include a marginalized matching over possible neighbouring Gaia sources. 
The model attempts to match each \emph{resolved} companion to a known Gaia source, after considering the source's $G$-band magnitude (Sec. \ref{sec:photometry}), position, proper motion, and parallax---after accounting for the local source density. Gaia sources may have either 2 parameters, giving position only, or a full 5-parameters, giving position, proper motion, and parallax. Because each candidate background or common-proper-motion companion's 2- or 5-parameter astrometric solution is compared against the model prediction, this is conceptually similar to performing multi-epoch direct imaging with image-orbit modelling \citep{octofitter} and a common-proper-motion check in order to reject background sources.

This likelihood term allows the model to attribute an astrometric acceleration to a specific stellar companion or, when no such companion is found, requires that the acceleration arise from a substellar object treated as dark by our model  ($<80\,\mjup$).

We do not incorporate constraints from the differential radial velocity between primary and companion out of concerns about calibrating the zero points between stars of different spectral types and converting spectroscopic to kinematic radial velocities.

We include a wide empirical uncertainty floor on the position and proper-motion differences that scales with orbital separation. This term allows companions to be matched even when the astrometric acceleration is somewhat affected by bias in the astrometric orbit model (e.g. for marginally resolved sources), since our goal is to identify plausible common-proper-motion companions rather than to maximally constrain their orbits.

To gather the data against which we compare the model, we start by performing a cone search in the Gaia DR3 archive, with an outer radius set to roughly the widest separation where an equal mass-ratio binary companion could induce a detectable astrometric acceleration. We also query an annulus surrounding the star to estimate the local background density. 
The cone search radius scales with the catalog parallax, $2154\,\mathrm{au}\times\varpi$, with a minimum of $10''$. Every DR3 source within $R_\mathrm{search}$ is retained as a candidate, with no magnitude cut. The background annulus spans $180''$--$600''$ ($285.9$\,arcmin$^2$) and is limited to $G<21$; its inner edge is placed outside the bright-primary incompleteness zone of the DR3 contrast curves ($\approx110''$ at the largest $\Delta G$), and its outer edge is small enough that the local density gradient across it is negligible even near the Galactic plane. Both queries use the DR3 (J2016.0) positions.

\subsubsection{Gaia Contrast Curve}\label{sec:contrast-curve}

For any given companion draw, we want to model whether a companion of that brightness and projected separation would be detectable as a separate source (\textit{resolved}) in Gaia or contributes instead to a blended photocentre \textit{(unresolved}). To determine the threshold at which Gaia resolves 50\% of companions for a given host magnitude and companion to host contrast, we calibrate an empirical ``contrast curve'' (in the direct imaging sense) for Gaia DR3.
In reality, such a distinction is not binary: a pair of closely separated sources may be combined or split different by Gaia per epoch and scan angle, and it may be that a resolved source still contributes to photocentre blending. A more faithful, per-epoch model must wait for Gaia DR4. Here, we proceed with this imperfect model as best we may.

In order to calibrate the separation and $\Delta\,\mathrm{mag}$ that a Gaia companion must have in order to be cataloged by DR3 as a separate source, we follow closely the procedure from \citet{brandeker_cataldi}, using data from DR3 instead of DR2. 
This empirical calibration begins by counting every DR3 neighbour of our target fields in bins of angular separation $\rho$ and contrast $\Delta G$. The counts are then converted to surface density, and divided by the maximum density for that $\Delta G$.  This gives a separation-dependent density-ratio, starting near zero where the bright primary makes the innermost annulus highly incomplete and rising towards unity at wide separation where the catalog's completeness is no longer impacted by the bright target.

Because that incompleteness depends strongly on the primary, we calibrate three magnitude bins separately: $G<6$ (5,234 targets, queried to $120''$), $6\le G<9$ (76,621, $36''$) and $G\ge9$ (24,608, $18''$). To each target of some $G$, we assign the curve with the matching $G$ range.

Within each bin and each $\Delta G$ column we fit a logistic in $\log_{10}\rho$,
\begin{equation}
\begin{split}
P_\mathrm{det}(\rho \mid \Delta G) &= \Bigl\{1 + \exp\Bigl[-\kappa(\Delta G) \\
&\quad \times \left(\log_{10}\rho - \log_{10}\rho_{50}(\Delta G)\right)\Bigr]\Bigr\}^{-1},
\end{split}
\label{eq:pdet}
\end{equation}
and interpolate $\rho_{50}$ and $\kappa$ linearly between $\Delta G$ nodes, holding them flat outside. 

In addition to this recipe from \citet{brandeker_cataldi}, we add additional constraints at the very bright end. In the $G<6$ bin, we also include 256 resolved binaries from ORB6, propagated to J2016. We find 58 detections and 198 non-detections, which help constrain the curve where the Gaia catalog alone is sparse. 

We finally impose Gaia's nominal $0.6''$ resolution for equal-brightness pairs, and force $\rho_{50}$ to be monotone in $\Delta G$ before smoothing. Finally, the DR3 faint limit is imposed as a logistic roll-off centred at $G = 20.7$ with a $0.3$\,mag width in the companion's predicted apparent magnitude.

\begin{figure*}
\centering
\includegraphics[width=\textwidth]{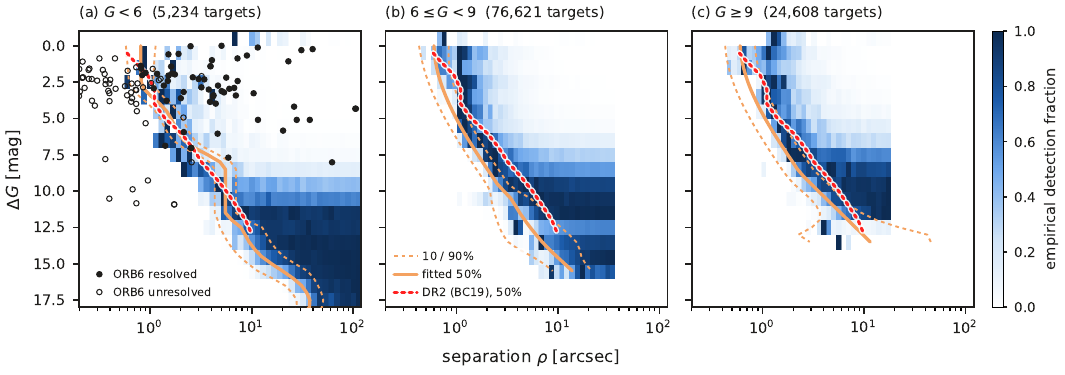}
\caption{Our Gaia DR3 contrast calibration for primary stars in different magnitude bins (panels (a)-(c)). The heatmap plots the DR3 neighbour surface density per separation $\rho$ and $\Delta G$, normalized by the cell with maximum density in that row. This deficit reveals the catalog incompleteness due to the bright primary.
The 50\% (solid) and 10\%-90\% range (dashed) of Equation~\ref{eq:pdet} are plotted in orange. For comparison, the 50\% separation derived for DR2 by \citet{brandeker_cataldi} is plotted as a bright red dotted curve (their fit is not separated by primary magnitude). 
Towards the bottom of each panel, the sources are primarily background objects, whereas towards lower $\Delta G$ the pairs are more often bound. The DR3 faint limit is applied separately as a roll-off in the companion's predicted apparent $G$.}
\label{fig:contrast-curves}
\end{figure*}

Figure~\ref{fig:contrast-curves} shows the three calibrations. Our 50\% separations are similar in shape to the DR2 map from \citet{brandeker_cataldi}, but DR3 resolves closer pairs. 
The vertical jump near 1" in panel (b) also appears in the previous DR2 analysis, suggesting it is a real feature of Gaia's processing. A similar vertical jump appears for bright stars in our calibration (panel a) near 5" and 10 mags of contrast, and may likewise be a genuine feature of Gaia's source processing. 

\subsubsection{Matching resolved companion slots to Gaia sources}
We now describe the way our model's likelihood function incorporates information from each potential neighbouring source. 
For each hypothesized companion $k$, a given draw specifies an orbit and a mass, and hence a predicted position, apparent $G$ magnitude, parallax and proper motion for that companion at J2016.0, together with a latent indicator of whether Gaia would have resolved it (above). Only companions that are luminous ($\ge0.08$\,M$_\odot$ in our grids), resolved, and projected inside the cone-search radius need to be accounted for by the Gaia catalogue; we call these the \emph{resolved companion draws}. Every DR3 source $i$ inside the cone is a \emph{candidate}. A typical draw has zero or one resolved companions and tens of candidates, nearly all of which are unrelated field stars. 

For each candidate--companion draw pair we ask whether the source's measured properties look more like the companion predicted by companion $k$ or like a field star that happens to lie in the cone. We express this as a likelihood ratio,
\begin{equation}
\mathrm{LR}_{ik} = \frac{p(\text{observables of } i \mid i \text{ is companion } k)}{p(\text{observables of } i \mid i \text{ is a field star})}. 
\label{eq:gcm_lr}
\end{equation}
The numerator is the probability density of the source's position, magnitude, astrometric solution type, parallax and proper motion if it is the companion predicted by companion $k$, 
\begin{align*}
p(&\text{observables of } i \mid i \text{ is companion } k) = \\
&\mathcal{N}(\Delta\boldsymbol\theta_{ik}\mid 0, \sigma_{\mathrm{pos},k})\; K_\mathrm{mag}(G_i - G_{\mathrm{pred},k}) \; \times \\ &P(\mathrm{type}_i\mid\mathrm{bound})\; \mathcal{N}(\tilde\varpi_i \mid \varpi, \sigma_{\varpi,i})\; \times \\ &\mathcal{N}(\tilde{\boldsymbol\mu}_i \mid \boldsymbol\mu_{\mathrm{pred},k}, \boldsymbol\sigma_{\mu,ik});
\end{align*}
the denominator is the same density if it is an unrelated star drawn from the local field: 
\begin{align*}
p(&\text{observables of } i \mid i \text{ is a field star}) = \\
&\rho_G(G_i)\; P(\mathrm{type}_i\mid\mathrm{field})\; f_\mathrm{field}(\tilde\varpi_i, \tilde{\boldsymbol\mu}_i).
\end{align*}

When comparing the magnitude of a resolved companion draw to the catalog value, we take into account that blending from the nearby bright primary may have biased the photometry of the catalogued source upwards but not downwards. On the faint side, we adopt a half-Gaussian of $0.4\,\mathrm{mag}$. On the bright side, we adopt a flat 2.5 mag plateau followed by a half-Gaussian of $1\,\mathrm{mag}$. We refer to this distribution as $K_\mathrm{mag}$.

Both are densities over the same observables, so the ratio is dimensionless: $\mathrm{LR}_{ik}\gg1$ means the source is more consistent with the predicted companion than a field star. In order, the factors in the ratio compare (1) the angular offset from the predicted companion position against the differential field density $\rho_G$ at the candidate's magnitude (2) the candidate's $G$ against the isochrone prediction (see Section \ref{sec:photometry}); (3) the candidate's astrometric solution type (5- or 6-parameter versus 2-parameter) through an empirical $P(\mathrm{5p}\mid G)$ curve, with a bound-side suppression at small separations where real companions of bright primaries often fail to obtain full solutions; and (4) for 5-parameter candidates, the parallax and proper motion against the system parallax and the companion's predicted absolute proper motion, relative to a per-target kernel-density estimate of the field $(\varpi, \mu_{\alpha*}, \mu_\delta)$ distribution built from the annulus sources. For 2-parameter candidates factor (4) is omitted.

We adopt conservative tolerances. On position, $\sigma_{\mathrm{pos},k}^2 = (30\,\mathrm{mas})^2 + (0.02\,\rho_{\mathrm{pred},k})^2$, the separation-scaled inflation mentioned above. On kinematics, we apply the \citet{hgca_dr3} factor of 1.37 to the formal DR3 uncertainties, floors of 0.3\,mas and 2\,mas\,yr$^{-1}$, confusion terms $\propto 1/\rho_i$ near the bright primary, and an orbital term $(0.5\,|\boldsymbol\mu_{\mathrm{orb},k}|)^2$ that widens the proper-motion tolerance for wide pairs with large orbital motion.

A single resolved companion draw has only two possible explanations: it is one of the candidates, or its counterpart is missing from the catalogue. We give the second, ``orphan'', outcome a fixed weight $\epsilon_\mathrm{orph} = 3\times10^{-3}$, so that
\begin{equation}
\mathcal{L}_\mathrm{CPM} \propto \epsilon_\mathrm{orph} + \sum_i \mathrm{LR}_{ik},
\end{equation}
where the proportionality constant does not depend on the parameters sampled by the MCMC chain.
$\epsilon_\mathrm{orph}$ is separate from whether Gaia would have resolved the companion at all, which is handled by the resolved indicator and  covers the residual chance that a source Gaia should have resolved is nevertheless absent from our candidate list (e.g.\ dropped for missing photometry), or is present but cannot be awarded a decisive likelihood ratio by the factors above. We adopt $\epsilon_\mathrm{orph} = 3\times10^{-3}$.

With two or more resolved companions in a given draw, we must also forbid a single Gaia source from being claimed as two different companions. An \emph{assignment} $\mathcal{A}$ gives each resolved companion either one candidate or the orphan outcome, with no candidate used more than once. 
In other words,  $\mathcal{A}$ is a bipartite matching valued random variable that we marginalize over. 
We place a prior over bipartite matchings between candidates and the drawn companion, with prior probability:
\begin{equation}
p(\mathcal{A}) \propto \epsilon_\mathrm{orph}^{\;\text{number of orphans in }\mathcal{A}}.
\end{equation}

The marginalized likelihood is the sum over all assignments,
\begin{equation}
\mathcal{L}_\mathrm{CPM} \propto \sum_\mathcal{A}\; p(\mathcal{A}) \prod_{(i,k)\in\mathcal{A}} \mathrm{LR}_{ik},
\label{eq:gcm_perm}
\end{equation}
where again the proportionality constant does 
not depend on parameters sampled by the MCMC 
algorithm. 
For two drawn companions this reads
\begin{multline}
  \mathcal{L}_\mathrm{CPM} \propto\\
   \epsilon_\mathrm{orph}^2 + \epsilon_\mathrm{orph}\sum_i \mathrm{LR}_{i1} + \epsilon_\mathrm{orph}\sum_j \mathrm{LR}_{j2} + \sum_{i\neq j}\mathrm{LR}_{i1}\,\mathrm{LR}_{j2},
\end{multline}
which is what one obtains by treating the two drawn companions independently, $(\epsilon_\mathrm{orph} + \sum_i \mathrm{LR}_{i1})(\epsilon_\mathrm{orph} + \sum_j \mathrm{LR}_{j2})$, minus the $i=j$ terms in which both drawn companions claim the same source. Equation~\ref{eq:gcm_perm} can be recognized as the permanent of the drawn companion $\times$ candidate matrix of likelihood ratios with a diagonal block of $\epsilon_\mathrm{orph}$ appended for the orphan outcomes. 

For resolved sources, we compare the companion to the Gaia $G$ magnitude. We do not consider \textit{BP} or \textit{RP} as they may be contaminated by close neighbours due to the larger Gaia BP / RP windows.  Due to this concern that companions may have their photometry contaminated \textit{upwards} by flux from the bright primary, we use a conservative half-Gaussian likelihood with  $\sigma = 0.4$\,mag on the faint side (the systematic error budget above) and a flat plateau on the bright side extending 2.5\,mag followed by a half-Gaussian tail with $\sigma = 1.0$\,mag, normalized to a proper density in $\Delta G$.

\subsection{Gaia RV variability modelling}

As in \paperI, we model excess scatter in the Gaia EDR3 radial velocities by comparing the \texttt{radial\_velocity\_error}, inverted to a per-transit sample variance via Eq.~A4 of \citet{paired}, against a model per-transit RV scatter $\sigma_\mathrm{RV}$ through a non-central $\chi^2$ likelihood.
The prior on $\sigma_\mathrm{RV}$ is the \texttt{paired} Gaussian-process prediction of the single-star per-transit RV uncertainty, $\ln\sigma_\mathrm{RV} \sim \mathcal{N}(\mu_\mathrm{paired},\,s_\mathrm{paired})$, with the mean $\mu_\mathrm{paired}$ and the standard deviation $s_\mathrm{paired}$, both in natural-log units, taken from the catalog columns \texttt{rv\_ln\_uncert\_dr3} and \texttt{rv\_ln\_uncert\_err\_dr3} respectively.

In early sampling runs this produced a population of spurious short-period companions near the RVS sensitivity floor concentrated on faint K dwarfs. We traced this to a Malmquist bias in the \texttt{paired} calibration: that catalog's uncertainty values were derived from the full, magnitude-limited RVS sample, whose bins in apparent $G$ and $BP{-}RP$ colour are dominated at the faint red end by distant giants with sharper lines, whereas our Hipparcos targets are dwarf-dominated at fixed $(G, BP{-}RP)$.
We measured the miscalibration by examining the inferred excess companions on two companion-free control samples (at the sensitivity of Gaia RVS): 5,115 externally RV-vetted quiet stars (NASA Exoplanet Archive RV hosts, \citealt{Christiansen2025}; CLS, \citealt{Rosenthal2021}; \citealt{isaacson_fischer_2010}; \citealt{luhn_2020}; and the RV standards of \citealt{soubiran_rv_standards}, with SB9 [\citealt{Pourbaix2004SB9}], NSS, and close-in WDS [\citealt{wds}] matches removed) and 63,325 astrometrically quiet stars (RUWE $<1.2$, no NSS/WDS/SB9 entry).
Fitting the lower half of the $z$-score distribution per cell (robust to residual companions) gave a multiplicative correction $f = \sigma_\mathrm{true}/\hat\sigma$ of 0.88 (B), 0.90 (A), 0.91 (F), 1.00 (G), 1.17 (K), and 1.09 (M) by host-mass class.
To be as conservative as possible, we adopted a broad global widening of the uncertainty: we kept the \texttt{paired} estimate and set the prior width to a minimum of 0.30, again in natural-log units, so a factor of $1.35$ in $\sigma_\mathrm{RV}$. Just over 0.02\% of stars with a \texttt{paired} entry had a quoted width larger than 0.30, which we kept.
This brackets the measured miscalibration of 95\% of the sample at $1\sigma$. The reduced RV sensitivity is still sufficient to rule out spectroscopic binaries. 

DR4 epoch radial velocities will remove the need for this proxy and calibration.

\subsection{Perspective acceleration and the catalog radial velocity}\label{sec:perspective-accel}

Perspective acceleration is a second order effect seen in absolute astrometry and radial velocity where stars appear to accelerate due only to the changing viewing angle between the system and the observer. We correct for perspective acceleration within the model itself, based on the DR3 catalog position, and sampled proper motions and absolute radial velocity. This correction applies to all proper motion data points from Hipparcos and Gaia. The impact of perspective acceleration is not considered for the DR3 UEVA channel, since it is handled for all nearby stars (with Gaia DR2 RVs) within the Gaia DR3 processing \citep{lindegren_dr2_astrometry,lindegren_edr3_astrometry}.
The absolute radial velocity of each system is a sensitive choice for nearby stars since biases here can masquerade as an astrometric acceleration \citep[see ][ for a discussion in the context of Hipparcos-Gaia proper motion anomaly]{2026arXiv260114459B}. As such, we adopt a prior for each system of $v_r \sim \mathcal{N}(\tilde v_r, 2.5\,\tilde\sigma_{v_r})$ centred on the Gaia DR3 mean RV. 

\begin{figure}
\centering
\includegraphics[width=\columnwidth]{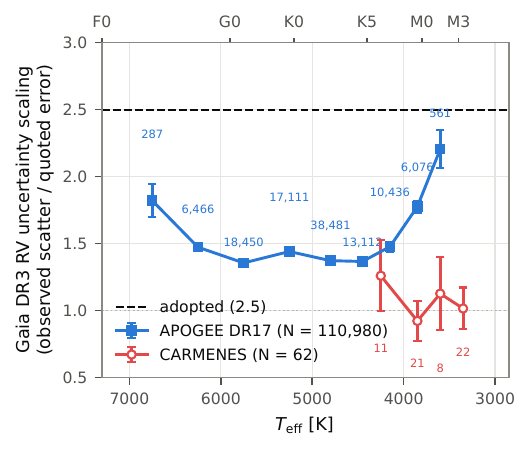}
\caption{Factor by which the Gaia DR3 \texttt{radial\_velocity\_error} must be inflated for the residuals against APOGEE DR17 (filled squares) and CARMENES (open circles) to have unit variance, as a function of effective temperature after iterative $5\sigma$ clipping of binaries. The dashed line is the global factor of 2.5 adopted for the prior on the systemic radial velocity, which enters the model only through perspective acceleration.}
\label{fig:rv-error-scaling}
\end{figure}

The factor 2.5 is from an empirical validation of the DR3 \texttt{radial\_velocity\_error} against APOGEE DR17 \citep[118,956 stars with STARFLAG $=$ ASPCAPFLAG $= 0$ and S/N $>20$;][]{apogee_dr17} and the CARMENES absolute RVs of \citet[][68 M dwarfs]{carmenes_rv}, with iterative $5\sigma$ clipping to remove binaries (Figure~\ref{fig:rv-error-scaling}). The factor by which the quoted uncertainty must be inflated for the normalized residuals to have unit variance is $\approx$1.4--1.5 for G and K stars, $\approx$1.8 for F stars ($N=287$), $2.20\pm0.14$ for M0--M3 dwarfs ($N=561$), and extrapolates to $\gtrsim$3 for M4 and later, for which neither catalog provides a usable sample; the bright, well-studied CARMENES M dwarfs alone give 1.0--1.3. Since the CARMENES RVs are well-centred on the DR3 values, it may be that the  mis-calibration actually originates in the APOGEE catalog. Nonetheless, we take the conservative option of inflating the Gaia DR3 RV uncertainty by a factor of 2.5 globally, loosening the RV constraints.

For the 10,989 stars in the recommended sample without a DR3 RV, we adopt a very conservative $\mathcal{U}(-500, 500)$\,km\,s$^{-1}$ prior. 
Downstream users who want to apply a more precise external absolute RV or test the posteriors' sensitivities to the adopted RV may cut or re-weight samples from our published posteriors.

\subsection{Priors}\label{sec:priors}

The priors we adopt for this model are tabulated in Table \ref{tab:priors}. The priors on some variables like parallax vary per star, in which case the table indicates the data source and any relevant formula. Some quantities present in the resulting chains are derived deterministically from other variables and are listed as well.

The number of companions is drawn from a categorical prior on the number of companions such that $P(N) = (0.500, 0.278, 0.144, 0.078)$ for $N = 0,1,2,3$: a geometric distribution $\mathrm{NegBin}(1, 1/2)$ truncated at $N=3$ and adjusted so that $P(N=0) = 1/2$ exactly, with the small excess redistributed equally over $N = 1$--$3$. Half the prior mass on the zero-companion model makes the posterior odds of ``at least one companion'' directly interpretable as a Bayes factor. 

The companions are activated in order: first $b$, then $c$, then $d$ for 1, 2, and 3 companion draws respectively. That said, the parameters assigned to the active companions are exchangeable---that is, there is no fixed meaning assigned to companion $b$ , $c$, or $d$ , and the parameters associated with any two may swap at each draw. 
Companions that are not active in a given draw are assigned zero mass and flux and simply follow their priors.

We adopt Jacobi coordinates, such that each companion sees all inner, active companions as the central gravitating body.

\begingroup
\setlength{\tabcolsep}{3pt}
\begin{deluxetable*}{llll}
\tabletypesize{\scriptsize}
\tablecaption{Parameters, priors, and derived expressions of the per-star G23H
multi-companion luminous-companion model. Normal distributions are written
$\mathcal{N}(\mu,\,\sigma)$ with $\sigma$ the standard deviation;
$\mathcal{TN}\!\left(\mu,\,\sigma;\;a < x < b\right)$ denotes that normal
truncated to $(a,b)$ and renormalised; $\mathcal{U}(a,\,b)$ and
$\mathrm{LogU}(a,\,b)$ are uniform and log-uniform on $(a,b)$; and
$\mathrm{LogNormal}(\mu,\,\sigma)$ has $\mu$ and $\sigma$ the mean and standard
deviation of the natural logarithm of the variate. Tildes denote
catalog-derived quantities that vary per star. \label{tab:priors}}
\tablehead{\colhead{Parameter} & \colhead{Type} & \colhead{Prior / expression} &
\colhead{Notes}}
\startdata
\multicolumn{4}{l}{\emph{Star}} \\
$M_\mathrm{pri}$ & Prior & $\mathcal{TN}\!\left(\mathrm{LogNormal}(0,\,0.5);\;0.08 < M < 10\right)\;M_\odot$ & $\ln M_\mathrm{pri} \sim \mathcal{N}(0,\,0.5)$ before truncation \\
$\mathrm{[Fe/H]}$ & Prior & $\mathcal{U}(-1.75,\,0.5)$ & MIST grid range \\
$\log_{10}(t/\mathrm{yr})$ & Prior & $\mathcal{U}(7.0,\,10.14)$ & shared by all luminous components\tablenotemark{a} \\
$A_0$ & Prior & $\mathcal{TN}\!\left(\tilde A_V,\,\tilde\sigma_{A_V};\;0 < A_0 < 15\right)$\;mag & \citet{edenhofer_dust} line-of-sight integral; $\tilde\sigma_{A_V}\ge0.005$ \\
 & & $\equiv 0$ & if $d < 70$\,pc (Local Bubble); \\
  \\
  \hline
\multicolumn{4}{l}{\emph{Astrometric reference}} \\
$\alpha$, $\delta$ & Fixed & $\alpha_\mathrm{cat}$, $\delta_\mathrm{cat}$ & Gaia DR3 catalog \\
$t_\mathrm{ref}$ & Fixed & Gaia DR3 reference epoch & MJD\,$57388.0$ ($\approx$J2016.0) \\
$\varpi$ & Prior & $\mathcal{TN}\!\left(\tilde\varpi,\,\tilde\sigma_\varpi;\;\max(0,\,\tilde\varpi - 10\tilde\sigma_\varpi) < \varpi\right)$\;mas & no upper bound \\
$\mu_{\alpha*}$, $\mu_\delta$ & Prior & $\mathcal{U}\!\left(\tilde\mu - \Delta\mu_\mathrm{hw},\;\tilde\mu + \Delta\mu_\mathrm{hw}\right)$\;mas\,yr$^{-1}$ & $\Delta\mu_\mathrm{hw} = \max(5,\;2\sqrt{10}\,\tilde\varpi)$; covers equal-mass companions \\
$v_r$ & Prior & $\mathcal{N}\!\left(\tilde v_r,\,2.5\,\tilde\sigma_{v_r}\right)$\;km\,s$^{-1}$ & Section~\ref{sec:perspective-accel} \\
 & & $\mathcal{U}(-500,\,500)$\;km\,s$^{-1}$ & if no DR3 RV \\
 \hline
\multicolumn{4}{l}{\emph{Companions, $k \in \{b,\,c,\,d\}$}} \\
$N$ & Prior & $(0.500,\;0.278,\;0.144,\;0.078)$ on $\{0,1,2,3\}$ & renormalised $\mathrm{NegBin}(1,\,0.5)$\tablenotemark{b} \\
$P_k$ & Prior & $\mathrm{LogU}(1/365.25,\;10^{5})$\;yr & \\
$q_k$ & Prior & $\mathrm{LogU}(10^{-5},\;5)$ & $q_k \equiv m_k / M_\mathrm{pri}$ \\
$u_{\mathrm{res},k}$ & Prior & $\mathcal{U}(0,\,1)$ & resolution draw \\
$i_k$ & Prior & $\mathrm{Sine}()$ & \\
$\Omega_k$, $\omega_k$, $\theta_k$ & Prior & $\mathcal{U}(0,\,2\pi)$ & $\theta_k$: position angle at the mean Gaia epoch $\bar t_\mathrm{G}$ \\
$e_k$ & Prior & $\mathcal{U}(0,\,0.99)$ & \\[4pt]
$z_k$ & Derived & $[N_\mathrm{p} \geq k]$ & companion active?\tablenotemark{c} \\
$m'_k$ & Derived & $q_k\,M_\mathrm{pri} / \mathcal{M}_\mathrm{Jup}$ & \\
$m_k$ & Derived & $m'_k \cdot z_k$ & \\
$M_{\mathrm{tot},k}$ & Derived & $M_\mathrm{pri}(1 + q_k z_k) + {\displaystyle\sum_{j:\,P_j < P_k}} q_j\,z_j\,M_\mathrm{pri}$ & mass enclosed by companion $k$ \\
$a_k$ & Derived & $\left(P_{\mathrm{Kep},k}^{2}\,M_{\mathrm{tot},k}\right)^{1/3}$ & $P_\mathrm{Kep}$ from Julian years via IAU conversion \\
$P_{\mathrm{det},k}$ & Derived & Gaia instrumental detection probability & \tablenotemark{d} \\
$r_k$ & Derived & $\left[u_{\mathrm{res},k} < \min\!\left(P_{\mathrm{det},k},\,0.997\right)\right]$ & resolved flag \\
$f_k$ & Derived & $G$-band flux ratio from MLR, $\times\,z_k(1 - r_k)$ & Gaia photocentre; $0$ if $m_k < 0.08\,M_\odot$ \\
$f^{H_p}_k$ & Derived & $H_p$-band flux ratio from MLR, $\times\,z_k$ & no $r_k$ gate \\
$\tau_k$ & Derived & $\texttt{$\theta$\_at\_epoch\_to\_tperi}(\theta_k,\,\bar t_\mathrm{G};\,a_k,e_k,i_k,\omega_k,\Omega_k,M_{\mathrm{tot},k})$ & time of periastron passage \\
\hline
\multicolumn{4}{l}{\emph{G23H nuisance (\paperI)}} \\
$\sigma_\mathrm{AL}$, $\sigma_\mathrm{att}$, $\sigma_\mathrm{cal}$ & Prior & $\mathcal{TN}\!\left(\tilde\sigma_X,\,\tilde\sigma_{X,\sigma};\;0 < \sigma_X < 10\right)$\;mas & \paperI{} per-star calibration \\
$\boldsymbol{u}$ & Prior & $u_j \sim \mathcal{N}(0,\,1)$ i.i.d., $j = 1\ldots N_\mathrm{tr}$ & transit priorities, one per GOST-forecast transit\tablenotemark{e} \\
$\mathcal{T}_\mathrm{ast}$ & Derived & top-$N_\mathrm{astrom}$ indices of $\boldsymbol{u}$ & selects observed epochs \\
$\mathcal{T}_{v_r}$ & Derived & top-$N_{v_r}$ indices of $\boldsymbol{u}$ & $N_{v_r} = \texttt{rv\_nb\_transits}$ \\
$u_\mathrm{dup,DR2}$ & Prior & $\mathcal{U}(0,\,1)$ & DR2 doubly-downlinked transit count (bright stars) \\
$\ln\sigma_{v_r/\mathrm{tr}}$ & Prior & $\mathcal{N}\!\left(\tilde\mu_\mathrm{paired},\,\max(\tilde s_\mathrm{paired},\,0.30)\right)$ & Section~\ref{sec:astrometry-rv}; if DR3 RV \\
\enddata
\tablenotetext{a}{additionally, $-\infty$ if the primary or any luminous companion is evolved past core helium burning at the sampled $(\mathrm{[Fe/H]},\,t)$.}
\tablenotetext{b}{$\mathrm{NegBin}(1,\,0.5)$ truncated at $N=3$ and renormalised so that $N=0$ is given $50\%$ of the overall weight.}
\tablenotetext{c}{Parameters of inactive companions ($z_k = 0$) are drawn from the prior and do not contribute to the likelihood.}
\tablenotetext{d}{From Gaia contrast sensitivity maps, modelled as a sigmoid in log-separation interpolated over $\Delta G$ nodes.}
\tablenotetext{e}{Marginalizes over transit assignment when GOST forecasts more epochs than Gaia reports as matched.}

\end{deluxetable*}
\endgroup

\subsection{Sampling}\label{sec:sampling}

We sample every system with non-reversible parallel tempering using the \texttt{Pigeons.jl} \citep{pigeons,non_rev_parallel_temp} Julia \citep{julia_bezanson} package with the slice sampler local explorer. The slice sampler correctly handles sampling from the discrete $N$ variable and resolved companion indicator variables. We sample each system for 12 rounds, resulting in 4096 samples after removal of burn-in. Gaia scan-law nuisance variables are sampled at a reduced rate, updated only every 8 MCMC iterations.
We initialized the sampler from the  best of 10,000 prior draws, and $\widehat{R}$ is computed over all parameters on the final round as a convergence diagnostic. 
We ran each star in an independent Julia \citep{julia_bezanson} process  with 6 threads. The bulk of the sample completed within a few hours per star, with a tail of $\sim$30--40\,h for the hardest systems, most of which turned out to be systems with an incorrect cross match between DR2 and DR3, which we exclude from the release (Section~\ref{sec:results}). Model summary statistics (log evidence ratio, global communication barrier, and timing) are stored in the header of each posterior file.

The survey was completed on the Trillium cluster (SciNet / Digital Research Alliance Canada) which consists of 192-core dual-socket AMD zen5 EPYC nodes. Each system was fit using an independent single-node Julia process, with 32 such workers per node (6 cores each). The production catalog of 106,508 systems consumed $2.34 \times 10^6$ allocated core-hours (12,197 node-hours), or 22 core-hours per system. The campaign briefly reached a peak concurrency of 1,012 nodes (194,304 cores). 

\section{Results}\label{sec:results}

We now discuss the results of this sample, beginning with a description of the hosts that sampled successfully.
We prepared 108,643 stars from the G23H catalog:
DR3 stars with a successful DR2--DR3 and Hipparcos cross match, brighter than 16th mag, within 1 kpc. 
Table~\ref{tab:selection} lists every subsequent selection step. Apart from the 54 incidental sampling failures, each step is defined on the catalog inputs rather than on the fitted posteriors.
Three requirements removed 2,081 stars before sampling: (i) all six Gaia noise-calibration columns of \paperI{} must be present (removing 1,074 stars); (ii) the Hipparcos-2 double-star solution had to be parseable by our pipeline (annex C/O solutions only; removing a further 629); and (iii) the catalog parallax signal-to-noise ratio must exceed $\varpi/\sigma_\varpi = 10$ (removing a further 378). A further 54 stars (0.05\%) did not produce a posterior for incidental reasons (transient file system errors, four stars with no IAD file, and a few unexplained failures), leaving 106,508 posteriors. 

We then removed 1,096 stars (1.0\%) whose G23H DR2 data are unreliable: either the G23H DR2 counterpart is not the source that Gaia's own DR2--DR3 cross-match (\texttt{gaiadr3.dr2\_neighbourhood}) associates with our DR3 source---usually the other component of a close Hipparcos pair---or the DR2 proper motion differs from the DR3 value by more than 20\,mas\,yr$^{-1}$, or the DR2--DR3 positional proper motion by more than $20\sigma$. Such inputs produce unphysical posteriors. Most of these stars (681) are Hipparcos annex C or O solutions, which the recommended sample below excludes in any case. This is a catalog-level cut only and will not capture every case: a few tens of posteriors affected by poor DR2 cross matches or G23H calibration likely remain in the release. The remaining 105,412 posteriors are published.

\begin{figure}
\centering
\includegraphics[width=\columnwidth]{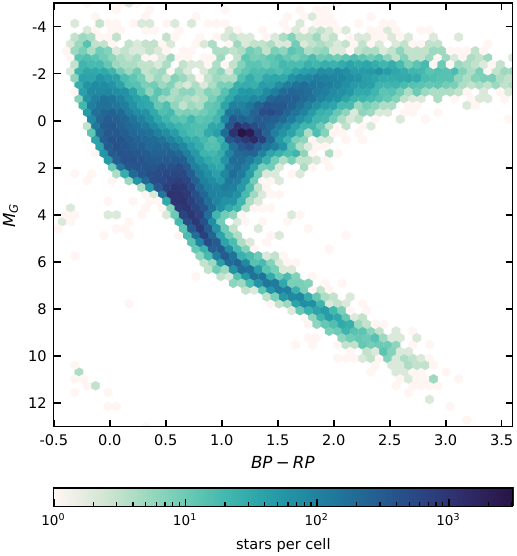}
\caption{Gaia colour-magnitude diagram of the 96,915 stars of the recommended sample (Table~\ref{tab:selection}). This figure is not corrected for extinction.}
\label{fig:cmd}
\end{figure}

Of the published posteriors, we recommend 96,915 for downstream use. The remaining 8,497 are stars whose Hipparcos astrometry comes from a Double and Multiple Systems Annex component (\texttt{MultFlag}~$=$~C) or orbital (O) solution, which our model does not describe faithfully (Section~\ref{sec:badfits}). These are binaries by construction, so this cut is a selection against binarity. Membership in the recommended sample is a flag carried by every published posterior, in its FITS header and in the summary table (Section~\ref{sec:data-availability}).
The Gaia colour-magnitude diagram for the recommended sample is presented in Figure \ref{fig:cmd}.

Of the recommended stars, 85,926 have a Gaia DR3 RV and RV uncertainty calibration from the \texttt{paired} catalog and 10,989 do not have RV constraints; all have the UEVA constraint and the Hipparcos--Gaia proper-motion data. The availability of RV constraints is impacted by binarity, since double lined spectroscopic binaries are selected against by Gaia DR3 itself. We dropped the DR2 and DR2 vs. DR3 scaled position differences for 3,367 of them where their catalog DR2--DR3 correlation coefficient $\rho_\mathrm{DR2,DR3} \ge 0.95$, a quality cut we recommended in \paperI. The median star has $G = 8.2$ and a parallax of 4.5\,mas ($\approx$220\,pc); 18,296 lie within 100\,pc. Every posterior has exactly 4,096 draws from 12 Pigeons rounds. 

We additionally cross match resolved common proper motion companions against the Gaia-identified white dwarf catalog of \citet{gentile_fusillo_wd}. Our model assumes that all companions $\geq80\mjup$ have not evolved past the core helium burning stage, and white dwarf companions violate that assumption. When these companions have detectable astrometric signals, the mass to light relation used would be incorrect and may lead the model to infer additional (potentially planetary-massed) companions to absorb the extra signal, and should accordingly be treated with caution.
We flag the 172 published systems with such a companion (Table~\ref{tab:selection}; some of which are flagged for other reasons as well).

\begin{deluxetable}{lrr}
\tablecaption{Sample selection\label{tab:selection} for the posteriors we publish and the subset we recommend for downstream use. Stars removed above the first rule are not part of the data release. The subsequent rows define flagged subsets.}
\tablewidth{0pt}
\tablehead{
\colhead{Step} & \colhead{Removed} & \colhead{Remaining}
}
\startdata
Hipparcos catalogue \citep{hipparcos_1,hipparcos_2} & & 118,218 \\
In HGCA eDR3 \citep{hgca_dr3} & $-2,872$ & 115,346 \\
In G23H (DR2 and DR3 cross-match, $G<16$, $d<1$\,kpc) & $-6,703$ & 108,643 \\
Has UEVA noise calibration & $-1,074$ & 107,569 \\
\textbf{Hipparcos-2 double-star solution not parsed by our pipeline (all annex C/O)} & $-629$ & 106,940 \\
Parallax S/N $>10$ & $-378$ & 106,562 \\
Sampling completed & $-54$ & 106,508 \\
Reliable G23H DR2 data: \textbf{published} & $-1,096$ & 105,412 \\
\hline
Not a Hipparcos annex C or O solution: \textbf{recommended} & $-8,497$ & 96,915 \\
\hline
Optional: also not a Hipparcos annex G, V or X solution (clean Hipparcos) & $-3,813$ & 93,102 \\
Optional: also no resolved white dwarf companion \citep[$P_\mathrm{WD}\geq0.5$;][]{gentile_fusillo_wd} & $-157$ & 92,945
\enddata
\end{deluxetable}

\subsection{Sensitivity}\label{sec:sensitivity}

\begin{figure}
\centering
\includegraphics[width=\columnwidth]{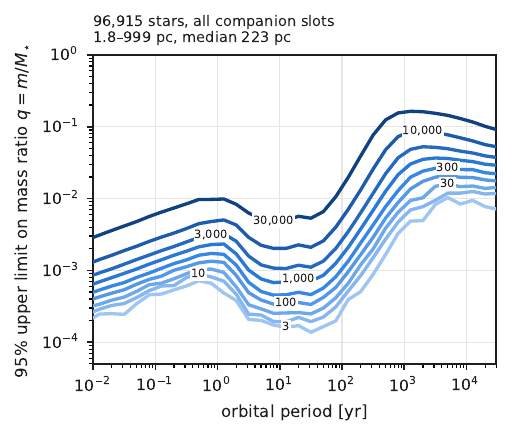}
\caption{
A summary of the catalogs companion sensitivity for all 96,915 sources of the recommended sample defined in Section \ref{sec:badfits} and Table~\ref{tab:selection}. The counts are of total companions across all stars whose 95th percentile upper limit is below each contour. The vertical axis is the mass-ratio  $q = m/M_\star$, which is comparable across stellar masses.  The curve labelled $N$ is the $N$-th smallest 95th percentile, so $N$ stars exclude a companion above the curve at that period. 
The sensitivity is largely a function of distance and the astrometric uncertainty for each star's colour, apparent magnitude, and sky position.}
\label{fig:sensitivity-q}
\end{figure}

We summarize the sensitivity of the full set of posteriors in Figure \ref{fig:sensitivity-q}, in mass-ratio vs orbital period. We calculate, for each bin in period, the companion draws across all stars that have a 95th percentile upper limit below each contour. At the most favourable periods, 30,000 stars constrain the mass ratio to $q < 4.9\times10^{-3}$, 3,000 stars to $1.1\times10^{-3}$, 300 stars to $4.5\times10^{-4}$, and the most sensitive few to $1.4\times10^{-4}$; 72,721 stars (75\%) reach $q < 10^{-2}$ at some period. 
The sensitivity at the short-period side of the plot is driven by the Gaia RVS excess variability, which is only included for the 85,926 stars in the sample with Gaia RV measurements present in the \texttt{paired} catalog \citep{paired}. 
The sensitivity close in comes from Gaia excess RV variability, then within-Gaia and Hipparcos-Gaia astrometric constraints, then finally flattens at very wide separations where luminous companions would be detected as separate, neighbouring sources by Gaia.

\begin{deluxetable*}{lrrrrrrrr}
\tablecaption{Number of companion identities with $P(\mathrm{exist})$ above three thresholds, by posterior median mass, for the 96,915 posteriors in the recommended sample (Section~\ref{sec:badfits}, Table ~\ref{tab:selection}). ``CPM'' counts the subset matched to a resolved Gaia DR3 source. Systems are the number of distinct stars contributing. The table summarizes 260,438 companion identities across the 96,915 systems. \label{tab:detection-counts}}
\tablehead{\colhead{$P(\mathrm{exist})$} & \colhead{$<13$\,M$_\mathrm{J}$} & \colhead{13--80} & \colhead{80--300} & \colhead{300--1000} & \colhead{1000\,M$_\mathrm{J}$--3\,M$_\odot$} & \colhead{$>3$\,M$_\odot$} & \colhead{CPM} & \colhead{Systems}}
\startdata
$\ge 0.5$  & 22,543 & 14,508 & 13,175 & 16,164 & 11,004 & 4,573 & 8,416 & 51,382 \\
$\ge 0.9$  &  1,288 &  4,580 & 11,596 & 15,426 & 10,499 & 3,337 & 7,591 & 42,443 \\
$\ge 0.99$ &    582 &  3,091 & 10,333 & 14,657 & 10,218 & 2,873 & 6,529 & 38,986 \\
\enddata
\end{deluxetable*}

\subsection{Recovered Companions}
The posteriors in this sample span the gamut from tightly concentrated posteriors for (sub-)stellar companions to ones consistent with upper limits only. In-between lie weakly constrained solutions spanning many orders of magnitude in period, compatible with either planetary or stellar objects.
The full set of all posteriors are viewable on the interactive website  \url{http://exopost.wthompson.space}, from which the following figures were retrieved.

We do not recommend any particular thresholds for interpreting the posteriors published in this work, and encourage the posteriors to be considered on a continuous scale.

\begin{figure*}
\centering
\includegraphics[width=0.32\textwidth]{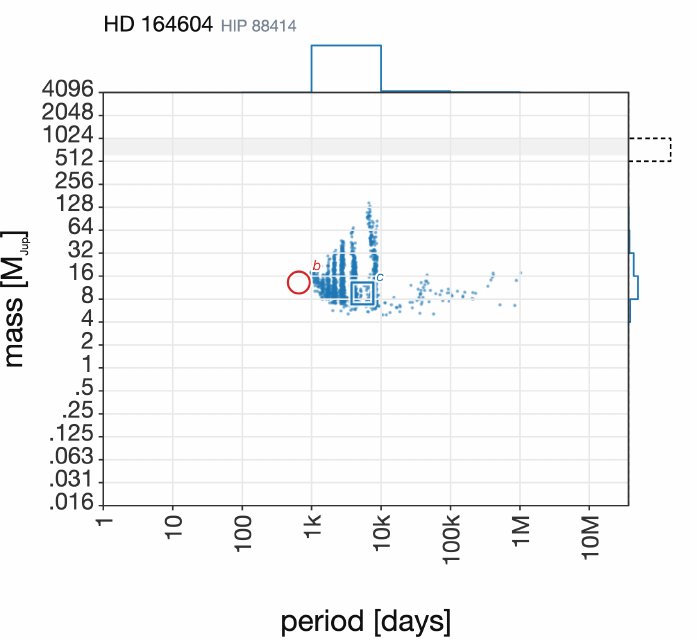}
\includegraphics[width=0.32\textwidth]{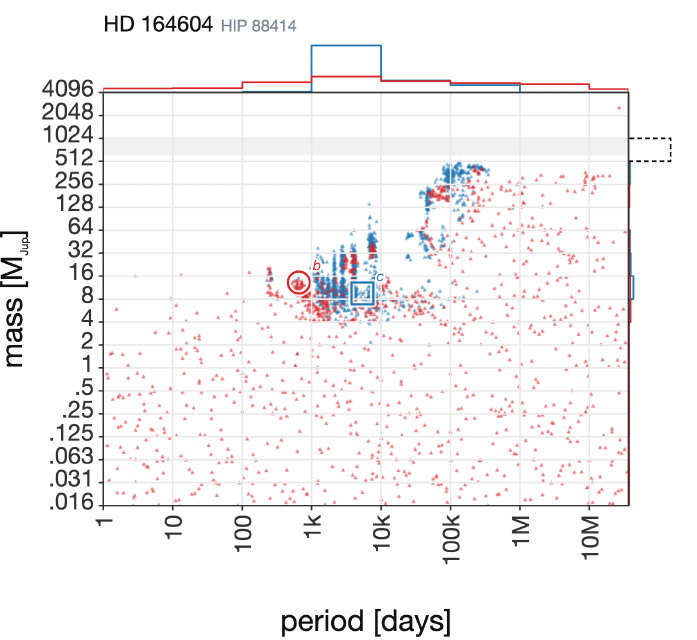}
\includegraphics[width=0.32\textwidth]{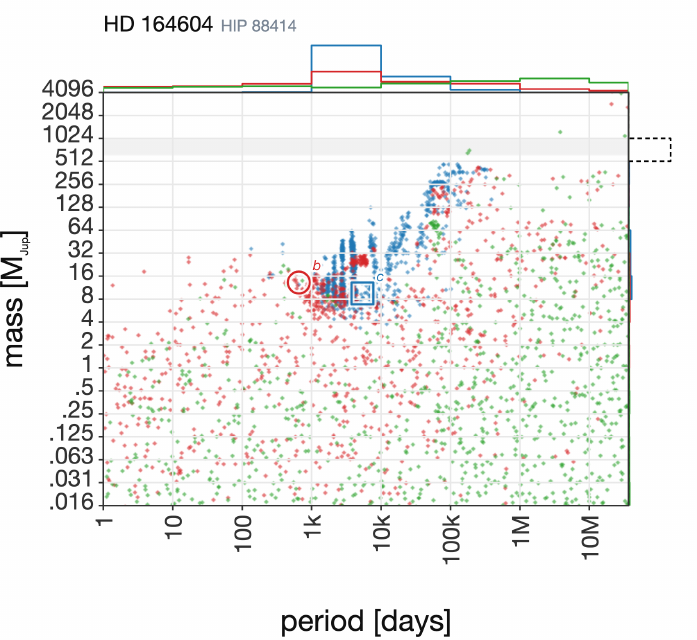}

\noindent\makebox[\textwidth][l]{%
\hspace{0.02\textwidth}%
\includegraphics[width=0.28\textwidth]{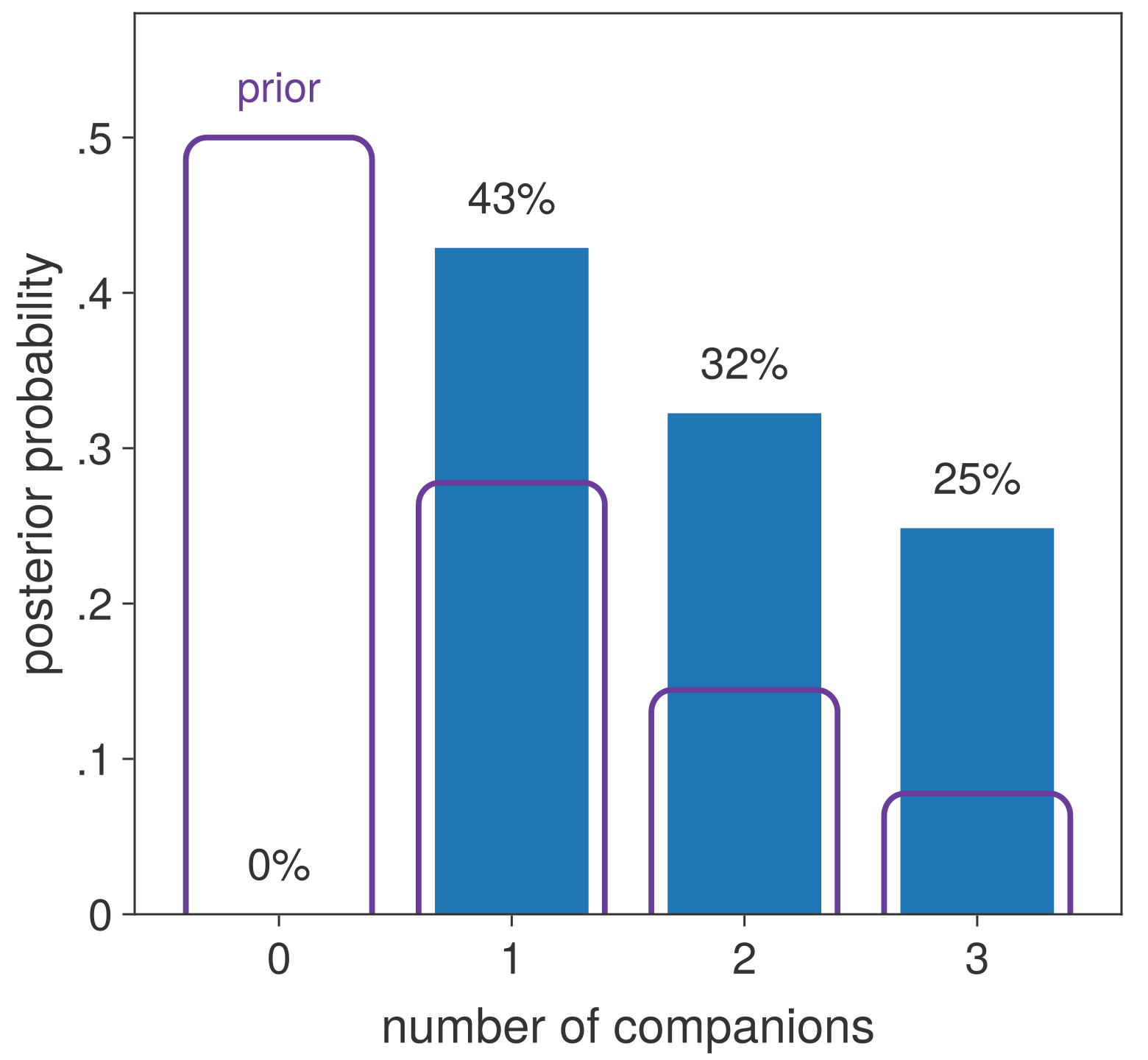}\hspace{0.04\textwidth}%
\includegraphics[width=0.32\textwidth]{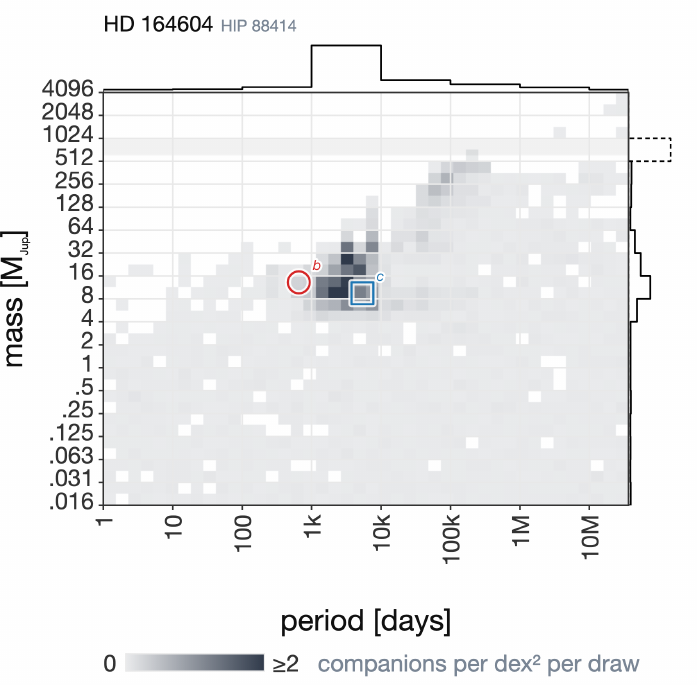}}

\caption{The posterior for HD~164604 (HIP~88414), illustrating marginalization over the number of companions (0--3). Left to right: draws with one, two, and three companions. For the panels with more than one companion, we have coloured the draws blue, red, then green according to the astrometric and RV SNR of the companion. We emphasize, however, that this ordering is not inherent to the posterior itself, and no ordering prior is applied during sampling. Bottom left: the marginal posterior over the number of companions, $P(N)$, i.e.\ the fraction of draws falling in each of the three top panels. The purple outline is the categorical prior on the integer-valued parameter $N$, $(0.500, 0.278, 0.144, 0.078)$ for $N=0$--$3$ (Section~\ref{sec:priors}). In this case, one or more companions are adopted by all draws in the posterior. Bottom right: the inferred companion rate binned over all draws, which is invariant to the companion ordering.
We draw the reader's eye to how the 2 and 3 companion draws loosen the constraints on the ``first'' companion; something that would not captured by subtracting off a best-fit solution and re-fitting.}
\label{fig:hd164604}
\end{figure*}

We now describe results for some particular illustrative systems.

\subsubsection{HD 164604}
The posterior for HD 164604 illustrates a strong astrometric detection of planetary massed companions, which we use as an example of how our trans-dimensional posteriors may be interpreted. 
The HD 164604 system consists of two known companions detected through radial velocity \citep{2010ApJ...711.1229A, 2026AJ....171..209X}, with dynamical masses constrained by joint astrometry and Gaia DR3 vs.\ Hipparcos proper motion anomaly \citep{2026AJ....171..209X}. 

As shown in Figure \ref{fig:hd164604}, our posteriors \textit{marginalize} over a variable number companions. 
Draws may have zero companions in the region of interest, implying a single star whose astrometry is consistent with straight line motion and Gaia photometry and parallax consistent with a single luminous star.
On the other hand, a draw may have 1--3 companions, each absorbing more of the signal from the proper-motion anomalies, and Gaia astrometric and RVS excess noise. 
In this case, 0\% of the posterior draws adopt a single-star solution; 100\% of the draws include 1 or more companions; 57\% include 2 or more companions; and 25\% include 3 companions. 

Beginning with the 1-companion solution, a strong banding pattern (a ``forest of modes'') is visible in the  marginal period histogram. These are caused both by period aliases that would all produce signals compatible with the astrometric measurements and by our remaining uncertainty in \textit{which} epochs and scan angles were used by the Gaia DR2 and DR3 catalog solutions (\paperI). The banding blends together towards lower masses, which in this case correspond to more-inclined orbits. These period aliases are seen in many posteriors consistent with strong detections, with the specific pattern depending strongly on the star's Gaia scan pattern. With only a single companion, the model must compromise to find solutions that must reproduce all measurements simultaneously. In this case, it finds solutions consistent with HD 164604 c.

Adding a second companion allows the model to reproduce both known companions HD 164604 b and c; however, it comes at the expense of weakening the constraints on draws that correspond to HD 164604 c, giving  some solutions that extend to stellar masses. This is the correct behaviour for a model that responds to inter-companion correlations and the additional uncertainty added by the new degrees of freedom. For this system, the second companion is not strongly detected: some draws match HD 164604 b, while many others are compatible with any mass below the detection limit.  Adding a third companion has essentially no impact on the solution, and the masses of said third companion would be too small to be detectable.

We remind the reader that the assignment of the companion parameters at each draw is not ordered in any way: the first, second, and third companions  may all swap at each draw between solutions compatible with  HD 164604 b, c, or a hypothetical undetectably-small companion. Each draw merely contains parameters for 0--3 companions with no meaning assigned.  For the sake of the visualizations in Figure \ref{fig:hd164604}, we have sorted the companion parameters by the approximate astrometric SNR its mass and semi-major axis would imply such that the highest SNR companions is consistently identified in blue across panels, second in red, and third in green. The simple proxy we used was $\frac{S}{N}\propto \frac{q a}{1 + (a/a_\mathrm{crit})}\frac{\varpi}{0.05 \, \mathrm{mas}}$ where $a_\mathrm{crit}\approx8.5 \, \mathrm{au}$, from the 24.75 yr Hipparcos-Gaia baseline. Other orderings are equally valid, such as sorting by semi-major axis or k-means clustering.

One may also summarize the posterior using quantities that are invariant to the labelling order: in the last panel of Figure \ref{fig:hd164604}, we plot the expected number of companions per square dex of period and mass---that is, the number of draws, pooled across all active companions. 
This visualization frames the posterior as an \textit{occurrence rate} model, conceptually, very similar to the marked Poisson processes used in the planet-occurrence literature \citep[e.g.,][]{tabachnik_tremaine_2002,cumming_2008,youdin_2011,foreman-mackey_2014,nielsen_2019,fulton_2021,vigan_2021}, applied to a single system only. The expected number of companions is then our updated expectation on rate of companions \textit{for this} system, as a function of mass and separation.

Towards the top-half of the plot, the data strongly rule out any companions, reducing our expected companion rate there to zero. In the centre, where the known companions b and c sit, the expected number of companions increases sharply. Towards the bottom-half of the plot, our expectations on the number of companions reverts to the prior.

Shifting from the astrometry constraints, we now discuss the photometric constraints that rule out many stellar companions in the system.
We find the photometry and parallax are well constrained, and additional unresolved near-equal-flux stellar companions are not supported. 
No neighbouring Gaia DR3 source is consistent with bound kinematics nor the astrometric constraints, so few companion draws are considered beyond $\approx 1\times10^6 \, \mathrm{d}$. The few draws that are present in this stellar-mass, ultra-long regime are all eccentric, with periastron near J2016, but with photometry that would not detectably impact the system's photometry if it were blended with the primary. A few samples are visible with masses even higher than the primary, but these are all extremely long period, high eccentricity solutions that would likewise have periastron slightly outside our cone-search. Calculating the projected separation near 2016 would separate these points cleanly.

\begin{figure*}
\centering
\includegraphics[width=0.32\textwidth]{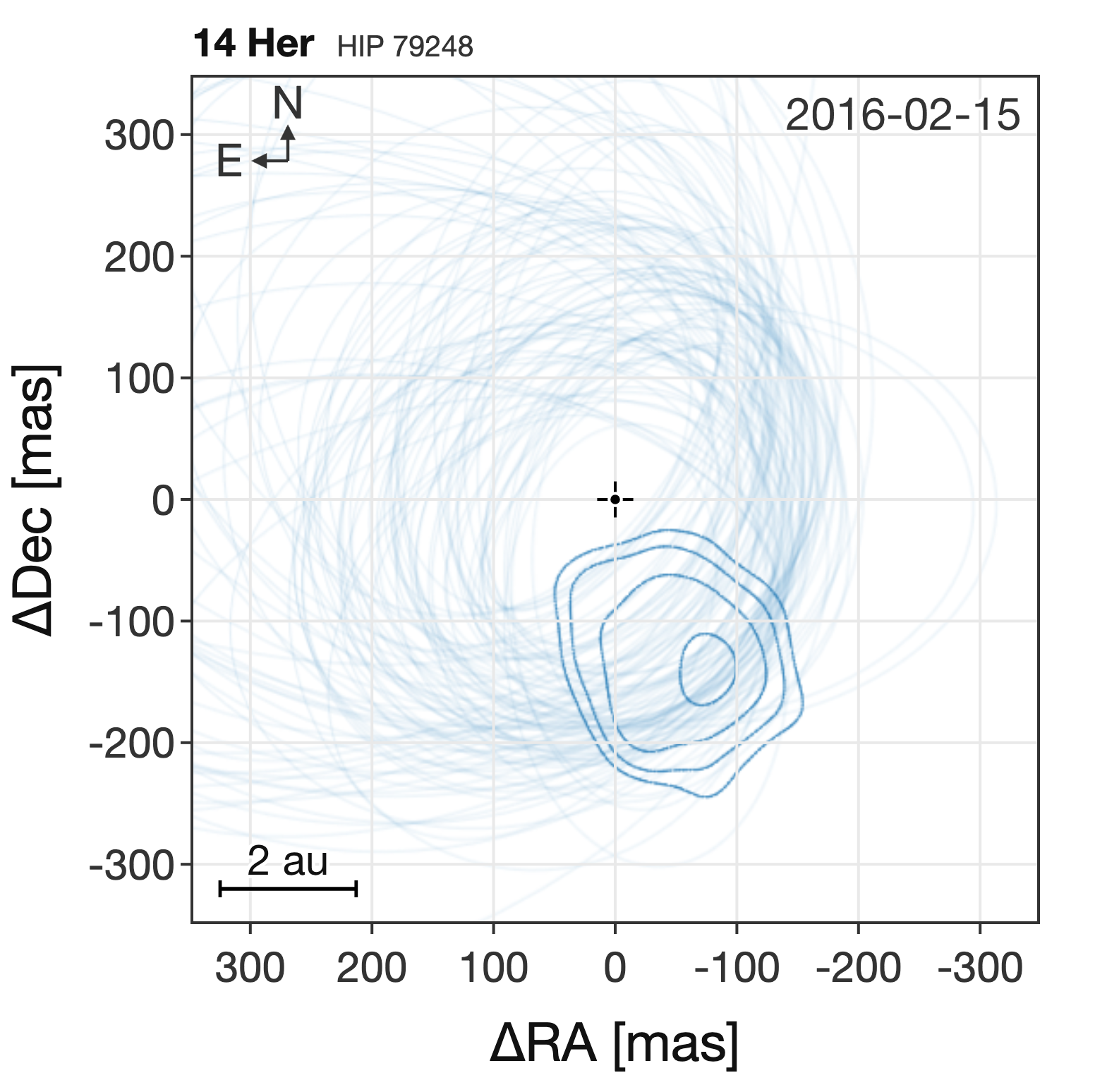}
\includegraphics[width=0.32\textwidth]{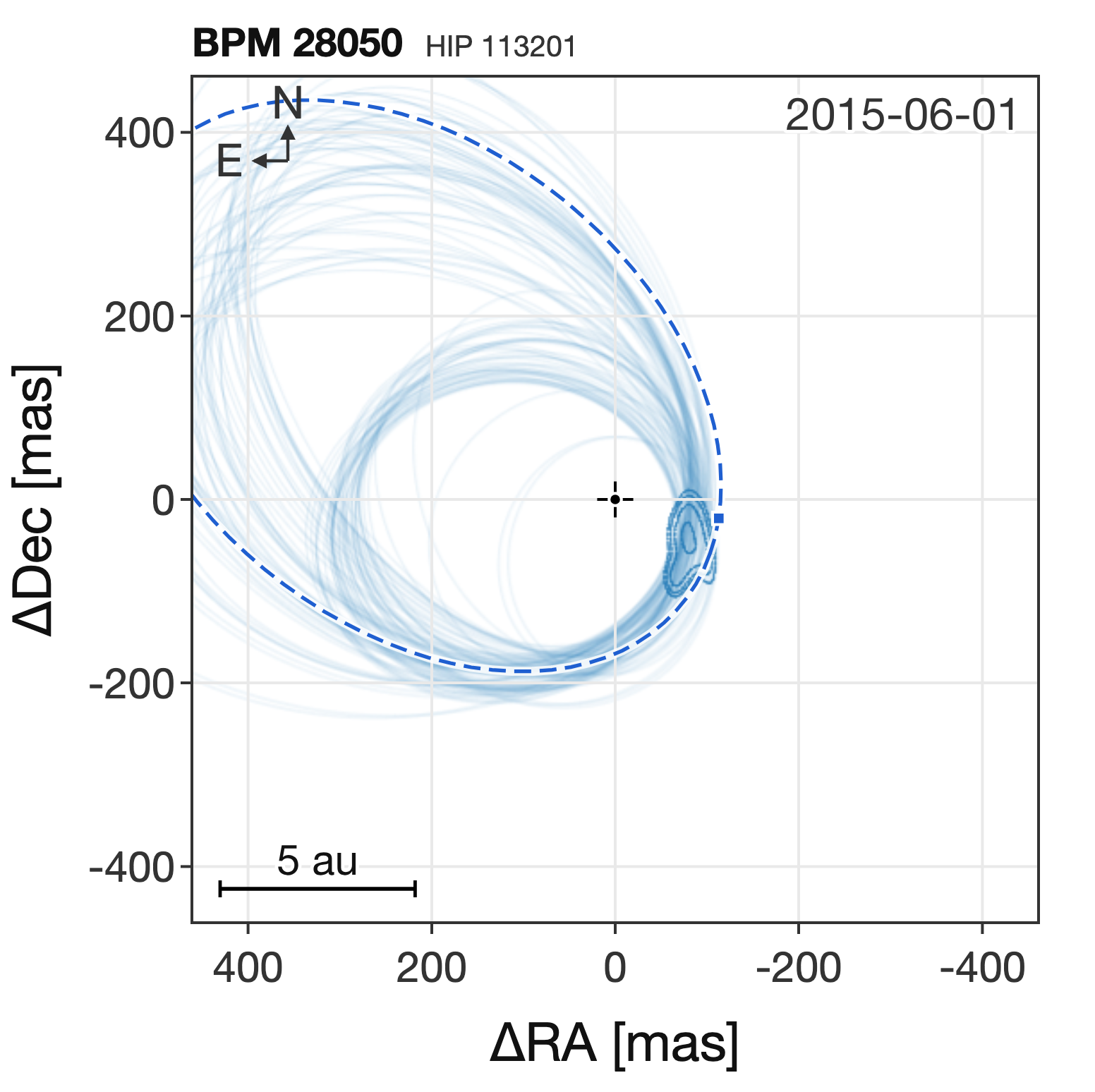}
\includegraphics[width=0.32\textwidth]{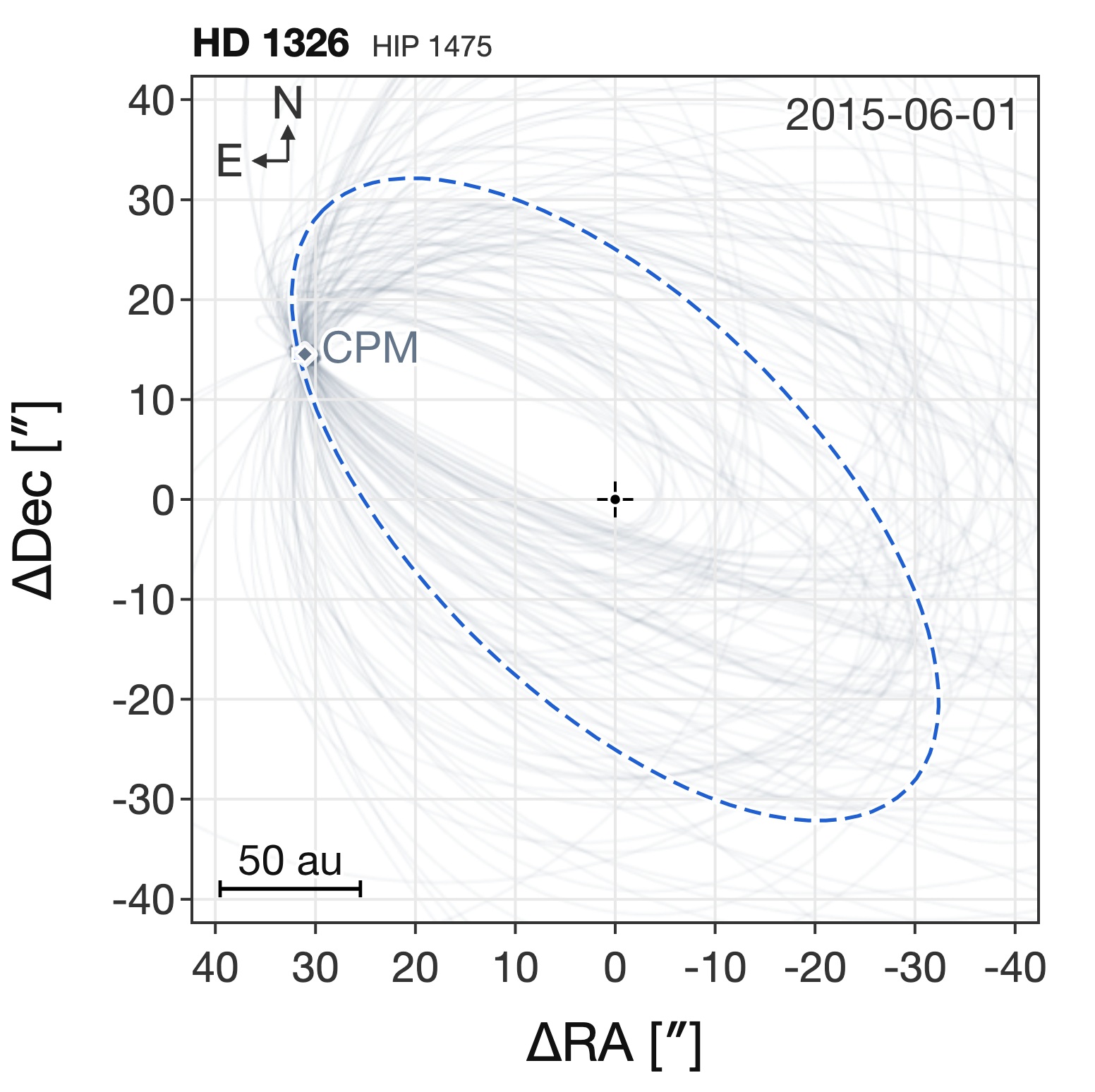}
\caption{Sky-projected orbit solutions for three well-detected companions, plotting only one-companion draws. Left: the planet 14~Her~b \citep{14herb}. Centre: Smethells~119 (HIP~113201), a well-characterized small-separation binary \citep{biller_hip113201}. Right: the wide-separation stellar companion HD~1326~B, detected as a common-proper-motion companion. The two stellar companions are compared with the ORB6 catalog orbit solution \citep{orb6}, informed by relative astrometry, radial velocities, etc. 14~Her~b has not yet been directly imaged.}
\label{fig:orbit-predictions}
\end{figure*}

\subsubsection{Orbit Predictions}

Unlike the detection maps presented in works such as \citet{pmex1}, these are complete posteriors containing values for every orbital parameter: eccentricity, inclination, and so on. 
For well-detected companions, this allows us to predict a sky-plane position at future epochs, as for example, in \citet{blakely_disks_g23h}. 

Figure~\ref{fig:orbit-predictions} presents sky-projected position contours and orbit tracks for single-companion draws of three different systems. In all three cases, we select epochs near the centre of the Gaia constraints. Selecting epochs further in the future requires propagating the orbit, which over a span of a decade, results in very loose constraints. Adding period information e.g., from radial velocities would improve the quality of future orbit predictions, as is well-understood in the joint  RV and proper-motion anomaly modelling literature.

Beginning with 14 Her (HIP 79248), we demonstrate that we are able to constrain the orientation and phase of 14 Her c to a region spanning $\sim100 \mathrm{\, mas}$ near the best epoch. This exercise is generally fruitful when there is significant motion over the DR2 vs. DR3 baseline such that the effective number of data points is greater than the effective number of degrees of freedom in the orbit.

The constraints for the low-mass stellar companion BPM 28050 (HIP 113201, or Smethells 119) are tighter, owing to its higher astrometric SNR. Here, we overplot the orbital solution from \citet{biller_hip113201} as collated by the ORB6 catalog. We find relatively good agreement, despite fitting largely different data: direct-imaging relative astrometry, radial velocities and the Hipparcos proper-motion anomaly in the literature case, against absolute astrometry from Gaia and Hipparcos alone in ours. 

HD1326 (HIP 1475) presents a different case: here the stellar companion is detected through our model as a resolved, common proper-motion source in Gaia DR3. Its orbit and position in 2016 is therefore well constrained by the catalog differences in position and proper motion between those two sources. 

We selected these examples to illustrate the ability of these posteriors to predict companion locations in the plane of the sky, but note that much of the time, the constraints are very weak and degenerate---sometimes only indicating a general position angle for acceleration solutions but not constraining the mass vs. separation. In the case of companions that are significantly luminous compared to the primary and who sit at projected separations between roughly $100  \, \mathrm{mas}$ and $800  \, \mathrm{mas}$, the position and orbit constraints are suspect due to the difficulty in accurately forward-modelling how Gaia treats marginally and intermittently resolved binaries. Our model reliably \textit{detects} these companions, but their inferred orbital parameters and multiplicity distribution should be treated with caution. This situation will improve with DR4, as the epoch can be used in place of the astrometric excess noise.

\subsubsection{Gallery}\label{sec:gallery}

\begin{figure*}
    \centering
    \includegraphics[width=0.88\linewidth]{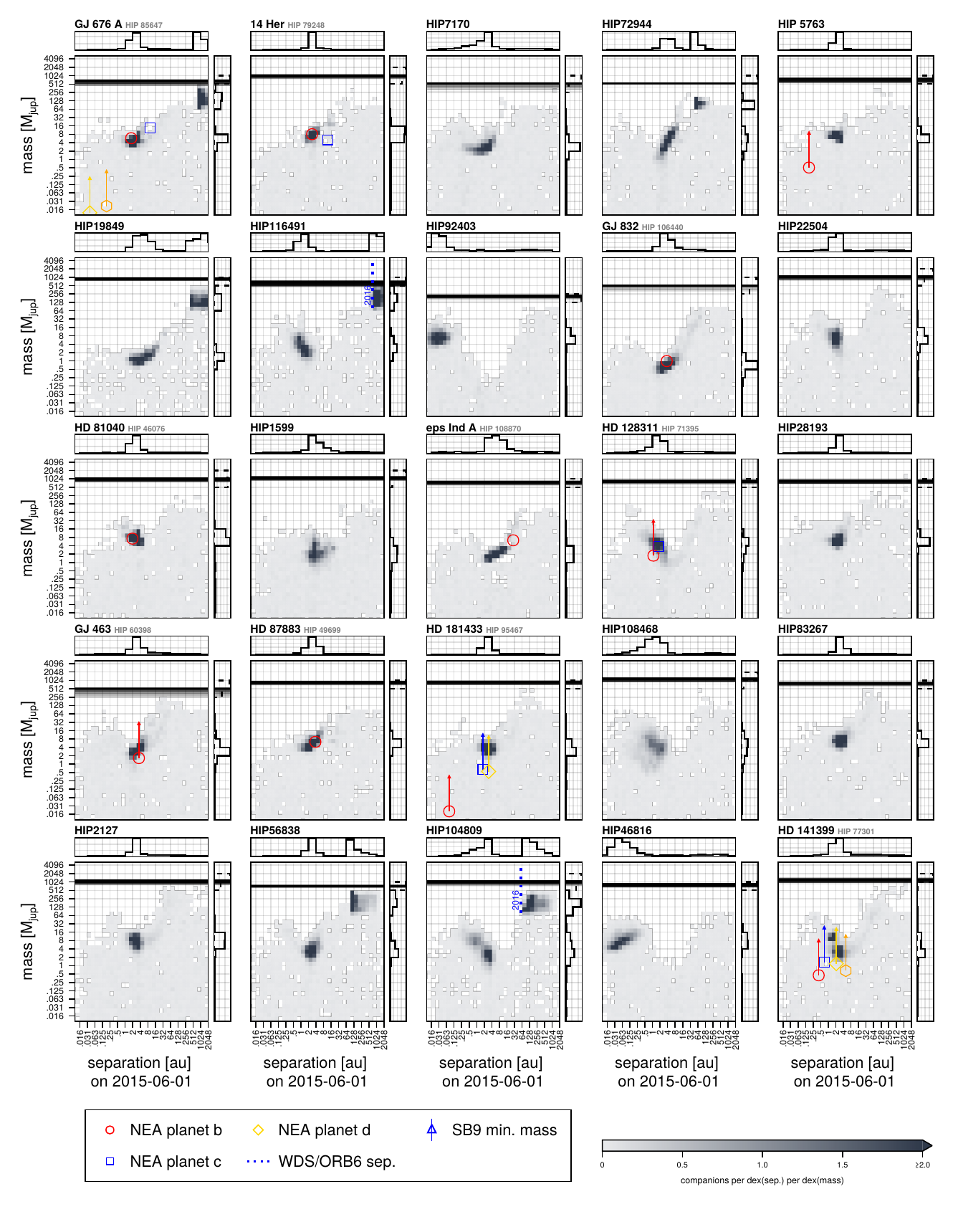}
    \caption{25 illustrative systems with planet-mass candidates out of the $\approx97\,\mathrm{k}$ star recommended Hipparcos sample. Each panel shows the posterior expected number of companions per dex in 3D separation (evaluated on 2015-06-01) per dex in mass. Integrating over an entire panel returns the posterior mean number of companions.  The horizontal lines are draws from the host mass. The black margin histograms summarize the expected companion density vs. separation and mass (solid: companions; dashed: primary). Overlays show known companions from the NASA Exoplanet Archive, WDS, ORB6, and SB9 catalogues.}
    \label{fig:gallery-1}
\end{figure*}

In Figure \ref{fig:gallery-1} we present an arbitrary sample of 25 systems that show evidence of companions that may be consistent with planetary masses.  We plot the ``3D separation'' on the horizontal axis which is the combined $x$, $y$, and $z$ separation from the star in au. We find that this quantity is often better constrained by the astrometry data than period or semi-major axis since it collapses the cloud of eccentric solutions with periastron near the Gaia observations onto the circular solutions that produce the same astrometric acceleration over this baseline.

As in \paperI, several posteriors are consistent with already-known companions collated from the NASA Exoplanet Archive and WDS visual binary catalog. For some known exoplanets, an upwards pointing-arrow is plotted to indicate the literature solution is for $m\sin(i)$ only and the true mass may be somewhat higher. Since our posteriors constrain the true mass and orientation of the orbit, it is also possible to compare the $m \sin(i)$ of each sample against the catalogued value for companions with only RV constraints.

\subsection{Validation}\label{sec:recovery}

We now discuss the validation we have performed on this model and posteriors and the quality cuts we recommend. 

\subsubsection{Internal Consistency}
We began by testing the internal consistency of our model and sampler through the use of simulation-based calibration \citep[SBC; ][]{talts_sbc}. To incorporate the real variance in data uncertainty between stars, we adopt the real measurement uncertainties 
and scan epochs for 1,000 Hipparcos stars within our larger sample, and replace their data with simulated companions and companion counts drawn from the priors. We perform a multiple-comparison corrected rank-uniformity test to check for issues with the sampler, or discrepancies between the model and simulator implementations. 

This test initially failed, revealing a bug in the G23H likelihood implemented in Octofitter, which we corrected and reported in an erratum to the original paper \citep{Thompson_2026_erratum}. This demonstrates this test has real utility for detecting subtle defects in statistical pipelines. 

The final model we adopted for this work passed these SBC criteria. Full results and description of the procedure are listed in Appendix \ref{sec:sbc}.

\subsubsection{Hipparcos Solution Types and the Recommended Sample}\label{sec:badfits}

The SBC tested that the model, simulator, and sampler were internally consistent; it cannot test whether the model is a faithful description of the catalog inputs. The main case where it is not concerns the Hipparcos solution type. Our model treats the catalogued Hipparcos position and proper motion as coming from a 5-parameter astrometric solution from a primary star and zero or more potentially luminous companions. The response of Hipparcos to luminous companions is forward-modelled, but the 5-parameter solution assumption is violated by a class of stars in the \citet{hipparcos_1} Double and Multiple Systems Annex. For a component solution (\texttt{MultFlag}~$=$~C) the catalogued astrometry refers to one resolved component, usually the primary, rather than to the photocentre; for an orbital solution (O) it is the barycentre, with the orbital motion removed at the catalogue stage. Our model is not faithful for these solution types. 

The \citet{hipparcos_2} reduction carried forward the reference points from this Annex, and so the two catalogues are typically  consistent. That said,   the G and V classes were reduced differently: some \citet{hipparcos_1} acceleration solutions changed to 5-parameter solutions in \citet{hipparcos_2}. The HGCA calibration \citep{hgca_dr3} (adopted by the G23H) combined both reductions in a 60/40 weighted average, mixing these two solution types in some cases, again making our model unfaithful to the underlying reduction process. 

The problem is most acute for the C pairs closer than $0\farcs4$, which Gaia does not resolve either, while for those at $0\farcs4$--$3''$ the mapping from Hipparcos component to DR3 source is itself ambiguous, and 676 such C entries are already removed by the DR2 data-quality cut (Section~\ref{sec:results}). We therefore exclude the C and O classes from the recommended sample: 8,497 of the 105,412 published posteriors (8.1\%), leaving 96,915 (Table~\ref{tab:selection}).

Unlike the DR2 data-quality cut, the stars removed by this selection consist of binary stars by construction.
For stellar-companion demographics those systems should be included as stars with strong evidence of binarity, but whose fitted period and mass ratio are not trustworthy. 

We flag but otherwise retain the remaining systems flagged by Hipparcos as acceleration (G), variability-induced mover (V), and stochastic (X) solutions, totalling 3,813 stars (3.9\% of the recommended sample), since in practice these appear to be well recovered.
Users who want a cleaner Hipparcos sample can drop them as well, which leaves 93,102 stars.

Per-star membership in every sample defined here is recorded in the header of each posterior file and in the summary table of the data release (Section~\ref{sec:data-availability}).

\begin{figure*}
\centering
\includegraphics[width=0.78\textwidth]{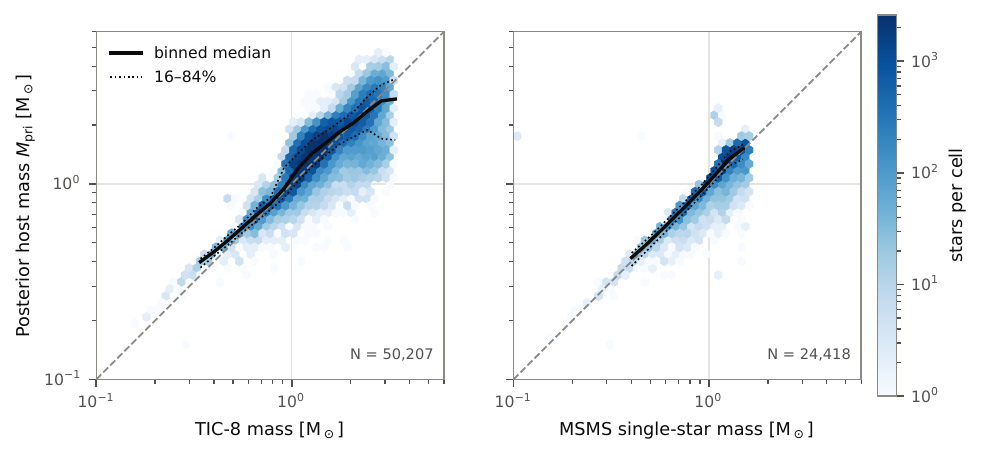}
\caption{Posterior median host mass compared against the TIC-8 mass (left; dwarfs only) and the MSMS BP/RP single-star mass (right) for the stars in common and within the ranges of each catalogue. Colour is the number of stars per cell; the solid line is the binned median of the posterior mass at fixed external mass and the dotted lines its 16th and 84th percentiles. We find adequate agreement, though where they disagree, the primary masses we infer may be more reliable thanks to the detection of astrometric and RV stellar companions, and treatment of their blended flux.}
\label{fig:host-mass}
\end{figure*}

\subsubsection{Host Masses}
We now examine if the inferred host-masses are consistent with external catalogs. The host masses in our model are constrained largely by Gaia photometry and parallax, though dynamical constraints do enter weakly for resolved, multiple systems. It is important to ensure the host masses are accurate because they are responsible for scaling the sampled companion mass-ratio into an absolute companion mass---especially for unresolved targets with no independent G-band photometry.

Figure~\ref{fig:host-mass} compares the posterior median primary mass with two external estimates that do not use our astrometric data: the TIC-8 mass \citep{tic}, available for the 50,207 stars that TIC classifies as dwarfs, and the MSMS single-star mass \citep{Li2025} from the Gaia BP/RP spectra. The MSMS masses saturate at the 1.6\,M$_\odot$ upper edge of that model's training set: 37,580 of the 61,998 stars with an MSMS mass sit within 0.01\,M$_\odot$ of it, 29,409 of these are reported at exactly 1.60. Those values are pinned at the grid edge, so we exclude them and compare against the remaining 24,418. Against TIC the median offset is $+0.026$\,dex with a robust scatter of 0.063\,dex; against MSMS it is $+0.013$\,dex with 0.020\,dex. Both comparisons are tightest below $\sim$1\,M$_\odot$, where the main sequence is unambiguous; the TIC comparison widens above $\sim$1.3\,M$_\odot$, where the TIC relation is $T_\mathrm{eff}$-based and does not resolve evolutionary state. 

We conclude that for stars meeting our assumed priors on age, mass, and metallicity, the host masses are sufficiently accurate that derived companion masses are not strongly biased. Regardless, the mass-ratios of dark companions are not impacted by the host mass.

\subsubsection{Known Companions}

We validate the posteriors' ability to recover companions across a wide range of orbital periods and masses by comparing against four reference catalogs:
ORB6 visual-binary orbits \citep{orb6}, SB9 spectroscopic orbits \citep{Pourbaix2004SB9}, detached eclipsing binaries \citep[DEBCat and TEPCat;][]{Southworth2015DEBCat,Southworth2026TEPCat}, and the NASA Exoplanet Archive \citep{Christiansen2025}.
We use the expected number of companions per square dex of period and mass as a way of testing the posteriors without imposing an ordering on the companions within each draw, or applying any detection threshold to the companions themselves.

We cross-matched our posteriors against these catalogues, keeping the systems in our recommended sample, and then retained only those systems for which our own posterior is confident that a companion of any kind is present, $P(N\ge1) > 0.95$. 
This cut leaves 2,011 catalogued entries over 1,854 systems: ORB6 379 (selected against by dropping the C/O Hipparcos solutions), SB9 single-lined 981, SB9 double-lined 315, NEA $m\sin i$ 150, NEA mass 127, and 59 eclipsing binaries. 

For each catalogued companion we recentre the intensity map on the catalogued values, working in $\Delta\log_{10} P = \log_{10}P - \log_{10}P_{\rm cat}$ and $\Delta\log_{10} y = \log_{10}y - \log_{10}y_{\rm cat}$, where $y$ is the companion mass when the catalog reports a mass and $m\sin i$ when it reports only an RV lower limit. ORB6 tabulates no companion masses, so for that catalog $y_{\rm cat}$ is the host mass scaled by the catalogued magnitude difference assuming $L\propto M^{3.5}$.

\begin{figure*}
\centering
\includegraphics[width=\textwidth]{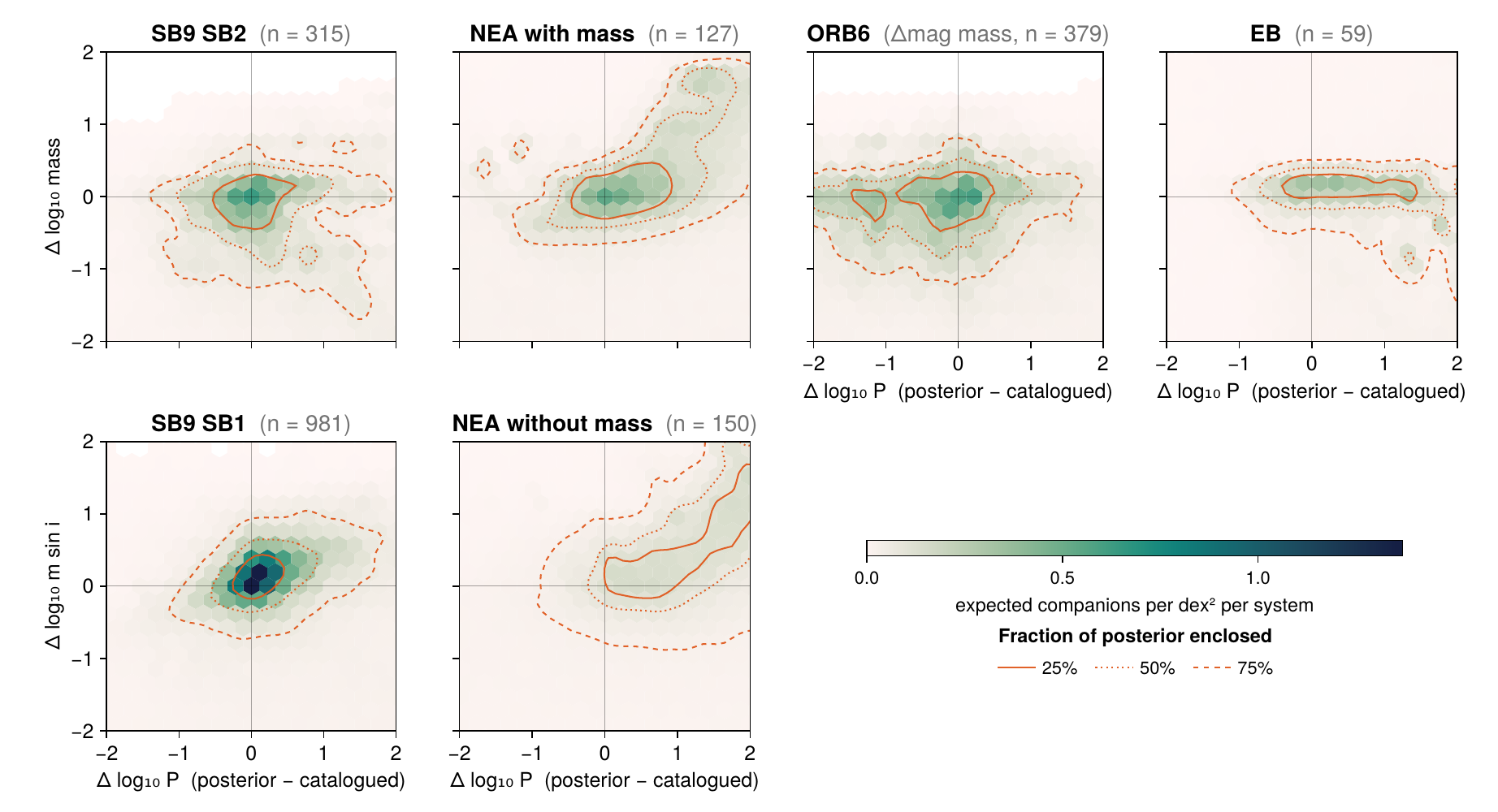}
\caption{
This figure shows the recovery of known companions from four catalogues. Each panel shows the expected number of companions per system, per square dex of period and mass (or $m \sin{i})$, shifted such that the catalog truth is (0,0) for each system. Systems with multiple catalogued companions are included multiple times with different shifts. 
Higher values near (0,0) are better, as that indicates that more companions are sensed (across stars) and that the posteriors are peaked towards the catalog values.
Entries are restricted to systems whose posterior is confident a companion exists, $P(N\ge1) > 0.95$. The contours enclose 25\%, 50\% and 75\% of the stacked posterior intensity lying inside the plotted window.
We find that recovered peak is well centred on the catalog values across populations. The posteriors for many NASA Exoplanet Archive (NEA) systems have long tails out to wider periods and higher masses, but averaged over the population and after shifting, the tails do not overlap as consistently as the peaks. 
Some amount of multi-modality is expected since some populations like ORB6 wide binaries may be correlated with having additional inner companions. The ``NEA without mass'' recovery is poor, because most companions that are known only through radial velocity fall below our sensitivity limits. The weight towards higher mass and longer periods may be the contribution of outer companions to cataloged inner companions.
ORB6 masses are estimated using only the $\Delta\,\mathrm{mag}$. }
\label{fig:known-recovery}
\end{figure*}

The resulting shifted, stacked posteriors are presented in Figure ~\ref{fig:known-recovery}.  The darker the heatmap, the greater the fraction of posteriors that recover the object, and the more posterior probability mass is placed on that $(\Delta \log_{10}P, \log_{10}y)$ bin across the population. In systems where we do not have sensitivity to the object, that posterior contributes only the conditional prior (smooth $\approx 0.02 \, \mathrm{dex^{-2}/star}$; zero where data rules out companions).

We find spectral (SB9), eclipsing (``EB''), and resolved binaries (ORB6) are all well recovered, as are exoplanets with catalogued true masses. Exoplanets (``NEA'') with only RV measurements available are not well-recovered as a whole, as this class of object is typically lower mass and closer to the star.

\subsubsection{Near-equal-flux binaries}\label{sec:equal-flux}

Near-equal flux binaries at moderate separations are one of the most challenging hazards for planet searches using astrometry. Here, we describe the challenge, sources of data that can disambiguate, why the challenge is perhaps less severe in practice, and assess the remaining failure cases in these posteriors.

For a given pair of bodies, Gaia and Hipparcos each measure the photocentre which is displaced according to
 $(q-f)/[(1+q)(1+f)]$ for mass ratio $q$ and flux ratio $f$.
For a dark companion like a planet, $f$ =0,  the motion of the photocentre is simply the motion of the primary.
When $f=1$  for exactly equal-flux binaries, the observed photocentre motion is zero. Finally, for a pair of stars with only a small difference in flux, the observed photocentre motion reduces to an amplitude that is entirely ambiguous between a planet and a star (e.g.: either $f=0$ and $q=0.01$, or $f=0.98$ and $q=1$). 

While these two cases are indistinguishable from astrometric position measurements alone, Gaia provides additional signals that can help to split these two cases. The first signal is the combination of photometry and parallax: for a given age, metallicity, and so-on, a source consisting of two unresolved near-equal-flux binaries should be twice as bright, at a given distance, but have otherwise the same colour as a single star. This information is incorporated into our models.

The second signal is Gaia radial velocities. By and large, this ambiguity does not exist for RV data, and so an elevated RV variability tilts the scale towards the stellar case. This information does, however, fall off  in our model for very short period near-equal-flux binaries. At short periods the orbital velocity separation is large and the two cross-correlation peaks well resolved. In Gaia DR3, sources with 10\% or more transits flagged as double-lined had their combined radial velocities discarded, as did those whose combined radial velocity uncertainty exceeded $40\,\mathrm{km\,s^{-1}}$ \citep{katz_dr3_rv}. \citet{paired} identify these same two root causes as the reason for the reduced sensitivity of their method in the short-period, near-unity mass-ratio corner. 

Of the SB9 double-lined systems in our cross-match---all of which are bright enough for the RVS, with a median $G = 7.6$---only 13\% of those with $P < 10$\,d have a published DR3 radial velocity, rising monotonically to 73\% beyond $P = 1000$\,d. Such systems therefore carry no RV datum in our model at all, rather than an uninformative one.  
The short period, however, means that astrometric positions would only be sensitive to stellar companions, and therefore the ambiguity between the two cases would not lead to the inference of a planetary massed companion.
This source of information finally fails at wider separations because, while astrometric sensitivity increases with period, the RV sensitivity does not, and eventually is insufficient to detect even a stellar companion.

Finally, a third signal is the ability of Gaia to resolve or at least detect that the image of the source consists of two separate objects. At very wide separations, this information enters into our model through separate, neighbouring Gaia sources.

For close separation objects inwards of Gaia's formal resolution limit, an additional channel of information reported by Gaia may be relevant:  the Gaia IPD multi-peak fraction which is the fraction of scans where more than one source was detected. This information is not used directly in our model but can be used as an external binarity indicator. In practice, however, elevated IPD multi-peak fraction tends to inflate the RUWE, which is included in our model and is interpreted as a Keplerian signal. Even when the IPD multi-peak fraction is 0, this additional source of peak-extraction noise may appear. 
Although the peak extraction ambiguity is a source of noise which does not match the Keplerian perturbation assumption,  it nonetheless tends to result in the posteriors inferring the presence of a stellar companion---albeit with incorrect parameters, often near 1-year periods so as to maximize the amount of RUWE explainable from the companion.
This means that the presence of companions even in this intermediate separation regime are often detectable in practice. 

For those near-equal-flux companions that are too widely separated to be detectable with RV and too close in to be detectable as either a separate source, through elevated IPD multi-peak fraction, or the peak-extraction related elevated RUWE, the only source of information remains the over-luminosity from blended photometry. Due to the way these sensitivity channels scale with distance, the ambiguity is least problematic for nearby stars
 while very distant stars do not have astrometric sensitivity to planetary companions in the first place. This leaves planetary massed solutions with moderate separations around moderate distance stars as the most susceptible regime. 

Thankfully, for systems closer than approximately 30 pc, the RV sensitivity to moderate separation equal-mass binaries extends out to or beyond the separation at which a separate source would be cataloged by Gaia, and this ambiguity disappears.

\subsection{Data Availability}\label{sec:data-availability}

The full posteriors of the 105,412 stars that pass the input-level cuts of Table~\ref{tab:selection} will be made available as one FITS file per star. These posteriors, along with the summary table, will be hosted at the Canadian Astronomy Data Centre upon acceptance. The same posteriors, reduced to a minimal set of columns, can be browsed and plotted interactively at \url{http://exopost.wthompson.space} under the label ``G23H''.

Each star is one file, \texttt{HIP<n>.fits}, with four HDUs. The primary HDU is a header only that lists identifiers and the per-star metadata: the Hipparcos number, the Gaia DR3 \texttt{source\_id} of the fitted source and the Gaia DR2 \texttt{source\_id} that supplied the DR2 data (both are given because the G23H DR2--DR3 cross-match is not one-to-one); the data included in the fit, as a compact label (\texttt{DATASET}, e.g.\ \texttt{hip+hg+dr3+dr2+dr32+ueva+rv}) and one boolean flag per optional element; the Hipparcos annex solution type and the recommended, clean-Hipparcos and resolved-white-dwarf flags of Table~\ref{tab:selection}; the sampler configuration; the final-round Pigeons log-evidence estimate and global communication barrier, and the start and stop times; and the numerical parameters of every prior that varies from star to star: the parallax, proper-motion and radial-velocity priors, the extinction prior, the three UEVA noise priors and the number of predicted Gaia transits.

The \texttt{POSTERIOR} extension is a binary table with one row per posterior draw (4,096 rows) and one column per parameter or derived quantity. The column count varies from about 180 to 640 with the number of Gaia transits. Every column is documented in Table~\ref{tab:posterior-columns}.

The \texttt{PIGEONS} extension holds the round-by-round parallel-tempering summary (scans, restarts, communication barrier, wall time and the stepping-stone evidence estimate after each round). 

The \texttt{MODEL} extension stores, verbatim, the model definition printed by the sampler for that star: every prior, the expression for every derived quantity, and the table of astrometric data that were fit.

Companions are exchangeable within a draw: because the model fills ``slot'' $b$ first, \texttt{b\_planet\_present} averages to $P(N\ge1)$, \texttt{c\_planet\_present} to $P(N\ge2)$, and so on, but the orbital parameters of the active companions have no intrinsic ordering between draws. Before summarizing per-companion quantities, an ordering should be applied by sorting by companions within each draw by any criteria of choice.

We store the log-prior and log-likelihood with every draw, so re-weighting the prior on $N$, mass, period, eccentricity, etc.\ by downstream users is straightforward. In the same way, additional constraints from direct imaging or precision radial velocity can be applied through reweighting/re-sampling.

The release also contains a single summary table, \texttt{summary\_table\_v27.csv}, with one row for each of the  published stars and the 49 columns listed in Table~\ref{tab:summary-columns} with source identities, flags, diagnostics, and some parameters summarized from the posterior for convenience.

Every posterior in the release was produced with \texttt{Octofitter} \citep{octofitter} v8.1.5 at commit \texttt{af7becc68359138a43738b767ec087d017366e62}, with \texttt{PlanetOrbits} v0.11.1 and \texttt{Pigeons} v0.4.11 \citep{pigeons} under Julia 1.11.3 \citep{julia_bezanson}. The simulation-based calibration of Appendix~\ref{sec:sbc} was run against this same commit.

\section{Discussion}\label{sec:discussion}\label{sec:limitations}

\begin{figure}
    \centering
    \includegraphics[width=\columnwidth]{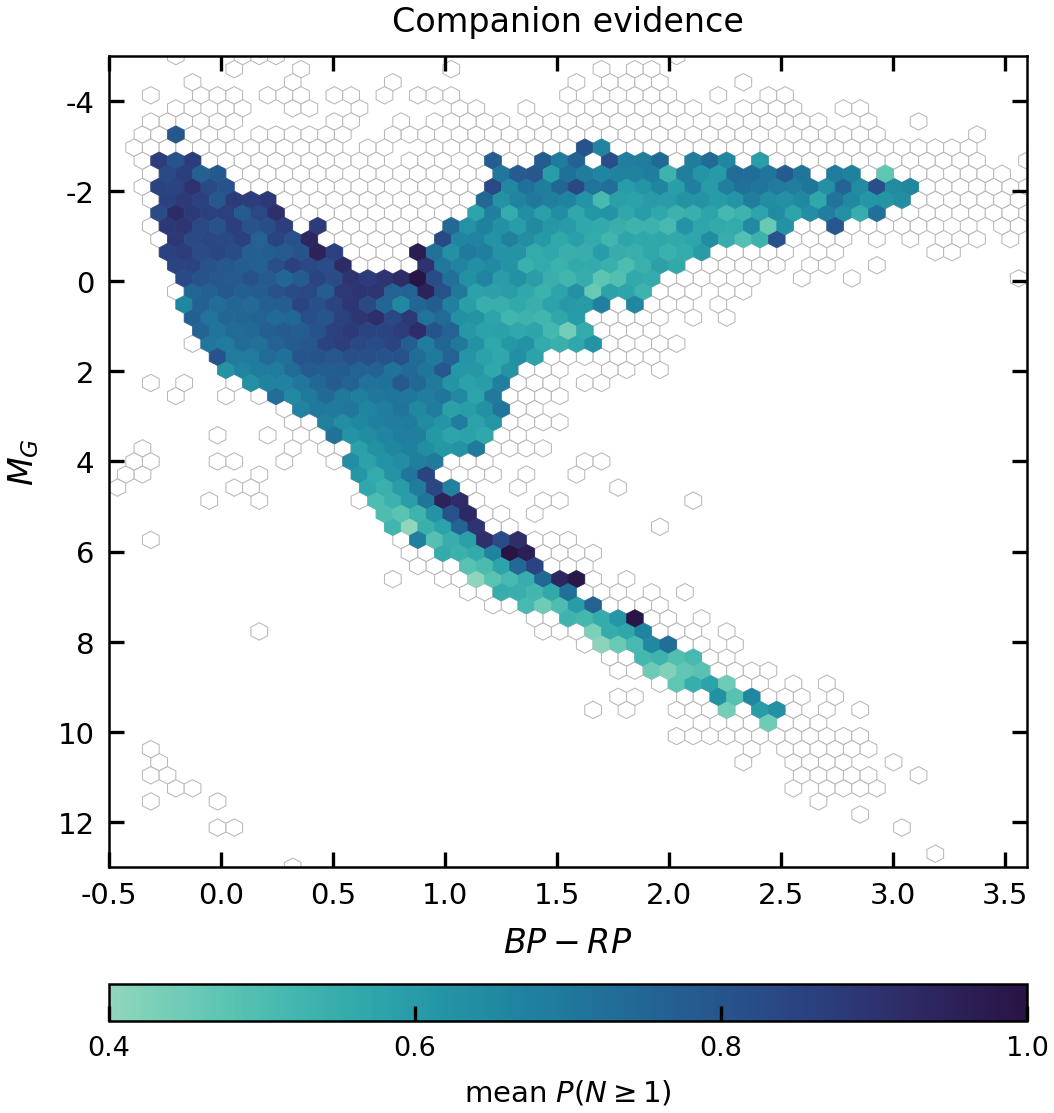}
    \caption{
        Colour--magnitude diagram of the recommended sample of 96,915 systems, coloured by the mean posterior probability that systems in that bin host at least one companion, $P(N\geq1)$. Figure~\ref{fig:cmd} reports instead the number of systems per bin.  Bins with fewer than 10 stars are excluded.
        Companion evidence is highest along the band $\sim$0.3--0.75\,mag above the main sequence, where an unresolved near-equal-luminosity companion displaces the primary. Earlier types also show higher rates of detected companions. This work and this figure do not consider sensitivity, and should not be read as an occurrence rate. This figure shows similar trends to that Figure 5 of \citet{paired}}
    \label{fig:cmd_companions}
\end{figure}

We now highlight limitations of the model.

\subsection{White dwarf companions} 
\begin{figure}
    \centering
    \includegraphics[width=0.32\textwidth]{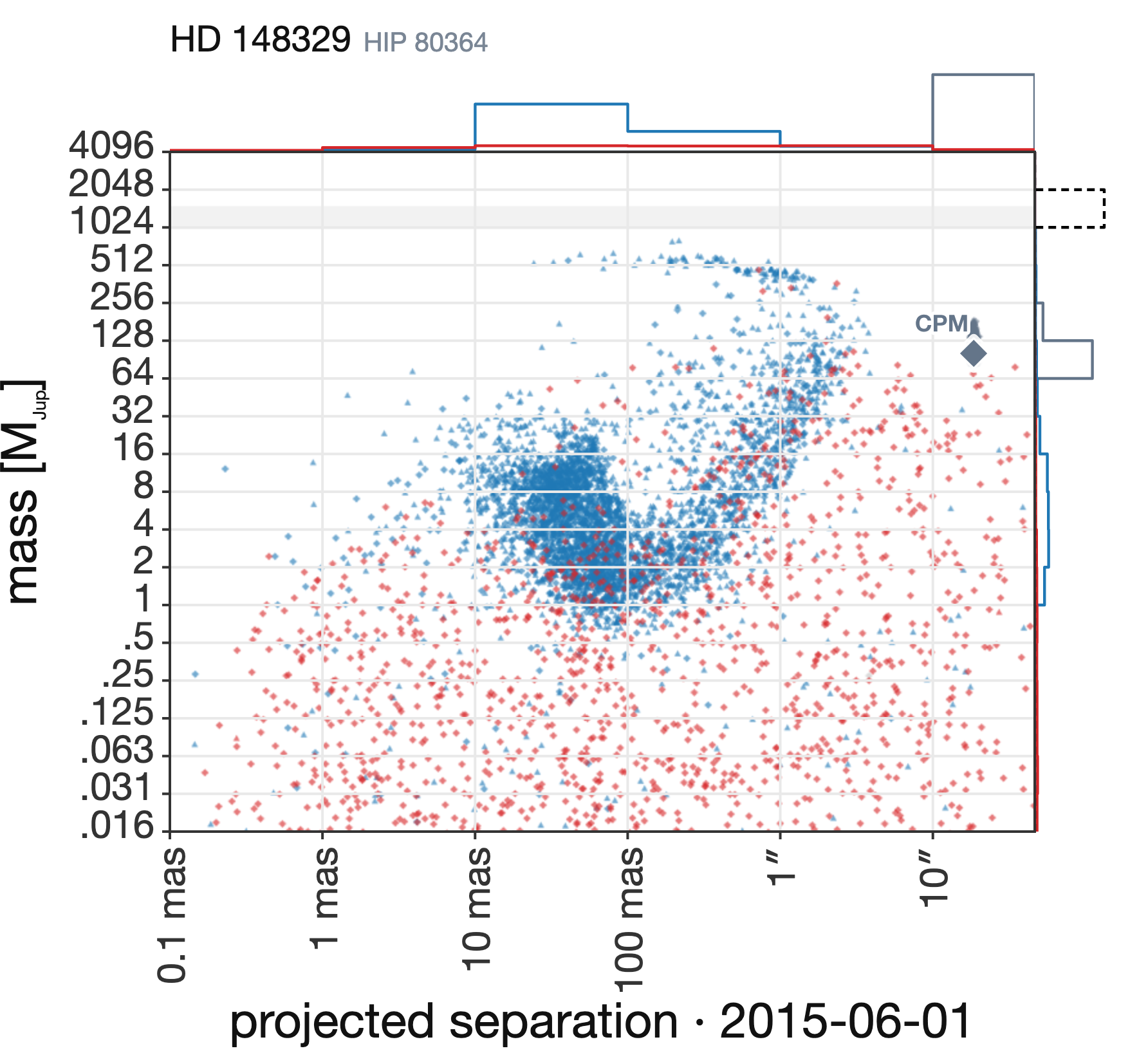}
    \includegraphics[width=0.32\textwidth]{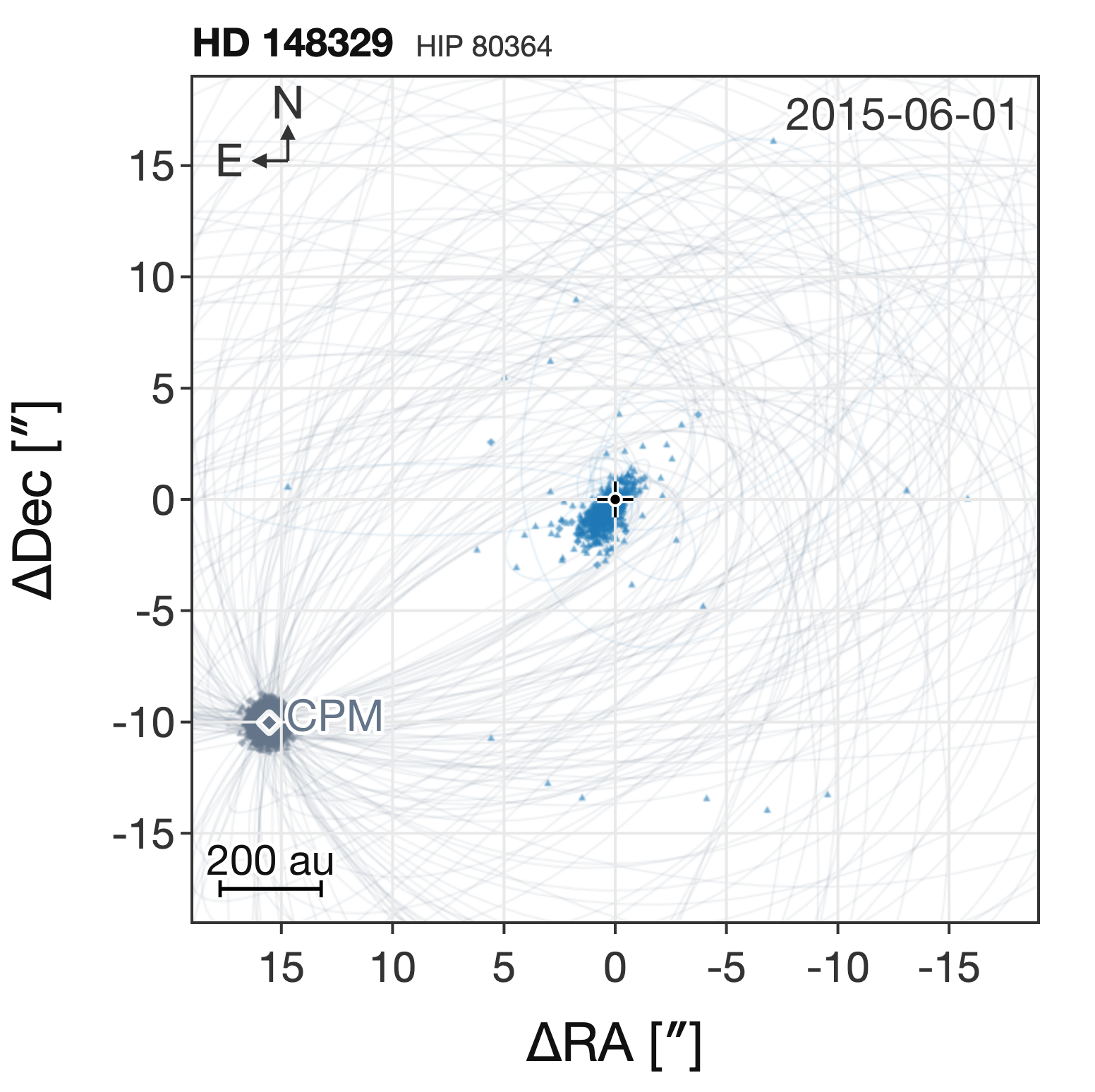}
    
    \caption{An example of an incorrect posterior due to the system falling outside our prior bounds. The star HD~148329 has a likely white dwarf companion, HD~148329~B \citep{gentile_fusillo_wd,vincent_wd_xp}, identified as a bound common-proper-motion pair by \citet{elbadry_binaries}, but our model only admits main sequence companions. As a result, under the model assumption, the companions Gaia G flux rules out a companion more massive than $ 80 \mjup$. The model then invokes an additional (unnecessary) inner companion, marked in blue, to soak up the remaining astrometric signal. This failure may be diagnosable by examining the position angle of the wide-separation tail of the additional companion, and noting that it points towards the neighbouring source. The Gaia BP - RP colour (not considered for companions in our model, out of concerns for blending) may also indicate a neighbour that does not meet our assumptions. }
    \label{fig:hd148329}
\end{figure}

The mass to light relation we use in our models assumes that the primary and all companions have the same age, [Fe/H], $A_0$, and have not evolved past the core-Helium burning stage. In most cases, a cut can be made on the brightest component's photometry to ensure this assumption is met for a given system, but white-dwarfs frustrate this simple approach: they may be present as companions, darker than the primary, massive enough to produce detectable accelerations, and yet darker than our mass-to-light model assumes. This is most impactful in the case of resolved companions where the object has an independent photometry constraint. Figure \ref{fig:hd148329} may be an example of one such system. In order to make up a shortfall in mass given the observed photometry of the resolved companion and mass-to-light assumptions, the model adds an additional companion. In sky-projected coordinates, the unnecessary companion is seen extending in the direction of the white dwarf. 

In practice, the posteriors for systems that violate this assumption are often not seriously impacted: resolved companions may be sufficiently distant from the primary to not produce any detectable astrometric signal. Or, when they are sufficiently distant to not have corrupted photometry, that independent photometry can be used to classify the companion as a potential white dwarf. We  cross match our sample against \citet{gentile_fusillo_wd} and flag a set of 157 such companions.
When unresolved, the incorrect mass-to-light model would be expected to lead to an inaccurate inference on the mass, but still above the planetary mass-regime.

\subsection{Companions that are themselves tight binaries or multiples} 
\begin{figure}
    \centering
    \includegraphics[width=0.32\textwidth]{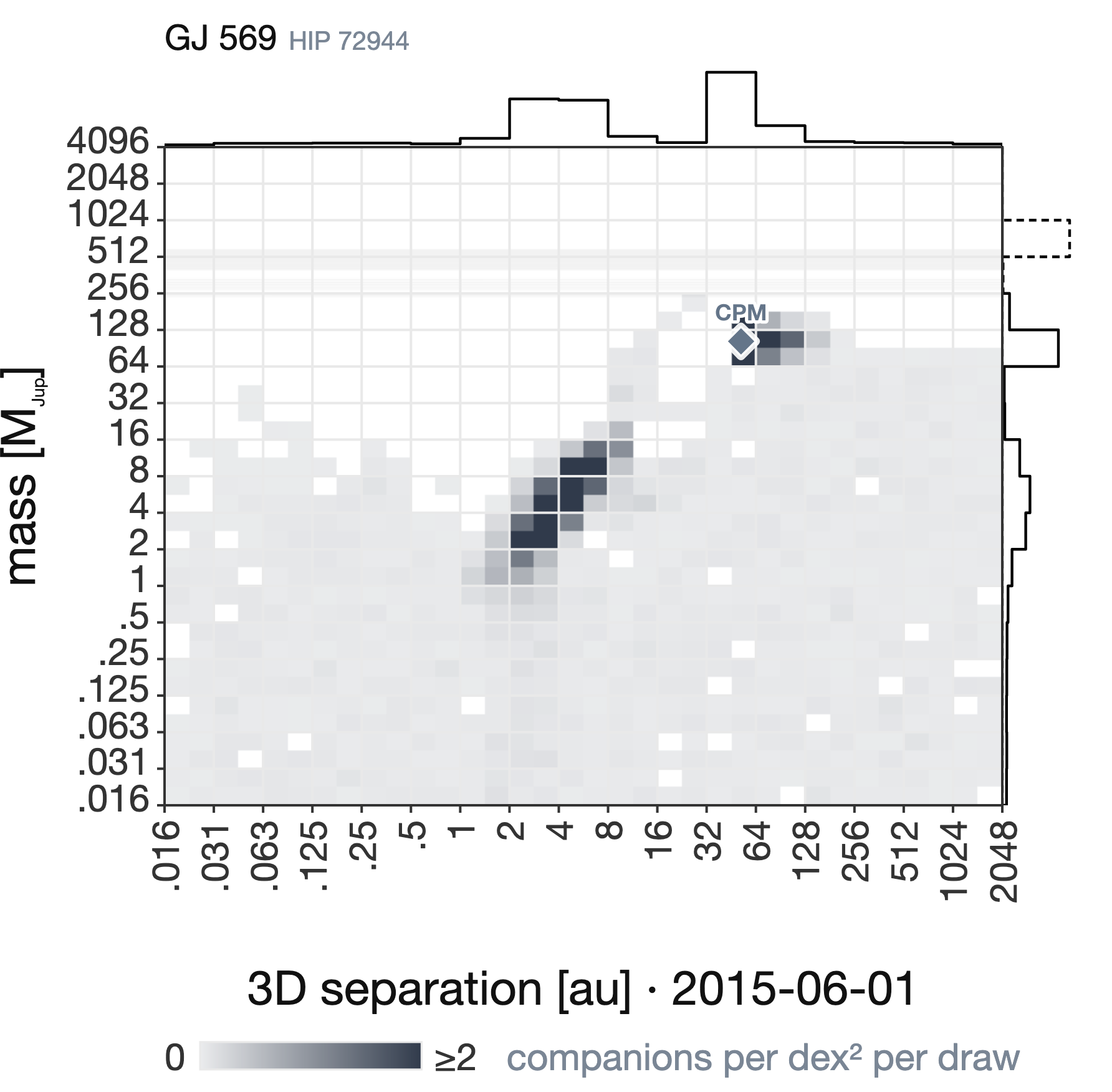}
    
    \includegraphics[width=0.32\textwidth]{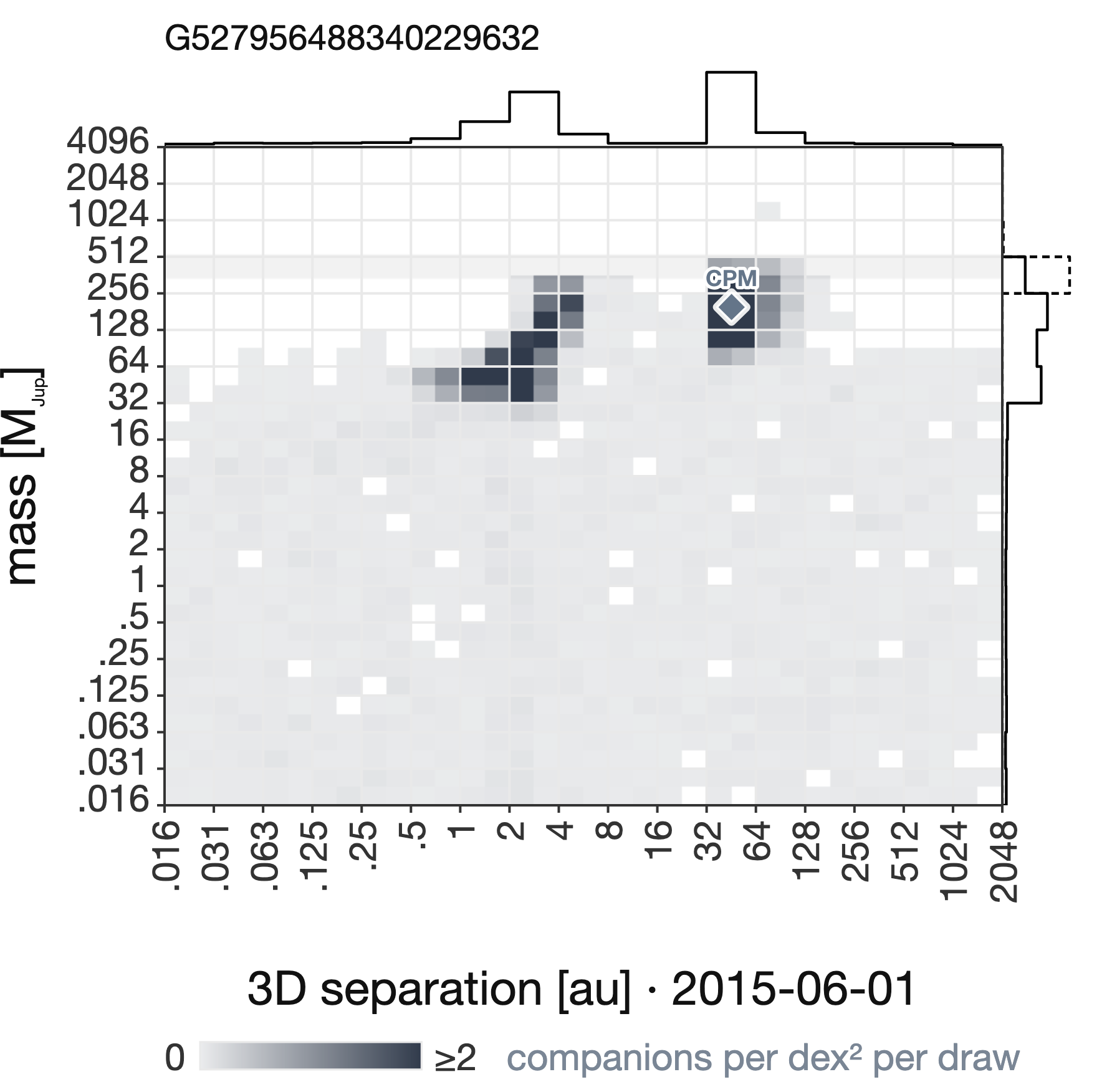}
    \caption{A failure and successful case of modelling hierarchical systems. Top: GJ 569 is an example where the common proper motion companion (flagged ``CPM'') is in fact an unresolved binary, too dark to explain the observed astrometry under the single-star mass-to-light model, which is compensated by an additional spurious dark companion. Bottom: GJ 22 is a nested hierarchical system with two companions that do not violate the model assumptions.}
    \label{fig:gj596_gj22}
\end{figure}

A similar limitation from our mass-to-light model: we assume that each companion has a dynamical mass that corresponds to its overall g band absolute magnitude, given the system-level age, parallax, [Fe/H], and $A_0$ parameters. Hierarchical systems violate this assumption by being too faint for their mass, the same direction as the white dwarf case. As a benchmark example, consider the posterior for GJ 569 in Figure \ref{fig:gj596_gj22}. The ``companion'' detected at a separation of roughly 64 au in fact consists of two unresolved M dwarfs. The mass-to-light model rules out the slightly higher-massed solutions that might have been sufficient to explain the observed astrometric signal, forcing the addition of a second, closer in companion. 
By contrast, hierarchical systems that form a Jacobi chain as in GJ 22 (second panel of Figure \ref{fig:gj596_gj22}) are well-modelled.\footnote{The posterior for GJ 22 itself is not part of this data release as it fails one or more cuts.}

\subsection{Companions that Saturate Gaia}
If a star were included in this sample that possessed a brighter companion that saturates Gaia and is therefore not catalogued, it would not appear in our list of potential common proper motion companions, again distorting the posterior. This would occur for instance with Sirius B: the much brighter Sirius A does not appear in Gaia DR3, and would appear ``dark'' to the companion matching likelihood. In practice, we have not found any instances of this occurring in this sample  (in which Sirius B does not appear) but we mention it here for completeness. 

\subsection{Non-unique crossmatches between Hipparcos, Gaia DR2, and Gaia DR3 }
In general, the cross matches between Hipparcos and Gaia DR3, and between Gaia DR2 and DR3, are not one-to-one. The G23H catalog lists 1,291 of our DR3 sources twice, with two candidate DR2 counterparts, and we adopted the first. The DR2 data-quality cut (Section~\ref{sec:results}) removes 659 of the 1,272 of these that we fit; for each of the 613 that remain in the release, the adopted DR2 counterpart is the one Gaia's own DR2--DR3 cross-match identifies, and 499 of them are in any case outside the recommended sample. Because the cut is defined on the catalog alone, it may not catch every incorrect match, and additional scrutiny is warranted for sources with multiple DR2 vs DR3 or DR3 vs Hipparcos cross-match pairs.

\subsection{Contaminated Gaia DR3 solutions}
\begin{figure}
    \centering
    \includegraphics[width=0.32\textwidth]{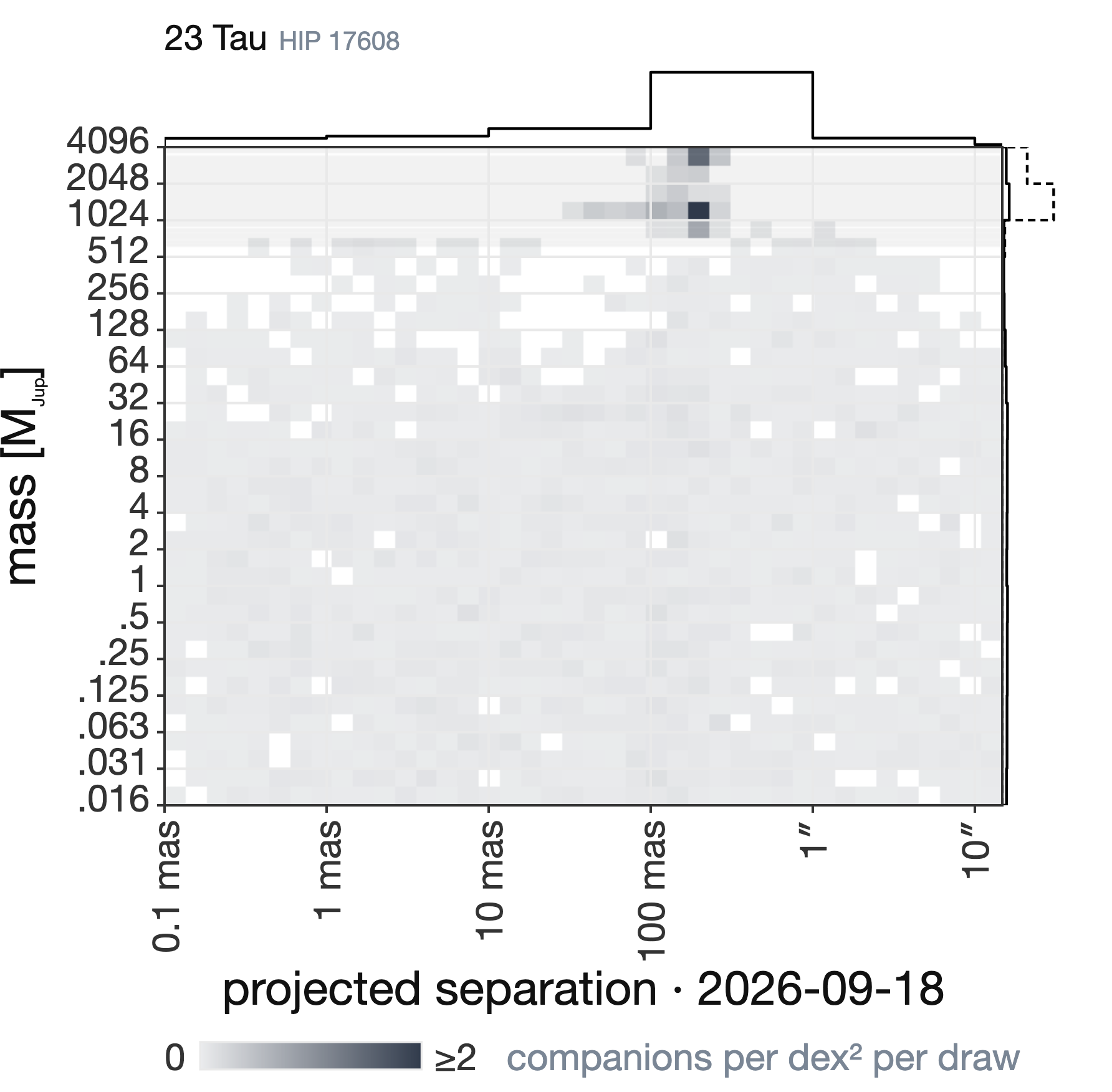}
    \includegraphics[width=0.32\textwidth]{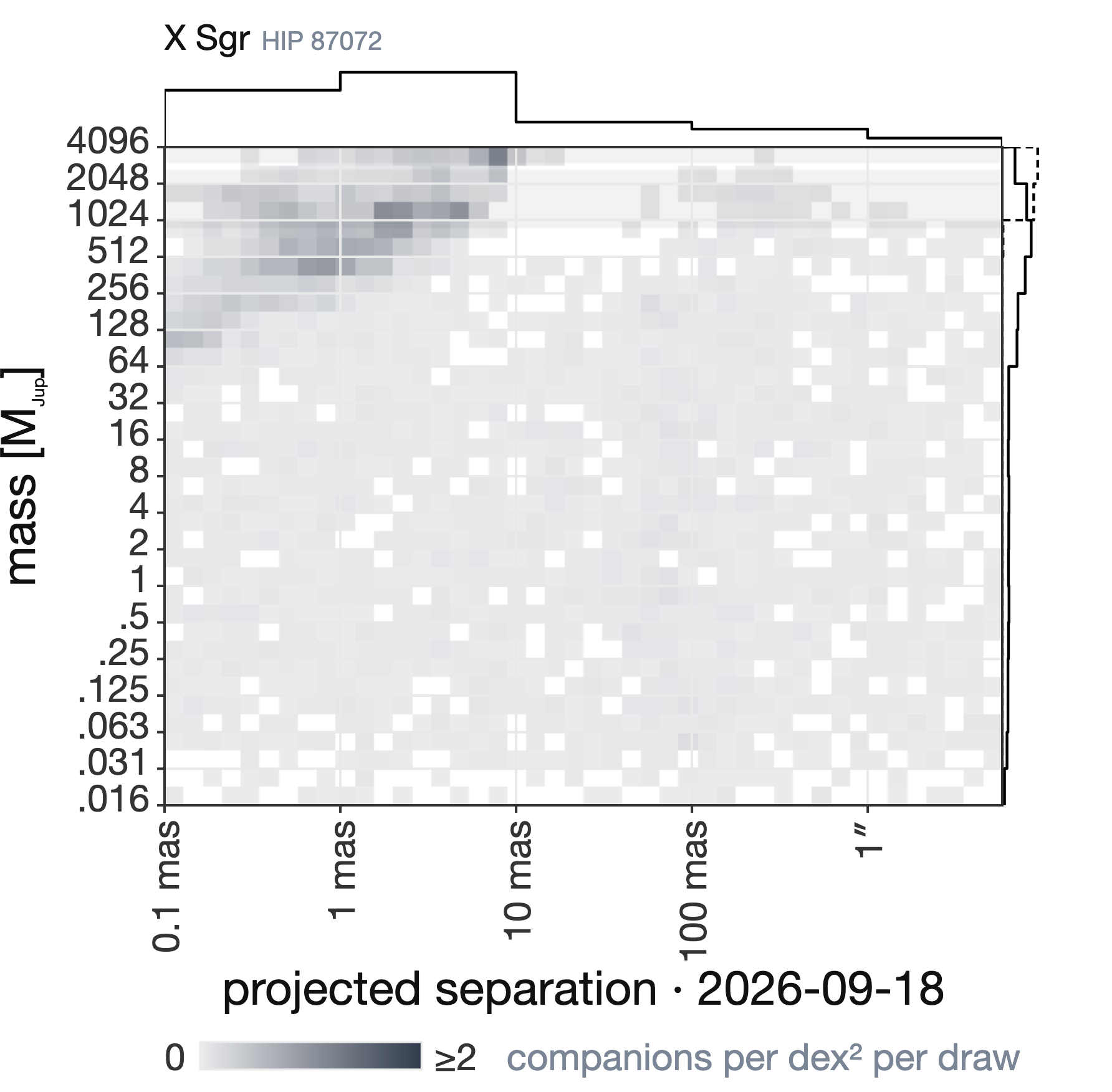}
    \caption{Companion signatures 23 Tau (Merope; HIP 17608) and 
 X Sgr (HIP 87072), systems.  These signatures may be genuine, but we highlight these two systems as ones which may be impacted by bright backgrounds or crowding, respectively. The source environment should be considered when evaluating candidates.}
    \label{fig:fit_environment}
\end{figure}

Some Gaia DR3 solutions may be contaminated by other sources in dense fields (over-crowding), proximity to bright stars, or bright backgrounds. These may inflate the Gaia RUWE, which in the G23H framework, is attributed entirely to Keplerian motion. As such, the posterior would infer a companion where none may exist. For example, the star 23 Tau (Merope; HIP 17608) appears in our posteriors to posses a companion at 0.1-0.3" separation, and the star X Sgr appears in our posteriors to possess a companion inwards of 10 mas. Both are plotted in  Figure \ref{fig:fit_environment}. These may either be genuine companions or the result of bright backgrounds or crowding respectively. This situation will become more common when considering fainter hosts accessible with Gaia DR4.

\subsection{Elevated RUWE from near-equal flux binaries}
\begin{figure}
    \centering
    \includegraphics[width=0.32\textwidth]{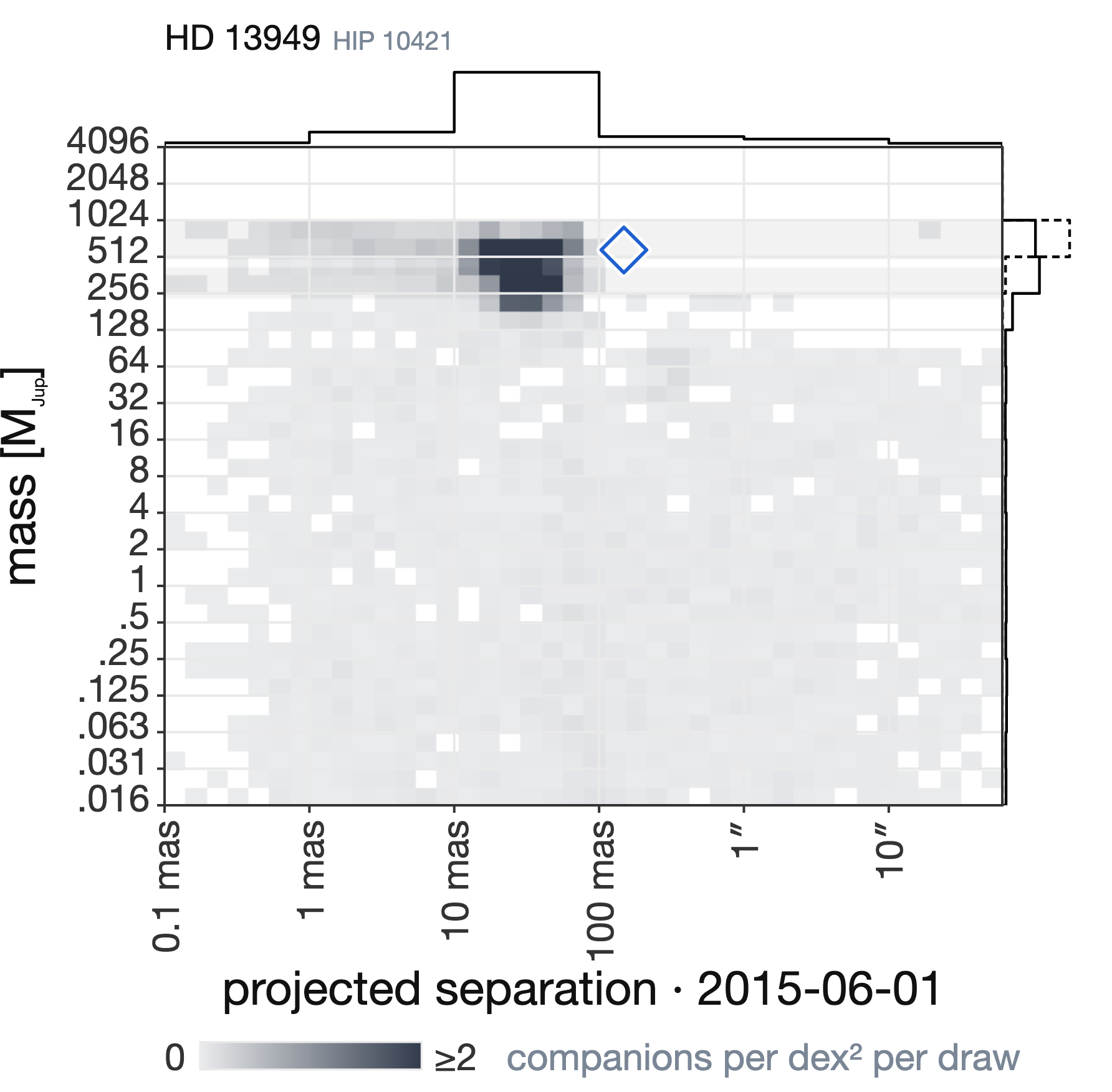}
    \caption{An example of an equal flux binary with an elevated IPD multi-peak fraction of 39, inflating the RUWE beyond what the UEVA model expects. This pushes the posterior towards solutions near 1-year periods.}
    \label{fig:hd13949}
\end{figure}

When Gaia observes a near-equal flux binary system  (see Section \ref{sec:equal-flux}), Gaia's peak-extraction may detect multiple peaks. Even when multiple peaks are not flagged, the centroiding may simply have increased scatter from a broadened, blended PSF. This appears to increase the RUWE in Gaia's fits, separately from the Keplerian motion induced by the companion. 
We see evidence that this increased RUWE tends to produce a companion in our model with a shorter period year than the true companion's orbit, often near 1-year. As such, while near-equal flux binaries are typically \textit{detected}, the accuracy of the orbit constraints are poor.  HD 13949 is presented in Figure \ref{fig:hd13949} as an example.

\subsection{Stellar Activity}
Astrometry measurements are not in general impacted by stellar activity, with the possible exception of bright inner disk material \citep{blakely_disks_g23h}; however, stellar activity can be seen to inflate the Gaia RV variability around active stars. 
For example, the system AF Lep contains one known, directly imaged planet \citep{AFLep2023,mesa_aflep_2023}. The posterior we present for this system includes an additional inner companion. While the RV variability by itself cannot distinguish these two cases, it is possible that other activity indicators (including those reported by Gaia) may be able to flag such systems.
\begin{figure}
    \centering
    \includegraphics[width=0.32\textwidth]{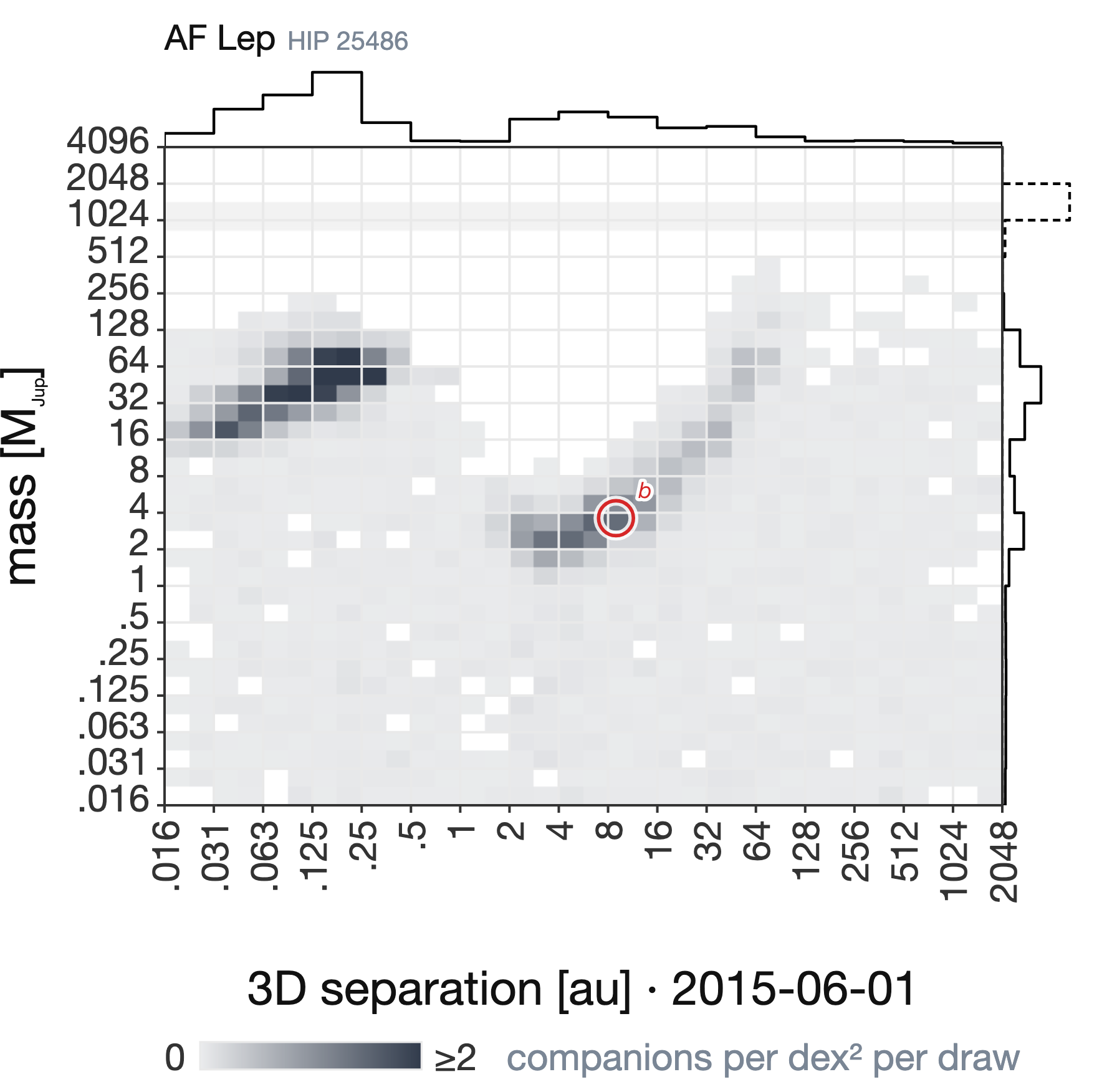}
    \caption{Posterior for AF Lep (HIP25486). The young system has stellar activity that inflates the Gaia RV variability constraint, leading the model to infer a second, inner companion. }
    \label{fig:aflep}
\end{figure}
\subsection{Gaia NSS}
We do not include constraints from the Gaia NSS table as we wished to have a sample with as-uniform constraints and model as possible. Systems with strong within-Gaia detections may have additional orbital constraints available.

\section{Conclusion}

We have presented posteriors that constrain the presence of 0--3 companions around a sample of $10^5$ stars in the Gaia and Hipparcos catalogues, with data consisting of G23H astrometric constraints, augmented with Gaia RV variability, a Gaia catalog search for wide companions, and photometric modelling of the primary or blended system. These posteriors represent the best available constraints for companions around many of these systems, and represent the deepest limits available ahead of Gaia DR4. When companions are present and detected, we find the inferred constraints to be consistent with existing detections in the literature from precision RV, direct imaging, and joint RV---Gaia-Hipparcos proper motion anomaly. 

Due to the inherent limitations in the data available ahead of the DR4 release, planetary massed companions, when detected, have very broad and non-Gaussian uncertainties. While limited in their use for dynamical studies and other applications requiring well known periods, eccentricities, and so-on, they nonetheless are an effective indicator of companion \textit{presence}. It is our goal that the posteriors presented here can be used for downstream tasks such as companion detection and vetting, demographic studies of companion rates, target planning, and preparation for Gaia DR4.  As such, we have endeavoured to describe the selection function, biases, and limitations as extensively as possible.

We structure the model as one that marginalizes over the presence and number of companions, an approach previously presented in \citet{kima} in the context of RV data. This model structure allows one to propagate uncertainty in the \textit{number} of companions, in addition to their orbital properties and correlations, and obviates the need to define a model comparison ladder to prefer a particular number of companions at the catalog level. 

With DR4 poised to release epoch astrometry, RV, and spectra, the quality of the data available to this and related projects will jump considerably. For detected planetary signals, uncertainties will tighten from multiple dex wide to percent levels. That said, for the targets in this sample, the sensitivity to the mere \textit{presence} of a companion may only improve by a modest amount from the increased number of observations and within-Gaia baseline. In some other cases, the best constraints on the presence of long-period accelerators will remain those from the Hipparcos-Gaia inter-mission baseline.

While the G23H astrometric and RV constraints will largely be superseded, the techniques presented in this work will not: the modelling of luminous, potentially blended sources; the matching of astrometric accelerations to observed, neighbouring sources; handling of perspective acceleration, and so on. 
Likewise, the model structure we have developed will remain an effective way to capture the constraints of new Gaia data. 

Similarly, the posteriors we produce for these available pre-DR4 constraints typically possess a more complex structure than what we expect for DR4 data, exhibiting more multi-modality and non-Gaussian parameter distributions. Still, since both these challenges will also be encountered at some level with DR4 epoch measurements, we took this opportunity to consider ways in which companion uncertainties can be propagated and summarized more robustly.

The full posteriors presented for all $10^5$ stars in machine readable form will be made available at the Canadian Astronomy Data Centre upon acceptance. They are also uploaded to the website \url{http://exopo.st/} for interactive access, filtering, and plotting.

\begin{acknowledgments}

This work has made use of data from the European Space Agency (ESA) mission Gaia (\href{https://www.cosmos.esa.int/gaia}{https://www.cosmos.esa.int/gaia}), processed by the Gaia Data Processing and Analysis Consortium (DPAC), \url{https://www.cosmos.esa.int/web/gaia/dpac/consortium}.
This work made use of the Gaia Observation Forecast Tool (GOST).

This research used the facilities of the Canadian Astronomy Data Centre operated by the National Research Council of Canada with the support of the Canadian Space Agency.

This research was enabled in part by support provided by the Digital Research Alliance of Canada (alliancecan.ca).

The Pigeons package's development and collaboration are supported by a Canadian Statistical Sciences Institute Collaborative Research Teams Grant.

This research has made use of the NASA Exoplanet Archive, which is operated by the California Institute of Technology, under contract with the National Aeronautics and Space Administration under the Exoplanet Exploration Program.

J.S. acknowledges financial support from the Natural Sciences and Engineering Research Council of Canada (NSERC) through the Canada Graduate Scholarships Doctoral (CGS D) program, and from the Heising–Simons Foundation through the 51 Pegasi b Fellowship (grant \#2025-5885).
J.W.X is thankful for support from the Heising-Simons Foundation 51 Pegasi b Fellowship (grant \#2025-5887).
D.J.\ is supported by NRC Canada and by an NSERC Discovery Grant.

This work benefited from the 2025 and 2026 Exoplanet Summer Program in the Other Worlds Laboratory (OWL) at the University of California, Santa Cruz, a program funded by the Heising-Simons Foundation and NASA.
\end{acknowledgments}

\facility{Gaia, Hipparcos, Exoplanet Archive}

\software{Octofitter \citep[v8.1.5;][]{octofitter},
PlanetOrbits (v0.11.1),
Pigeons \citep[v0.4.11;][]{pigeons},
Julia \citep[v1.11.3;][]{julia_bezanson}}

\appendix

\section{Simulation-based calibration}\label{sec:sbc}

We perform simulation-based calibration \citep{talts_sbc} to test that the model's simulator and likelihood agree and that the sampler behaves as intended. This does not test whether the likelihood agrees with reality or whether the priors are well chosen, which we consider in Section~\ref{sec:recovery}. We select 1000 real systems from our sample to obtain realistic noise realizations, scan laws, and so on. For each system we draw values of all variables from the priors, including \texttt{n\_planets}, and simulate the corresponding data.

The presence of a real companion can increase the uncertainties of the Gaia and Hipparcos 5-parameter solutions. To avoid selecting stars that have unusually large uncertainties for our simulation, we select hosts from a pool of real stars that are astrometrically quiet
The model is built using each host's real catalog row and Gaia neighbours so that the per star uncertainties, scan law, transit counts, and background sources are realistic. 
The full parameter vector, including the number of companions $N$, all orbit parameters, the latent uncertainty variables, resolved indicators, and priorities vector are then drawn from the prior. 

Every data point in the model is then simulated from the drawn parameters using the same noise model the likelihood assumes: 
catalog astrometry and UEVA, Gaia photometry, and the RV statistic from the non-central $\chi^2$.

We employ a simplification with respect to the Gaia neighbouring source match likelihood. 
We inject a synthetic Gaia source for each active, resolved, in-range stellar companion, but rather than generating a fully synthetic sky,  we merely append the simulated source to the real table of Gaia sources. This maintains the real field contamination we would otherwise have to simulate. Any genuine common-proper-motion companion present in the real catalog remains.

We then sample a posterior for each simulated system and compute the rank of the true, simulated value of each parameter among its posterior draws. Under correct calibration this rank is uniformly distributed \citep{talts_sbc}.

The uniformity of the rank holds for independent posterior draws, whereas each of our posteriors is a single autocorrelated chain of 4,096 draws. For each parameter and simulated system we therefore compute the rank-normalized bulk effective sample size \citep[ESS;][]{vehtari_rhat} of the chain, thin the chain to $L = \lfloor\mathrm{ESS}\rfloor$ evenly spaced draws (with $4 \le L \le 4{,}096$), and convert the rank of the truth among the thinned draws into a randomized probability integral transform,
\begin{equation}
u = \frac{r_< + U\,(r_= + 1)}{L+1}, \qquad U\sim\mathcal{U}(0,1),
\end{equation}
where $r_<$ and $r_=$ count the thinned draws strictly below and tied with the true value. Under correct calibration $u$ is exactly uniform for any $L$, so systems with different ESS can be pooled, and the randomization handles the ties that arise for the discrete variables ($N$ and the resolved indicators). The median ESS is $\approx3{,}900$ for most parameters, lower for the host mass, metallicity and age (median 330, 440 and 510) and $\approx800$ for the transit priorities.

\begin{figure*}
\centering
\includegraphics[width=\textwidth]{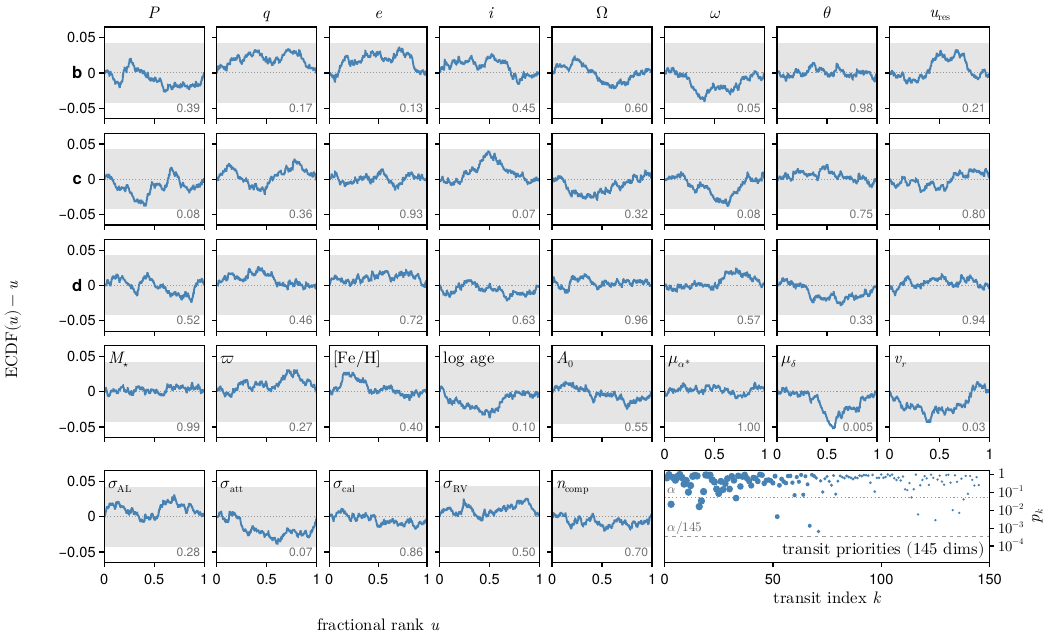}
\caption{Simulation-based calibration: rank-uniformity (ECDF-difference) plots for all free model variables, using ESS-thinned randomized probability-integral-transform ranks. The grey band shows the 95\% simultaneous null band and the number in the lower right of each panel is the Monte Carlo $p$-value. The bottom-right panel summarizes the 145-dimensional \texttt{transit\_priorities} nuisance block as the $p$-value of each dimension against transit index $k$, with marker area increasing with the number of systems that carry that dimension; the dotted and dashed lines mark $\alpha=0.05$ and the Bonferroni threshold $\alpha/145$.}
\label{fig:sbc}
\end{figure*}

Pooling $u$ over the simulated systems, we test each parameter for uniformity with the Kolmogorov--Smirnov statistic, the largest absolute deviation of the empirical CDF of $u$ from the uniform CDF. We obtain its null distribution by Monte Carlo, from 20,000 sets of uniform draws of the same size, so that $p = (n_\ge + 1)/(20{,}001)$, where $n_\ge$ is the number of null sets with a deviation at least as large as the observed one. The grey band in Figure~\ref{fig:sbc} is the 95th percentile of the same null distribution, i.e.\ a 95\% simultaneous band on the ECDF difference. We correct for multiple comparisons in two groups, matching the layout of the figure. The 37 scalar parameters plotted in individual panels form one Bonferroni family. The transit priorities are corrected separately as one group. Parameters that are fixed in a given system are excluded from that system's ranks: $A_0$ for the 119 hosts inside the Local Bubble, where it is fixed to zero, and $\sigma_\mathrm{RV}$ for the 13 hosts without a Gaia RV datum.

Because we include nuisance parameters tied to the number of potential Gaia epochs, which varies from star to star, we compute ranks only for the nuisance parameters present for a given star. The interpretation of this test for the transit-priority nuisances is less clear than for the orbital and other system parameters, since which scan angle and epoch corresponds to which nuisance is star-dependent; we therefore report these separately in Figure~\ref{fig:sbc}. The simulated hosts have between 17 and 145 predicted transits (median 43), and the $k$-th transit priority exists only in systems with at least $k$ transits. While the ESS within each chain is adequate, the number of systems contributing to each test therefore falls from 1,000 for $k\le17$ to fewer than 50 for 70 of the 145 dimensions, and to a single system for the last five, so the tests of the late transit priorities have little power and their $p$-values scatter more widely than those of the early ones.

We find that 2 out of the 37 scalar parameters fall below an uncorrected $p<0.05$ ($\mu_\delta$, $p=0.005$; $v_r$, $p=0.03$), consistent with the 1.85 failures expected at $p<0.05$ applied to 37 largely independent tests. After Bonferroni correction, the smallest adjusted $p$-value is 0.18 ($\mu_\delta$), so no scalar parameter is rejected at a family-wise $\alpha = 0.05$. Among the transit priorities, 12 of the 145 dimensions fall below an uncorrected $p<0.05$ (7.25 expected), the smallest $p$-value is $6.5\times10^{-4}$ ($k=71$, 75 systems), and the group-corrected $p$-value is 0.09, so this group also passes. In a repeated application of this test, the $\approx2$ failures among the scalar parameters occurred on a different pair of variables.

\section{Data-release table description}\label{sec:columns}
Tables~\ref{tab:summary-columns} and \ref{tab:posterior-columns} list every column of the summary table and of the per-star posterior files described in Section~\ref{sec:data-availability}. Both dictionaries are also distributed with the data as machine-readable CSV files.
\startlongtable
\begin{deluxetable*}{llll}
\tabletypesize{\scriptsize}
\tablecaption{Columns of the per-star summary table (\texttt{summary\_table\_v27.csv}; one row per published star, 105,412 rows).\label{tab:summary-columns}}
\tablehead{\colhead{Column} & \colhead{Unit} & \colhead{Type} & \colhead{Description}}
\startdata
\parbox[t]{0.20\textwidth}{\raggedright \texttt{hip}} & \nodata & int & \parbox[t]{0.60\textwidth}{\raggedright Hipparcos identifier. The star name in the paper and on the website is HIP$<$hip$>$.} \\
\parbox[t]{0.20\textwidth}{\raggedright \texttt{gaia\_dr3\_source\_id}} & \nodata & int & \parbox[t]{0.60\textwidth}{\raggedright Gaia DR3 source\_id of the fitted source (G23H cross-match).} \\
\parbox[t]{0.20\textwidth}{\raggedright \texttt{gaia\_dr2\_source\_id}} & \nodata & int & \parbox[t]{0.60\textwidth}{\raggedright Gaia DR2 source\_id whose position and proper motion supplied the DR2 data (G23H cross-match). The cross-match is not one-to-one: 1,291 DR3 sources have two DR2 candidates in G23H (the first row was used); the wrong-component cases were removed from the release by the DR2 pre-filter.} \\
\parbox[t]{0.20\textwidth}{\raggedright \texttt{posterior\_file}} & \nodata & str & \parbox[t]{0.60\textwidth}{\raggedright File name of the per-star posterior FITS file in the data release.} \\
\parbox[t]{0.20\textwidth}{\raggedright \texttt{ra}} & deg & float & \parbox[t]{0.60\textwidth}{\raggedright Gaia DR3 right ascension (ICRS, epoch J2016.0).} \\
\parbox[t]{0.20\textwidth}{\raggedright \texttt{dec}} & deg & float & \parbox[t]{0.60\textwidth}{\raggedright Gaia DR3 declination (ICRS, epoch J2016.0).} \\
\parbox[t]{0.20\textwidth}{\raggedright \texttt{parallax}} & mas & float & \parbox[t]{0.60\textwidth}{\raggedright Gaia DR3 catalog parallax (prior mean; not zero-point corrected).} \\
\parbox[t]{0.20\textwidth}{\raggedright \texttt{parallax\_error}} & mas & float & \parbox[t]{0.60\textwidth}{\raggedright Gaia DR3 catalog parallax uncertainty.} \\
\parbox[t]{0.20\textwidth}{\raggedright \texttt{pmra}} & mas/yr & float & \parbox[t]{0.60\textwidth}{\raggedright Gaia DR3 catalog proper motion in RA (pmra*cos(dec)).} \\
\parbox[t]{0.20\textwidth}{\raggedright \texttt{pmra\_error}} & mas/yr & float & \parbox[t]{0.60\textwidth}{\raggedright Gaia DR3 catalog pmra uncertainty.} \\
\parbox[t]{0.20\textwidth}{\raggedright \texttt{pmdec}} & mas/yr & float & \parbox[t]{0.60\textwidth}{\raggedright Gaia DR3 catalog proper motion in Dec.} \\
\parbox[t]{0.20\textwidth}{\raggedright \texttt{pmdec\_error}} & mas/yr & float & \parbox[t]{0.60\textwidth}{\raggedright Gaia DR3 catalog pmdec uncertainty.} \\
\parbox[t]{0.20\textwidth}{\raggedright \texttt{radial\_velocity}} & km/s & float & \parbox[t]{0.60\textwidth}{\raggedright Gaia DR3 catalog radial velocity (empty when DR3 has none).} \\
\parbox[t]{0.20\textwidth}{\raggedright \texttt{radial\_velocity\_error}} & km/s & float & \parbox[t]{0.60\textwidth}{\raggedright Gaia DR3 catalog radial velocity uncertainty.} \\
\parbox[t]{0.20\textwidth}{\raggedright \texttt{phot\_g\_mean\_mag}} & mag & float & \parbox[t]{0.60\textwidth}{\raggedright Gaia DR3 mean G magnitude.} \\
\parbox[t]{0.20\textwidth}{\raggedright \texttt{bp\_rp}} & mag & float & \parbox[t]{0.60\textwidth}{\raggedright Gaia DR3 BP-RP colour.} \\
\parbox[t]{0.20\textwidth}{\raggedright \texttt{ruwe}} & \nodata & float & \parbox[t]{0.60\textwidth}{\raggedright Gaia DR3 renormalised unit weight error.} \\
\parbox[t]{0.20\textwidth}{\raggedright \texttt{n\_transits\_dr3}} & \nodata & int & \parbox[t]{0.60\textwidth}{\raggedright Gaia DR3 astrometric\_matched\_transits; sets the number of simulated transits and hence the number of nuisance columns in the chain.} \\
\parbox[t]{0.20\textwidth}{\raggedright \texttt{rv\_nb\_transits}} & \nodata & int & \parbox[t]{0.60\textwidth}{\raggedright Gaia DR3 rv\_nb\_transits.} \\
\parbox[t]{0.20\textwidth}{\raggedright \texttt{data\_label}} & \nodata & str & \parbox[t]{0.60\textwidth}{\raggedright Data included in the fit, as a + joined list: hip = Hipparcos position (1991.25), hg = Hipparcos-Gaia mean proper motion, dr3 = Gaia DR3 position/PM/UEVA, dr2 + dr32 = Gaia DR2 position and DR2-DR3 positional PM (present unless rho\_dr2\_dr3 $\ge$ 0.95), ueva = DR3 UEVA datum, rv = DR3 RV-variability datum. The Hipparcos IAD were never included.} \\
\parbox[t]{0.20\textwidth}{\raggedright \texttt{used\_rv}} & \nodata & bool & \parbox[t]{0.60\textwidth}{\raggedright True when the Gaia DR3 RV-variability datum was fit (requires finite rv\_ln\_uncert; 537 stars have a catalog RV but no usable variability statistic).} \\
\parbox[t]{0.20\textwidth}{\raggedright \texttt{used\_dr2\_dr32}} & \nodata & bool & \parbox[t]{0.60\textwidth}{\raggedright True when the DR2 position and DR2-DR3 positional PM data were fit; False for the stars whose G23H DR2-DR3 error correlation was $\ge$ 0.95 at job preparation (3,803 of the 106,508 fitted stars).} \\
\parbox[t]{0.20\textwidth}{\raggedright \texttt{hip\_annex\_flag}} & \nodata & str & \parbox[t]{0.60\textwidth}{\raggedright Hipparcos Double and Multiple Systems Annex solution type (field H59): C component, G acceleration, O orbital, V variability-induced mover, X stochastic; empty when the star has a standard 5-parameter solution.} \\
\parbox[t]{0.20\textwidth}{\raggedright \texttt{hip\_annex\_co}} & \nodata & bool & \parbox[t]{0.60\textwidth}{\raggedright True when hip\_annex\_flag is C or O (a resolved-component or orbital solution: a binary by construction). 8,497 stars; removed from the recommended sample.} \\
\parbox[t]{0.20\textwidth}{\raggedright \texttt{hip\_annex\_gvx}} & \nodata & bool & \parbox[t]{0.60\textwidth}{\raggedright True when hip\_annex\_flag is G, V or X. 3,813 stars. Kept in the recommended sample and carried as an optional cut.} \\
\parbox[t]{0.20\textwidth}{\raggedright \texttt{recommended}} & \nodata & bool & \parbox[t]{0.60\textwidth}{\raggedright Recommended sample of the paper: not hip\_annex\_co. 96,915 stars.} \\
\parbox[t]{0.20\textwidth}{\raggedright \texttt{clean\_hipparcos}} & \nodata & bool & \parbox[t]{0.60\textwidth}{\raggedright Clean Hipparcos subset of the paper: recommended and not hip\_annex\_gvx. 93,102 stars.} \\
\parbox[t]{0.20\textwidth}{\raggedright \texttt{cpm\_white\_dwarf}} & \nodata & bool & \parbox[t]{0.60\textwidth}{\raggedright True when a resolved common-proper-motion companion attributed to a Gaia DR3 source in the posterior is a white dwarf in the Gaia EDR3 white dwarf catalog of Gentile Fusillo et al. 2021 (P\_WD $\ge$ 0.5). The model admits only main-sequence companions, so the fit is unreliable for these systems (sec:limitations). Kept in the recommended sample and carried as an optional cut.} \\
\parbox[t]{0.20\textwidth}{\raggedright \texttt{p\_n\_ge1}} & \nodata & float & \parbox[t]{0.60\textwidth}{\raggedright Posterior probability of at least one companion, P(N $\ge$ 1) = 1 - fraction of draws with n\_planets = 0.} \\
\parbox[t]{0.20\textwidth}{\raggedright \texttt{p\_n\_ge2}} & \nodata & float & \parbox[t]{0.60\textwidth}{\raggedright Posterior probability of at least two companions.} \\
\parbox[t]{0.20\textwidth}{\raggedright \texttt{p\_n\_ge3}} & \nodata & float & \parbox[t]{0.60\textwidth}{\raggedright Posterior probability of three companions (the model maximum).} \\
\parbox[t]{0.20\textwidth}{\raggedright \texttt{n\_planets\_mean}} & \nodata & float & \parbox[t]{0.60\textwidth}{\raggedright Posterior mean of n\_planets.} \\
\parbox[t]{0.20\textwidth}{\raggedright \texttt{log10\_odds\_n\_ge1}} & \nodata & float & \parbox[t]{0.60\textwidth}{\raggedright log10 posterior odds P(N$\ge$1)/P(N=0), both clipped to [1/4096, 1-1/4096]. The prior odds are exactly 1, so this equals the log10 Bayes factor for N $\ge$ 1 versus N = 0.} \\
\parbox[t]{0.20\textwidth}{\raggedright \texttt{m\_pri\_50}} & M$_\odot$ & float & \parbox[t]{0.60\textwidth}{\raggedright Posterior median host-star mass.} \\
\parbox[t]{0.20\textwidth}{\raggedright \texttt{m\_pri\_05}} & M$_\odot$ & float & \parbox[t]{0.60\textwidth}{\raggedright Posterior 5th percentile host-star mass.} \\
\parbox[t]{0.20\textwidth}{\raggedright \texttt{m\_pri\_95}} & M$_\odot$ & float & \parbox[t]{0.60\textwidth}{\raggedright Posterior 95th percentile host-star mass.} \\
\parbox[t]{0.20\textwidth}{\raggedright \texttt{plx\_50}} & mas & float & \parbox[t]{0.60\textwidth}{\raggedright Posterior median parallax.} \\
\parbox[t]{0.20\textwidth}{\raggedright \texttt{plx\_05}} & mas & float & \parbox[t]{0.60\textwidth}{\raggedright Posterior 5th percentile parallax.} \\
\parbox[t]{0.20\textwidth}{\raggedright \texttt{plx\_95}} & mas & float & \parbox[t]{0.60\textwidth}{\raggedright Posterior 95th percentile parallax.} \\
\parbox[t]{0.20\textwidth}{\raggedright \texttt{pmra\_50}} & mas/yr & float & \parbox[t]{0.60\textwidth}{\raggedright Posterior median barycentric proper motion in RA.} \\
\parbox[t]{0.20\textwidth}{\raggedright \texttt{pmdec\_50}} & mas/yr & float & \parbox[t]{0.60\textwidth}{\raggedright Posterior median barycentric proper motion in Dec.} \\
\parbox[t]{0.20\textwidth}{\raggedright \texttt{log\_age\_50}} & $\log_{10}$(yr) & float & \parbox[t]{0.60\textwidth}{\raggedright Posterior median log10 age.} \\
\parbox[t]{0.20\textwidth}{\raggedright \texttt{fe\_h\_50}} & dex & float & \parbox[t]{0.60\textwidth}{\raggedright Posterior median metallicity [Fe/H].} \\
\parbox[t]{0.20\textwidth}{\raggedright \texttt{a\_0\_50}} & mag & float & \parbox[t]{0.60\textwidth}{\raggedright Posterior median extinction A\_0 (0 for stars inside the Local Bubble, where it is fixed).} \\
\parbox[t]{0.20\textwidth}{\raggedright \texttt{m\_comp\_total\_50}} & M$_\mathrm{Jup}$ & float & \parbox[t]{0.60\textwidth}{\raggedright Posterior median of the summed mass of all active companions, in Jupiter masses (0 when N = 0).} \\
\parbox[t]{0.20\textwidth}{\raggedright \texttt{loglike\_50}} & \nodata & float & \parbox[t]{0.60\textwidth}{\raggedright Posterior median log-likelihood.} \\
\parbox[t]{0.20\textwidth}{\raggedright \texttt{log\_evidence\_ratio}} & \nodata & float & \parbox[t]{0.60\textwidth}{\raggedright Pigeons stepping-stone log(Z\_1/Z\_0) after the final round (log evidence relative to the reference distribution).} \\
\parbox[t]{0.20\textwidth}{\raggedright \texttt{global\_barrier}} & \nodata & float & \parbox[t]{0.60\textwidth}{\raggedright Pigeons global communication barrier after the final round.} \\
\parbox[t]{0.20\textwidth}{\raggedright \texttt{n\_restarts\_final\_round}} & \nodata & int & \parbox[t]{0.60\textwidth}{\raggedright Number of tempered restarts in the final Pigeons round (4,096 scans).} \\
\parbox[t]{0.20\textwidth}{\raggedright \texttt{walltime\_final\_round\_s}} & s & float & \parbox[t]{0.60\textwidth}{\raggedright Wall time of the final Pigeons round.} \\
\enddata
\end{deluxetable*}

\startlongtable
\begin{deluxetable*}{llll}
\tabletypesize{\scriptsize}
\tablecaption{Columns of the \texttt{POSTERIOR} extension of each per-star FITS file (one row per draw, 4,096 rows). \texttt{$<$k$>$} stands for a companion slot \texttt{b}, \texttt{c} or \texttt{d}; \texttt{$<$i$>$} for a 1-based index. The number of transit-indexed columns varies from star to star and is given by the header cards \texttt{NTRANSIT} (predicted DR3 scans), and by \texttt{n\_transits\_dr3} and \texttt{rv\_nb\_transits} of the summary table. Angles are in radians; epochs are MJD. ``Type'' is the storage type: integer-valued quantities are packed as \texttt{uint8}/\texttt{int16}, the log-densities, epochs and coordinates are kept in \texttt{float64}, and everything else is stored as \texttt{float32}.\label{tab:posterior-columns}}
\tablehead{\colhead{Column} & \colhead{Unit} & \colhead{Type} & \colhead{Description}}
\startdata
\parbox[t]{0.22\textwidth}{\raggedright \texttt{iteration}} & \nodata & int16 & \parbox[t]{0.56\textwidth}{\raggedright Draw index, 1..4096.} \\
\parbox[t]{0.22\textwidth}{\raggedright \texttt{chain}} & \nodata & uint8 & \parbox[t]{0.56\textwidth}{\raggedright Chain index (always 1: the 4,096 draws are the final Pigeons round of the single target chain).} \\
\parbox[t]{0.22\textwidth}{\raggedright \texttt{loglike}} & \nodata & float64 & \parbox[t]{0.56\textwidth}{\raggedright Log-likelihood of the draw (all data terms).} \\
\parbox[t]{0.22\textwidth}{\raggedright \texttt{logprior}} & \nodata & float64 & \parbox[t]{0.56\textwidth}{\raggedright Log-prior density of the draw, including the discrete n\_planets prior.} \\
\parbox[t]{0.22\textwidth}{\raggedright \texttt{logpost}} & \nodata & float64 & \parbox[t]{0.56\textwidth}{\raggedright Log-posterior density = loglike + logprior.} \\
\parbox[t]{0.22\textwidth}{\raggedright \texttt{pigeons\_logpotential}} & \nodata & float64 & \parbox[t]{0.56\textwidth}{\raggedright Pigeons log-potential at the target (beta = 1); equals logpost.} \\
\parbox[t]{0.22\textwidth}{\raggedright \texttt{M\_pri}} & M$_\odot$ & float32 & \parbox[t]{0.56\textwidth}{\raggedright Host-star mass.} \\
\parbox[t]{0.22\textwidth}{\raggedright \texttt{fe\_h}} & dex & float32 & \parbox[t]{0.56\textwidth}{\raggedright Host-star metallicity [Fe/H].} \\
\parbox[t]{0.22\textwidth}{\raggedright \texttt{log\_age}} & $\log_{10}$(yr) & float32 & \parbox[t]{0.56\textwidth}{\raggedright Host-star (system) log10 age.} \\
\parbox[t]{0.22\textwidth}{\raggedright \texttt{n\_planets}} & \nodata & uint8 & \parbox[t]{0.56\textwidth}{\raggedright Number of active companions N in this draw, 0..3.} \\
\parbox[t]{0.22\textwidth}{\raggedright \texttt{P\_$<$k$>$}} & yr & float32 & \parbox[t]{0.56\textwidth}{\raggedright Orbital period of slot $<$k$>$. Sampled for every slot; only meaningful when the slot is active (pp\_$<$k$>$ = 1).} \\
\parbox[t]{0.22\textwidth}{\raggedright \texttt{q\_$<$k$>$}} & \nodata & float32 & \parbox[t]{0.56\textwidth}{\raggedright Companion-to-host mass ratio of slot $<$k$>$. Sampled for every slot; meaningful only when active.} \\
\parbox[t]{0.22\textwidth}{\raggedright \texttt{plx}} & mas & float32 & \parbox[t]{0.56\textwidth}{\raggedright System parallax.} \\
\parbox[t]{0.22\textwidth}{\raggedright \texttt{A\_0}} & mag & float32 & \parbox[t]{0.56\textwidth}{\raggedright Monochromatic interstellar extinction at 541.4 nm; fixed at 0 for stars inside the Local Bubble (header A0\_FIX).} \\
\parbox[t]{0.22\textwidth}{\raggedright \texttt{i\_$<$k$>$}} & rad & float32 & \parbox[t]{0.56\textwidth}{\raggedright Orbital inclination of slot $<$k$>$.} \\
\parbox[t]{0.22\textwidth}{\raggedright \texttt{Omega\_$<$k$>$}} & rad & float32 & \parbox[t]{0.56\textwidth}{\raggedright Longitude of the ascending node of slot $<$k$>$ (the raw sampler output names this column in LaTeX).} \\
\parbox[t]{0.22\textwidth}{\raggedright \texttt{theta\_$<$k$>$}} & rad & float32 & \parbox[t]{0.56\textwidth}{\raggedright Orbital-phase parameter of slot $<$k$>$: the position angle at the mean epoch of the Gaia DR3 scans used in the fit. Use tp\_$<$k$>$ for the epoch of periastron.} \\
\parbox[t]{0.22\textwidth}{\raggedright \texttt{e\_$<$k$>$}} & \nodata & float32 & \parbox[t]{0.56\textwidth}{\raggedright Orbital eccentricity of slot $<$k$>$.} \\
\parbox[t]{0.22\textwidth}{\raggedright \texttt{omega\_$<$k$>$}} & rad & float32 & \parbox[t]{0.56\textwidth}{\raggedright Argument of periastron of slot $<$k$>$ (the raw sampler output names this column in LaTeX).} \\
\parbox[t]{0.22\textwidth}{\raggedright \texttt{u\_resolved\_$<$k$>$}} & \nodata & float32 & \parbox[t]{0.56\textwidth}{\raggedright Uniform(0,1) auxiliary variable that realises the Bernoulli resolved\_$<$k$>$ draw.} \\
\parbox[t]{0.22\textwidth}{\raggedright \texttt{pmra}} & mas/yr & float32 & \parbox[t]{0.56\textwidth}{\raggedright Barycentric proper motion in RA (pmra*cos(dec)) of the system.} \\
\parbox[t]{0.22\textwidth}{\raggedright \texttt{pmdec}} & mas/yr & float32 & \parbox[t]{0.56\textwidth}{\raggedright Barycentric proper motion in Dec of the system.} \\
\parbox[t]{0.22\textwidth}{\raggedright \texttt{rv}} & m/s & float32 & \parbox[t]{0.56\textwidth}{\raggedright Barycentric radial velocity of the system (enters through perspective acceleration).} \\
\parbox[t]{0.22\textwidth}{\raggedright \texttt{pp\_$<$k$>$}} & \nodata & uint8 & \parbox[t]{0.56\textwidth}{\raggedright Presence indicator of slot $<$k$>$: 1 when n\_planets $\ge$ slot index (b = 1, c = 2, d = 3).} \\
\parbox[t]{0.22\textwidth}{\raggedright \texttt{mass\_prime\_$<$k$>$}} & M$_\mathrm{Jup}$ & float32 & \parbox[t]{0.56\textwidth}{\raggedright Companion mass of slot $<$k$>$ before the presence gate, q\_$<$k$>$ * M\_pri, regardless of whether the slot is active (the raw sampler output names this column in LaTeX).} \\
\parbox[t]{0.22\textwidth}{\raggedright \texttt{mass\_$<$k$>$}} & M$_\mathrm{Jup}$ & float32 & \parbox[t]{0.56\textwidth}{\raggedright Companion mass of slot $<$k$>$: mass\_prime\_$<$k$>$ when active, exactly 0 when inactive.} \\
\parbox[t]{0.22\textwidth}{\raggedright \texttt{M\_total\_$<$k$>$}} & M$_\odot$ & float32 & \parbox[t]{0.56\textwidth}{\raggedright Total mass governing the orbit of slot $<$k$>$: M\_pri plus the masses of the active companions with periods shorter than P\_$<$k$>$ (the companions are nested Keplerians).} \\
\parbox[t]{0.22\textwidth}{\raggedright \texttt{a\_$<$k$>$}} & au & float32 & \parbox[t]{0.56\textwidth}{\raggedright Semi-major axis of slot $<$k$>$ from Kepler's third law with P\_$<$k$>$ and M\_total\_$<$k$>$.} \\
\parbox[t]{0.22\textwidth}{\raggedright \texttt{tp\_$<$k$>$}} & d & float64 & \parbox[t]{0.56\textwidth}{\raggedright Epoch of periastron of slot $<$k$>$ (MJD), derived from theta\_$<$k$>$ and the orbital elements.} \\
\parbox[t]{0.22\textwidth}{\raggedright \texttt{pdet\_$<$k$>$}} & \nodata & float32 & \parbox[t]{0.56\textwidth}{\raggedright Probability that the companion in slot $<$k$>$ would be resolved as a separate Gaia DR3 source, from the contrast-sensitivity model at the J2016.0 projected separation; 0 for inactive or sub-stellar slots.} \\
\parbox[t]{0.22\textwidth}{\raggedright \texttt{resolved\_$<$k$>$}} & \nodata & uint8 & \parbox[t]{0.56\textwidth}{\raggedright Realised resolved indicator of slot $<$k$>$: 1 when the companion is modelled as a separate Gaia source (no photocentre dilution in the DR3 astrometry; must match a catalogued neighbour), 0 when blended.} \\
\parbox[t]{0.22\textwidth}{\raggedright \texttt{fluxratio\_$<$i$>$}} & \nodata & float32 & \parbox[t]{0.56\textwidth}{\raggedright G-band flux ratio companion/host of slot $<$i$>$ (1 = b, 2 = c, 3 = d) as applied to the Gaia photocentre; 0 when the slot is inactive or resolved.} \\
\parbox[t]{0.22\textwidth}{\raggedright \texttt{fluxratio\_hip\_$<$i$>$}} & \nodata & float32 & \parbox[t]{0.56\textwidth}{\raggedright Hp-band flux ratio companion/host of slot $<$i$>$ as applied to the Hipparcos photocentre; 0 when inactive.} \\
\parbox[t]{0.22\textwidth}{\raggedright \texttt{ref\_epoch}} & d & float64 & \parbox[t]{0.56\textwidth}{\raggedright Gaia DR3 reference epoch, MJD 57388.5 (J2016.0); constant.} \\
\parbox[t]{0.22\textwidth}{\raggedright \texttt{ra}} & deg & float64 & \parbox[t]{0.56\textwidth}{\raggedright Gaia DR3 catalog right ascension at J2016.0 (constant; the fitted position offsets are internal to the likelihood).} \\
\parbox[t]{0.22\textwidth}{\raggedright \texttt{dec}} & deg & float64 & \parbox[t]{0.56\textwidth}{\raggedright Gaia DR3 catalog declination at J2016.0 (constant).} \\
\parbox[t]{0.22\textwidth}{\raggedright \texttt{G23H\_sigma\_AL}} & mas & float32 & \parbox[t]{0.56\textwidth}{\raggedright Gaia along-scan per-transit noise (UEVA calibration nuisance parameter).} \\
\parbox[t]{0.22\textwidth}{\raggedright \texttt{G23H\_sigma\_att}} & mas & float32 & \parbox[t]{0.56\textwidth}{\raggedright Gaia attitude noise term (UEVA calibration nuisance parameter).} \\
\parbox[t]{0.22\textwidth}{\raggedright \texttt{G23H\_sigma\_calib}} & mas & float32 & \parbox[t]{0.56\textwidth}{\raggedright Gaia calibration noise term (UEVA calibration nuisance parameter).} \\
\parbox[t]{0.22\textwidth}{\raggedright \texttt{G23H\_transit\_priorities\_$<$i$>$}} & \nodata & float32 & \parbox[t]{0.56\textwidth}{\raggedright Standard-normal nuisance variable for predicted Gaia transit $<$i$>$; the ranking of these values selects which predicted transits are treated as matched (see G23H\_transits\_$<$i$>$). Count = NTRANSIT (predicted DR3 scans) and varies per star.} \\
\parbox[t]{0.22\textwidth}{\raggedright \texttt{G23H\_sigma\_rv\_per\_transit}} & km/s & float32 & \parbox[t]{0.56\textwidth}{\raggedright Per-transit RV scatter of the DR3 RV-variability datum (nuisance parameter; present only when USED\_RV = T).} \\
\parbox[t]{0.22\textwidth}{\raggedright \texttt{G23H\_u\_dup\_dr2}} & \nodata & float32 & \parbox[t]{0.56\textwidth}{\raggedright Uniform(0,1) auxiliary variable that marginalizes over the fraction of DR2 transits that are duplicated in the Gaia solution. Present only for bright stars where duplication is observed.} \\
\parbox[t]{0.22\textwidth}{\raggedright \texttt{G23H\_fluxratio\_$<$i$>$}} & \nodata & float32 & \parbox[t]{0.56\textwidth}{\raggedright Copy of fluxratio\_$<$i$>$ as seen by the G23H likelihood.} \\
\parbox[t]{0.22\textwidth}{\raggedright \texttt{G23H\_fluxratio\_hip\_$<$i$>$}} & \nodata & float32 & \parbox[t]{0.56\textwidth}{\raggedright Copy of fluxratio\_hip\_$<$i$>$ as seen by the G23H likelihood.} \\
\parbox[t]{0.22\textwidth}{\raggedright \texttt{G23H\_transits\_$<$i$>$}} & \nodata & int16 & \parbox[t]{0.56\textwidth}{\raggedright Index (into the predicted DR3 scan list) of the $<$i$>$-th selected DR3 transit in this draw; count = astrometric\_matched\_transits (n\_transits\_dr3 in the summary table).} \\
\parbox[t]{0.22\textwidth}{\raggedright \texttt{G23H\_transits\_dr2\_$<$i$>$}} & \nodata & int16 & \parbox[t]{0.56\textwidth}{\raggedright Index of the $<$i$>$-th selected DR2 transit; count = the DR2 matched-transit number. Absent when USED\_DR2 = F.} \\
\parbox[t]{0.22\textwidth}{\raggedright \texttt{G23H\_transits\_rv\_$<$i$>$}} & \nodata & int16 & \parbox[t]{0.56\textwidth}{\raggedright Index of the $<$i$>$-th selected RVS transit; count = rv\_nb\_transits. Present only when USED\_RV = T.} \\
\parbox[t]{0.22\textwidth}{\raggedright \texttt{$<$k$>$\_planet\_present}} & \nodata & uint8 & \parbox[t]{0.56\textwidth}{\raggedright Alias of pp\_$<$k$>$.} \\
\parbox[t]{0.22\textwidth}{\raggedright \texttt{$<$k$>$\_P}} & yr & float32 & \parbox[t]{0.56\textwidth}{\raggedright Alias of P\_$<$k$>$.} \\
\parbox[t]{0.22\textwidth}{\raggedright \texttt{$<$k$>$\_q}} & \nodata & float32 & \parbox[t]{0.56\textwidth}{\raggedright Alias of q\_$<$k$>$.} \\
\parbox[t]{0.22\textwidth}{\raggedright \texttt{$<$k$>$\_mass}} & M$_\mathrm{Jup}$ & float32 & \parbox[t]{0.56\textwidth}{\raggedright Alias of mass\_$<$k$>$.} \\
\parbox[t]{0.22\textwidth}{\raggedright \texttt{$<$k$>$\_mass\_prime}} & M$_\mathrm{Jup}$ & float32 & \parbox[t]{0.56\textwidth}{\raggedright Alias of mass\_prime\_$<$k$>$.} \\
\parbox[t]{0.22\textwidth}{\raggedright \texttt{$<$k$>$\_M}} & M$_\odot$ & float32 & \parbox[t]{0.56\textwidth}{\raggedright Alias of M\_total\_$<$k$>$.} \\
\parbox[t]{0.22\textwidth}{\raggedright \texttt{$<$k$>$\_a}} & au & float32 & \parbox[t]{0.56\textwidth}{\raggedright Alias of a\_$<$k$>$.} \\
\parbox[t]{0.22\textwidth}{\raggedright \texttt{$<$k$>$\_i}} & rad & float32 & \parbox[t]{0.56\textwidth}{\raggedright Alias of i\_$<$k$>$.} \\
\parbox[t]{0.22\textwidth}{\raggedright \texttt{$<$k$>$\_Omega}} & rad & float32 & \parbox[t]{0.56\textwidth}{\raggedright Alias of Omega\_$<$k$>$.} \\
\parbox[t]{0.22\textwidth}{\raggedright \texttt{$<$k$>$\_theta}} & rad & float32 & \parbox[t]{0.56\textwidth}{\raggedright Alias of theta\_$<$k$>$.} \\
\parbox[t]{0.22\textwidth}{\raggedright \texttt{$<$k$>$\_e}} & \nodata & float32 & \parbox[t]{0.56\textwidth}{\raggedright Alias of e\_$<$k$>$.} \\
\parbox[t]{0.22\textwidth}{\raggedright \texttt{$<$k$>$\_omega}} & rad & float32 & \parbox[t]{0.56\textwidth}{\raggedright Alias of omega\_$<$k$>$.} \\
\parbox[t]{0.22\textwidth}{\raggedright \texttt{$<$k$>$\_tp}} & d & float64 & \parbox[t]{0.56\textwidth}{\raggedright Alias of tp\_$<$k$>$.} \\
\enddata
\end{deluxetable*}

\bibliography{ms}{}
\bibliographystyle{aasjournalv7}

\end{document}